\documentclass[journal,11pt,twocolumn]{IEEEtran}
\usepackage[cmex10]{amsmath}
\usepackage{cite}
\usepackage{amssymb}
\usepackage{stackrel}
\usepackage{subcaption}

\usepackage[table]{xcolor}
\usepackage[pdftex]{graphicx}

\usepackage{array}
\usepackage{tabulary}

\usepackage{chemarrow}
\usepackage{color}
\usepackage{algorithmic}
\usepackage{algorithm}
\usepackage{clrscode}
\usepackage{amsmath, amssymb, mathrsfs, amsfonts, amsthm}
\usepackage{epsfig, epstopdf}
\usepackage{multirow}

\allowbreak
\usepackage{caption}
\usepackage{stfloats}
\usepackage{enumerate}
\usepackage{float}
\usepackage{enumitem}
\usepackage{hyperref}

\hypersetup{colorlinks=true,citecolor=black,linkcolor=black, urlcolor=black}

\newcolumntype{K}[1]{>{\centering\arraybackslash}p{#1}}
\usepackage[top=0.75in,bottom=1in,left=0.625in,right=0.625in]{geometry}
\usepackage{tikz}
\usetikzlibrary{%
  calc,
  arrows,%
  shapes.misc,% wg. rounded rectangle
  shapes.arrows,%
  chains,%
  matrix,%
  positioning,% wg. " of "
  scopes,%
  decorations.pathmorphing,% /pgf/decoration/random steps | erste Graphik
  shadows,%
  decorations.text
}

\newcommand{\vect}[1]{\boldsymbol{#1}}

\usepackage{longtable}
\definecolor{light-gray}{HTML}{E5E4E2}
\definecolor{light-cyan}{HTML}{E0FFFF}
\definecolor{light-green}{HTML}{d1e4d0}
\newtheorem{exmp}{Example}

\usepackage{tikz}
\usetikzlibrary{shapes,snakes,positioning,shapes.misc}

\usepackage{siunitx}

\usepackage{makecell}
\theoremstyle{definition}

\usepackage{chemformula}
\usepackage{chemfig}
\usepackage{chemmacros}

\usepackage{tikz}
\usetikzlibrary{mindmap}
\usepackage{lipsum,adjustbox}

\usetikzlibrary{arrows}
\usetikzlibrary{through}

\usepackage{graphicx}
\newcommand\Mark[1]{\textsuperscript#1}
\newcommand\blfootnote[1]{%
  \begingroup
  \renewcommand\thefootnote{}\footnote{#1}%
  \addtocounter{footnote}{-1}%
  \endgroup
}

\date{}
\title{Abnormality Detection and Localization Schemes using Molecular Communication Systems: A Survey}

\author{\IEEEauthorblockN
{Ali Etemadi\Mark{1}\IEEEauthorrefmark{1}, Maryam Farahnak-Ghazani\Mark{1}\IEEEauthorrefmark{1}, Hamidreza Arjmandi\Mark{2}, Mahtab Mirmohseni\Mark{3}, \\ and Masoumeh Nasiri-Kenari\Mark{1}
 \\
}
\IEEEauthorblockA{\Mark{1}Sharif University of Technology, \Mark{2}University of Warwick}, \Mark{3}University of Surrey}
\begin{document}
	\maketitle

%%%%%%%%%%%%%%%%%%%%%%%%%%%%%%%%%%%%%%%%%555

\begin{abstract}\blfootnote{\IEEEauthorrefmark{1}Co-first authors}
Abnormality detection and localization (ADL) have been studied widely in wireless sensor networks (WSNs) literature, where the sensors use electromagnetic waves for communication. Molecular communication (MC) has been introduced as an alternative approach for ADL in particular areas such as healthcare, being able to tackle the shortcomings of conventional WSNs, such as invasiveness, bio-incompatibility, and high energy consumption. In this paper, we introduce a general framework for MC-based ADL, which consists of multiple tiers for sensing the abnormality and communication between different agents, including the sensors, the fusion center (FC), the gateway (GW), and the external node (e.g., a local cloud), and describe each tier and the agents in this framework. We classify and explain different abnormality recognition methods, the functional units of the sensors, and different sensor features. Further, we describe different types of interfaces required for converting the internal and external signals at the FC and GW. Moreover, we present a unified channel model for the sensing and communication links. We categorize the MC-based abnormality detection schemes based on the sensor mobility, cooperative detection, and cooperative sensing/activation. We also classify the localization approaches based on the sensor mobility and propulsion mechanisms and present a general framework for the externally-controllable localization systems. Finally, we present some challenges and future research directions to realize and develop MC-based systems for ADL. The important challenges in the MC-based systems lie in four main directions as implementation, system design, modeling, and methods, which need considerable attention from multidisciplinary perspectives.
\end{abstract} 
	
%%%%%%%%%%%%%%%%%%%%%%%%%%%%
%%%%%%%%%%%%%%%%%%%%%%%%%%%%
	
%%%   Authors: Dr. Farahnak	

\section{Introduction}  
Abnormality detection and localization (ADL)
have been studied widely and have established literatures in different application areas such as fraud detection in credit cards, healthcare,  insurance \cite{abdallah2016fraud}, intrusion detection in cyber security \cite{khraisat2019survey}, fault detection in safety-critical systems \cite{miljkovic2011fault}, 
injection  and denial-of-service attacks in networks \cite{thatte2008detection,thatte2010parametric, chattopadhyay2018attack}, detection of physical entity malfunctions in smart grids \cite{levorato2012fast}, underwater target detection and localization \cite{hu2020decentralized, chen2017underwater}, cancer detection and localization in medicine \cite{sanderson1975early, islam2013survey}, 
and odor source detection and localization \cite{matthes2005source,kuzu2008survey}. 

\subsection{Motivation}
Abnormality, defined as any abnormal feature in the system, is also referred to as \emph{anomaly}, \emph{target}, \emph{outlier}, \emph{exception}, \emph{surprise}, \emph{contaminant} and \emph{peculiarity} in different domains \cite{chandola2009anomaly}.
In one hand, due to the emergence of the Internet of things (IoT) and artificial intelligence (AI) into many applications in industry, education, and medicine, resulting in a large amount of data be gathered from the devices, the systems are more exposed to the abnormalities, and  the ADL have become even more important problems \cite{zhu2021anomaly}. On the other hand, the new advances in technology have paved the way for new techniques to address the problem.

Conventionally, wireless sensor networks (WSNs) are used to detect and localize the abnormality in a variety of healthcare, environmental, and industrial applications \cite{xie2011anomaly, karapistoli2013anomaly, fu2011wireless, meesookho2007energy, akyildiz2010wireless}. In particular, wireless body area networks (WBANs), which are one of the important classes of WSNs, are used for health monitoring
 \cite{yuce2012introduction, movassaghi2014wireless, hasan2019comprehensive, negra2016wireless, anwar2017wireless, thatte2009energy, thatte2009optimal, thatte2011optimal, zois2013energy}. 
In \cite{yuce2012introduction, movassaghi2014wireless, hasan2019comprehensive, negra2016wireless, anwar2017wireless}, WBANs are introduced to detect various abnormal conditions of physical health. In \cite{zois2013energy, thatte2011optimal, thatte2009energy, thatte2009optimal}, physical activity detection is investigated using WBANs. 

WSNs employ electromagnetic (EM) waves to convey information with some disadvantages in some environments such as inside structural metal pipes \cite{farsad2016comprehensive}. Also, invasiveness, high energy consumption, and incompatibility with the environment limit the applications of WSNs in some areas, such as inside living bodies. 
	Moreover, the conventional approaches are not
	reliable in applications that demand fast detection and accurate localization of the abnormality.
	For instance, the light symptoms at the early stage of cancer may not be measured by these methods  \cite{mosayebi2018early, tarro2005early}. In these applications, the combination of molecular communication (MC) nano-networks inside the body and wireless communication networks outside the body are promising systems \cite{qiu2014molecular}. Considering these types of applications, we focus on the MC-based systems for abnormality detection and localization in this paper.

MC is a new paradigm, which is inspired by the communication between living cells in nature, and it is biocompatible and energy-efficient, and hence more suitable for in-body applications \cite{nakano2013molecular, AtakanBook, felicetti2016applications}. 
MC systems use molecules as information carriers with many healthcare, industrial, and environmental applications. In bio-medicine, MC can be used for targeted drug delivery, health monitoring, and immune system support \cite{AtakanBook, Akyildiz2008}. The industrial applications of MC include pattern and structure formation, water and food quality control, and functionalized materials. The environmental applications include environment monitoring, air pollution control, and biodiversity control \cite{AtakanBook, Akyildiz2008}. In most of these applications, it is required to monitor the environment to detect and localize the occurrence of  abnormal changes, e.g., the presence of tumor cells inside the body, pollution in the air or water, leakage or pressure drop in oil pipelines. MC systems are envisioned to detect abnormalities in both micro-scale (range of a nano-meter to a micro-meter) applications, such as early detection of cancer cells inside the body \cite{felicetti2014molecular}, and macro-scale (range of a micro-meter to a meter) applications, such as detection of virus in the air \cite{amin2021viral}, and detection of leakage or pressure drop in oil-pipelines \cite{khaloopour2021theoretical}.

%------------------------------------------------
\begin{figure}
	\centering
	\includegraphics[scale=0.4]{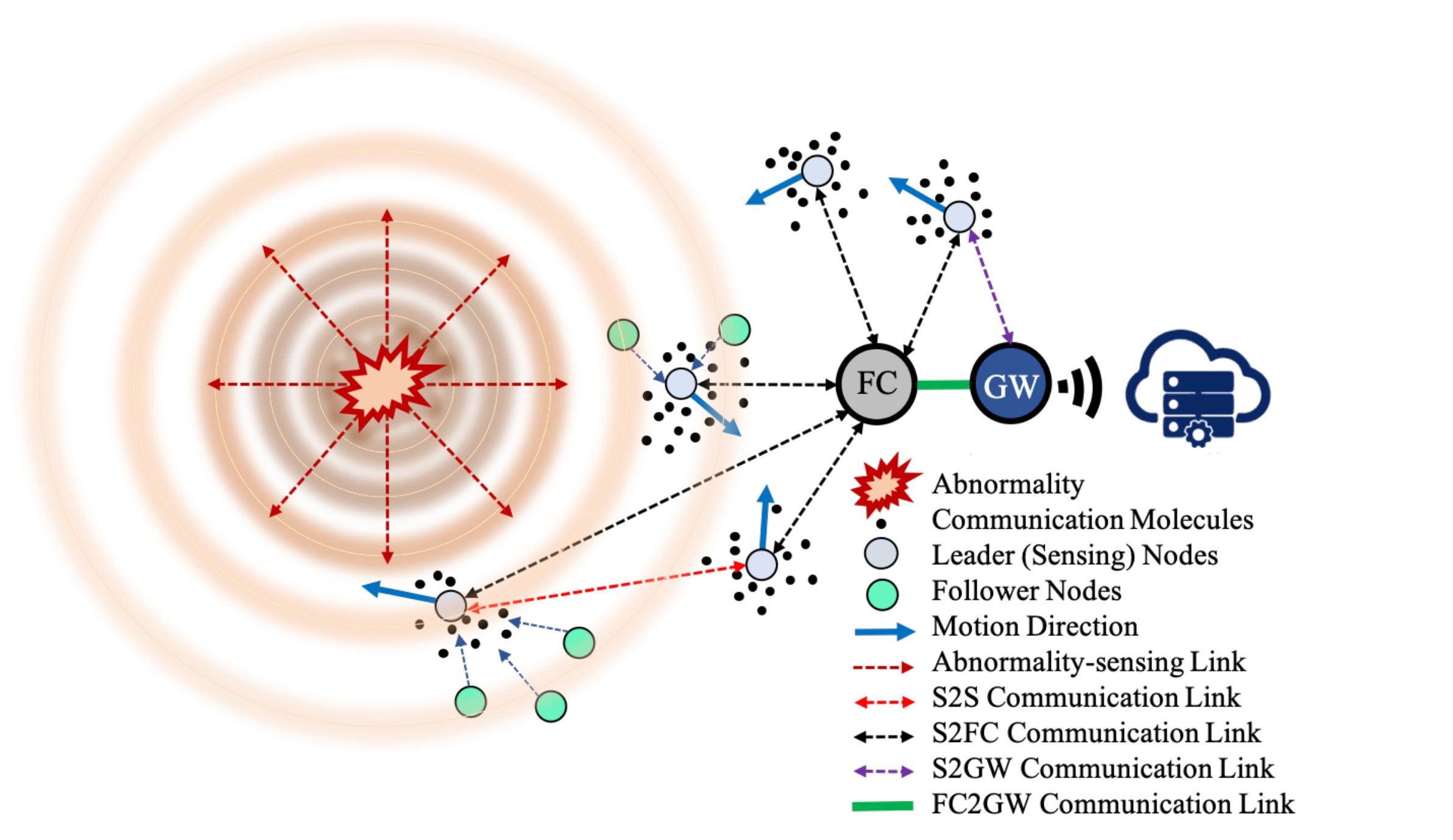}
	\setlength{\abovecaptionskip}{0.2 cm}
	\vspace{-1.5em}
	\caption{Schematic view of MC-based ADL system.}
	\label{fig_network}
	\vspace{-1em}
\end{figure}
%------------------------------------------------

\subsection{Multi-tier MC-based ADL Layout}
An MC-based ADL system may consist of different tiers as shown in Fig.~\ref{fig_network}. Here, we introduce a general multi-tier framework to comprehensively consider the existing works. However, some tiers may not appear in some applications and use-cases. In the first tier, there is an abnormal entity that changes a parameter in the medium.
The abnormality may affect the environment in different ways,
including molecule release (such as releasing biomarkers), medium effect (such as changing temperature or pressure), and molecule absorption that we describe in this paper. The sensors sense some variables in the medium to capture one or multiple effects, to obtain information about the abnormality.
The abnormality sensing may be molecular or non-molecular. For example, pressure measurement is non-molecular and biomarker concentration measurement is molecular. More specifically, we consider a molecule release/absorption method for the abnormality recognition as molecular sensing,  because the sensors measure the concentration of molecules, released by the abnormality source, and then diffused in the environment. Thus, it requires to analyze the movement of molecules from the molecule source to a molecular receiver at the sensors.
However, detecting medium effects such as the pressure or temperature measurements are non-molecular since 
they are not based on direct  concentration measurements related to the abnormality source.
We explain the functional units of a sensor including the recognition layer, transducer, processor, memory, and power units in this paper. Moreover, we describe different sensor features for abnormality detection and localization, including sensing, transduction, processing, memory, communication, molecule release, mobility, propulsion, manufacturing technology, and energy. 

In abnormality detection, after obtaining the information about different variables
that may be affected by an abnormality, the sensors can make the decision about the existence of the abnormality themselves (called non-cooperative detection). Alternatively, they can send their information (either the raw information or their local decision about the abnormality) to an FC to make the final decision (called cooperative detection). In the existing works, for non-cooperative detection, only one sensor is used, which makes the final decision about a single abnormal event \cite{khalid2020modeling, amin2021viral}.

In the second tier, for cooperative abnormality detection, stationary or mobile sensors may send their decisions to FC, through the sensor-to-FC (S2FC) communication link, to overcome the uncertainty in sensor decision  \cite{ghavami2017abnormality, varshney2018abnormality}. As another approach, the sensors may move towards the FC, where 
 their activation levels could be read \cite{mosayebi2018early}. Mobile sensors can be useful when there is no physical access to the sensing area and the sensing variables are very weak outside this area. Moreover, if the sensing area is wide, it is more feasible to use mobile sensors instead of multiple stationary sensors.

The sensors may also communicate with other sensors, through the sensor-to-sensor (S2S) communication link, and cooperatively activate each other upon detecting the abnormality  \cite{khaloopour2021theoretical} (which we call cooperative sensing/activation). 
In non-cooperative sensing/activation, 
the sensors rely on their own measurements to be activated and do not cooperate or interact with each other in being activated. The FC processes the information and makes the final decision itself, or transfers its raw samples or (soft/hard) decisions to a GW through the FC-to-GW (FC2GW) communication link (third tier). 
The sensors may also send their information directly to the GW through a sensor-to-GW (S2GW) communication link  \cite{gomez2021machine}; see Fig.~\ref{fig_network}. The GW may send these information to an external agent for complex processes \cite{stelzner2016precise}. The GW or external agent may also control the sensors based on the received information \cite{nakano2014externally}. The priority of the information stream identifies the level of communication tier in which the decision should be made. 
In addition to priority, introduction of cross-tier complexities (e.g., data processing and signal conversion capabilities  of the FC and GW) is pivotal to make such a decision.
For low-priority events, the decision is made at the FC or possibly GW. However, for high-priority events, which may need extensive global processing, the information is transmitted to the cloud computing system. 

The information gathered from the sensors can also be used to localize the abnormality by the FC, GW, or external agents \cite{khaloopour2021theoretical}. 
In contrast to the  stationary localization approaches, the sensors may move towards the abnormality using different mobility mechanisms, e.g., chemotaxis, or using external forces. The sensors may perform an action on the abnormality themselves such as releasing drugs. Also, the leader sensors may release some attractant molecules to passively attract follower sensors to the location of the abnormality for a particular action \cite{raz2015bioinspired, nakano2016performance}. 
The leader sensors may also communicate with the external environment through a GW to actively localize the abnormality \cite{okaie2014cooperative}.

The communication in each tier may be based on wireless links (radio-frequency (RF), optical, and acoustic) or molecular links. An ADL approach is called MC-based if it uses molecules as carriers of information in at least one communication link (mostly in the second communication link) or it uses molecule release/absorption method  in abnormality sensing link.
Hence, the MC-based layout consists of at least one molecular sensing or communication link. 

For transferring the information to an external node in MC-based layout, there should be an interface to convert the internal signals to the external signals. We explain different options for constructing \emph{wireless to wireless} and \emph{molecular to wireless} interfaces. 
For example, carbon nano-tubes (CNT) and Graphene nano-ribbons (GNR), embedded with molecular nano-sensors, can be used as promising molecular to wireless interfaces in the FC or GW. 
Other examples of interfaces to convert molecular signals to EM waves or vice versa are constructed using genetically engineered cells and artificially synthesized materials \cite{nakano2014externally}. 

Moreover, we summarize the modeling of sensing and communication channels and signal reception for molecular and non-molecular channels (including magnetic, ultrasonic, or RF-based channels). Molecular channels can be generally modeled by advection-reaction-diffusion partial differential equations (PDEs) with some boundary conditions according to the medium structure and receiver type \cite{jamali2019channel}. Magnetic, ultrasonic, or RF-based channels can also be modeled by wave equations following the second-order linear PDEs \cite{hoelen1999new,santagati2013opto}.

\subsection{State-of-the-Art MC-based ADL Schemes}

\raggedbottom
ADL have been studied widely in the literature of MC.
We first formulate the abnormality detection problem, in which the detection of the abnormality is investigated and the location of the abnormality is not of interest. Different abnormality detection
schemes will be discussed later in detail. These schemes are categorized based on using stationary or mobile sensors. 
For schemes with stationary sensors, the abnormality may be detected using either a single sensor (non-cooperative detection) or multiple sensors that send their decisions to an FC for final decision (cooperative detection).
As mentioned above, the sensors may cooperatively activate each other (cooperative sensing/activation). However,  this possibility for stationary sensors has not been studied yet. The schemes using stationary sensors are categorized based on cooperative or non-cooperative detection. 
For example, in \cite{khalid2018system}, a single stationary sensor placed in a room is used for detecting virus-laden aerosols in the room, and in \cite{mosayebi2018advanced}, multiple stationary sensors and an FC are placed near the suspicious tissue and the target detection is investigated at the sensors and the FC.
On the other hand, all detection schemes using mobile sensors in the MC literature use cooperative detection (multiple sensors and an FC). Hence, the schemes using mobile sensors are categorized  into schemes with non-cooperative or cooperative sensing/activation. For example, in \cite{mosayebi2018early}, multiple mobile sensors with non-cooperative sensing/activation are inserted into the circulatory system for early cancer detection. In \cite{khaloopour2021theoretical}, the abnormality detection inside pipelines is accomplished using multiple mobile sensors with cooperative sensing/activation, where the pipeline is divided into multiple regions with an FC placed at the end of each region. 

The next step after the abnormality detection is to localize the abnormality. The MC literature has recently proposed schemes to estimate the location of the transmitter as a source of releasing molecules that leads to a diffusion field \cite{kumar2020nanomachine, baidoo2020channel, zhu2021target,miao2019cooperative,yetimoglu2021multiple}.
In many abnormality localization applications, the navigation of a swarm of sensors towards the intended region is of high interest, e.g., in the targeted drug delivery systems  \cite{chahibi2013molecular} and macro-scale robot-assisted surveillance systems \cite{dunbabin2012robots}.  Similar to abnormality detection, the abnormality can be localized using either mobile or stationary sensors. 
The stationary sensors may employ classification, discrete-model, or continuous-model-based approaches to infer the location of the abnormality \cite{matthes2005source}.
Moreover, the mobile sensors are divided into non-propelled or propelled sensors.  The propelled sensors are influenced by  specific internal and external forces which are referred to as self-propulsion and external propulsion, respectively.
The self-propelled sensors can be potentially employed in a variety of applications, e.g., water decontamination, water remediation, cargo transport, and biomedical applications \cite{tsang2020roads}.
The self-propulsion mobilities are originated  in response to the natural mechanisms (i.e., using taxis-based approaches) or relying on synthetic motors (i.e., using control-driven approaches). 
Taxis-based approaches, e.g., chemotaxis \cite{roussos2011chemotaxis}, optotaxis \cite{de2018optogenetic}, anemotaxis \cite{jatmiko2007pso}, and infotaxis \cite{vergassola2007infotaxis}, are well-studied self-propulsion mechanisms in the literature on physical chemistry.
In biological environments, biocompatibility is vital for self-propelled sensors which means that the sensors should emulate biological locomotion mechanisms to efficiently localize an abnormality.
On the other hand, control-driven approaches \cite{sinha2018consensus} may perform more efficiently than simple taxis-based approaches at the cost of more computational complexity. The important challenge in self-propulsion mechanisms is that the sensors have to autonomously locate the abnormality taking into account the limited resources (e.g., energy, memory, etc) of the sensors.
As another approach, external triggers may be employed to control the locomotion of the swarm of sensors. To this end, the communication between sensors, FC, GW, and possibly the cloud computing servers is required to localize the abnormality externally. Thereby, data gathered by the sensors from the abnormality micro-environment may be employed locally by the sensors or globally by either FC, GW, or cloud computing servers to make a local or global decision. The proposed multi-tier framework for the abnormality detection schemes is still  valid for abnormality localization schemes. It should be notified that the third tier (including S2GW and FC2GW links), which enables controlling  of the externally-propelled sensors, are more important for abnormality localization schemes compared to the abnormality detection schemes.
To motivate the role of the externally-controllable systems, a variety of application-centric frameworks are developed, e.g., a novel touch communication (TouchCom) framework for drug-delivery systems \cite{chen2015touch} and a novel tumor detection and localization framework using external devices \cite{shi2020nanorobots}, which are clarified throughout the paper.
Recent advancements in the field of medical imaging and actuating systems in line with network cloudification promises future externally-controllable health monitoring systems.

\subsection{Contribution Summary and Paper Organization}
In this paper, we provide a comprehensive survey on the MC-based ADL  systems. We describe the most important challenges in modeling, designing, and implementing the MC-based ADL systems and provide future research and development directions. 
The MC-based systems are envisioned to be mostly used in nano-scale in-body applications. Therefore, the most important challenges to implement these systems are to design bio-nano-sensors with multiple functionalities (such as mobility mechanisms, sensing methods, communication strategies, storage capabilities, and energy harvesting mechanisms), and interfaces with different types of signal conversion, for connecting the internal  and external networks \cite{felicetti2016applications}.
The modeling of these systems is another important challenge that requires a well understanding of biological systems and environments.
The existing models introduced for MC-based systems are simplified models, which are far from reality in some cases. Therefore more realistic, yet tractable models, are needed for these systems. Further, the detection and localization approaches, which can be categorized in general into classic, bio-inspired, and machine learning methods, can be improved by appropriately combining different approaches. 

The main contributions of the paper are summarized as:
	\begin{itemize}
		\item An end-to-end MC-based ADL framework considering a multi-tier layout is proposed to encompass most of the models in the ADL literature. The proposed multi-tier layout consists of at least one molecular link, typically considered the first or second tier in our model.
		\item A detailed  description of the  sensors, interfaces, and external networks including the functionality and inter-connectivity options, customized for the ADL framework,  is explained.
		\item A general channel model describing the physical layer of each single tier (molecular and non-molecular)   based on the existing literature is presented.
		\item A comprehensive classification of the MC-based ADL schemes based on the mobility and propulsion of the sensor type is  presented.
	\item Finally, practical challenges and important future research directions in the area of MC-based ADL systems are clarified.
		\end{itemize}

This paper is organized as follows: in Section~\ref{sec:system}, the ADL problem and the important elements of the systems including abnormality, sensors, FC and GW (interfaces), and sensing/communication links are described. In Section~\ref{sec:detection}, the abnormality detection schemes and in Section~\ref{localization}, abnormality localization schemes are reviewed. Finally, in Section~\ref{sec:challenges}, the important challenges of the MC-based ADL systems are explained and some future research directions are introduced.

\section{Problem and System Description}\label{sec:system}
In this section, we describe the MC-based ADL systems, represented schematically in Fig.~\ref{fig_network}.
In the following, we describe the important elements of this system including abnormality, sensors, FC/GW (interfaces), and sensing/communication channels. 
We first describe different abnormality recognition approaches and explain the ADL problems in Subsection~\ref{sec:abnormality}. In Subsection~\ref{sec:sensors}, we illustrate different features for the sensors in these systems and in Subsection~\ref{sec:interfaces}, we describe the possible technical options for FC/GW (interfaces). Finally, in Subsection~\ref{eq}, we provide the modeling of sensing and communication channels and signal reception for molecular and some non-molecular channels.

\subsection{Abnormality}\label{sec:abnormality}
Abnormality is defined as any abnormal feature, characteristic, or occurrence that interferes with the normal status of a system \cite{khaloopour2021theoretical}.

\subsubsection{Abnormality Recognition Methods} 
\begin{figure}[t]
\centering
\includegraphics[scale=0.59]{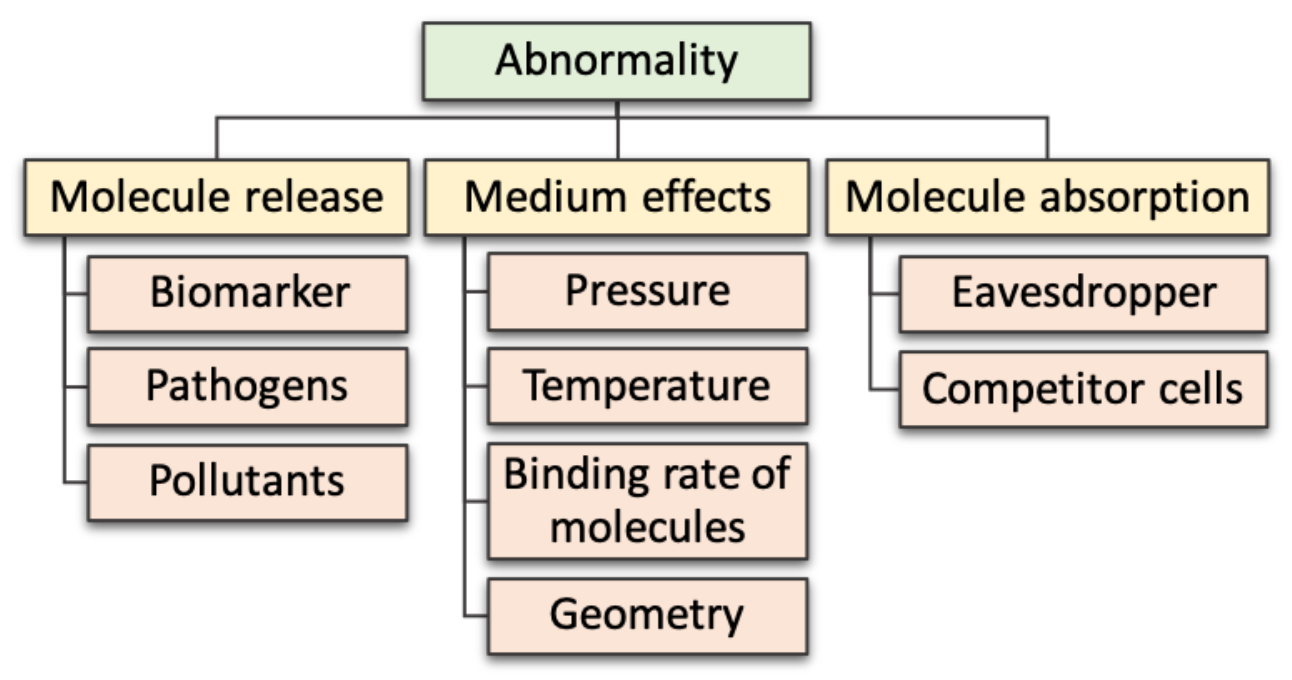}
\caption{Different abnormality recognition methods.}
\label{fig_abnormality}
\vspace{-1em}
\end{figure}

The abnormal entities can be recognized in different ways. We consider the following abnormality recognition methods, which are mostly considered in  MC-based systems (see Fig.~\ref{fig_abnormality}):

\begin{itemize}[leftmargin=11pt]
\item{ \textit{Molecule release}: an abnormal entity may release certain molecules into the medium. For example, cancer cells as abnormal entities in the body may release biomarkers, which can be used to detect and localize the cancerous site \cite{solak2020rnn, kerry2021comprehensive}. Further, pathogens (viruses, bacteria, etc) in the air or inside the body may be detected by the sensors \cite{khalid2020modeling, gomez2021machine}. Viruses may be released to the air by an infected person (as an abnormal entity). On the other hand, bacteria (as abnormal entities) may release quorum-sensing molecules inside the body. Further, antibodies inside the body may be released by the immune system in the result of viral or bacterial infections. Moreover, pollutants may be present in the air, water, or oil pipelines (released by cars, factories, etc, as abnormal entities), which can be detected by certain sensors \cite{khaloopour2021theoretical}. The authors in \cite{kerry2021comprehensive} introduce the biomarkers released by different cancers and the corresponding sensors to detect them. 
The biomarker release rate over time, noted by $x_{\textrm{B}}(t)$, can be modeled by Weibull function as follows \cite{ghavami2020anomaly}:
\begin{align}
x_{\textrm{B}}(t)=x_{\textrm{B},0}(1-\exp{(-kt^b)}),
\end{align}
where $b$ and $k$ are biomarker power-law and biomarker release coefficients, respectively; and $x_{\textrm{B},0}$ is the steady-state release rate when $t \rightarrow \infty$.
{The authors in} \cite{hori2011mathematical} {provide} a more realistic model for biomarker release rate from healthy and cancerous cells. According to this model, biomarker release rates from healthy and cancerous cells that reach the blood vessels, noted by $x_\textrm{H}(t)$ and $x_\textrm{C}(t)$, respectively, are obtained as follows \cite{mosayebi2018early}:
\begin{align}
\begin{cases}
x_\textrm{H}(t)=f_\textrm{H}r_\textrm{H}(t)n_\textrm{H}(t), \hspace{0.45 cm} \text{for healthy cells,}\\
x_\textrm{C}(t)=f_\textrm{C}r_\textrm{C}(t)n_\textrm{C}(t), \hspace{0.5 cm} \text{for cancerous cells,}
\end{cases}
\end{align}
where $n_\textrm{H}(t)$ and $n_\textrm{C}(t)$ are the number of healthy and cancerous cells over time, respectively; $r_\textrm{H}(t)$ and $r_\textrm{C}(t)$ are the biomarker release rates from each healthy and cancerous cell (assumed to be the same for all cells), respectively; and $f_\textrm{H}$ and $f_\textrm{C}$ are the fractions of released biomarkers from healthy and cancerous cells that reach the blood vessels (while the rest of them remain in tissues), respectively. It is assumed that biomarker release rates from each cancerous and healthy cell are constant, i.e., $r_\textrm{C}(t)=r_\textrm{C}$ and $r_\textrm{H}(t)=r_\textrm{H}$. Further, the number of healthy cells is assumed to be constant over time, i.e., $n_\textrm{H}(t)=n_{\textrm{H},0}$, and the number of cancerous cells is modeled by Gompertzian function as follows \cite{norton1976predicting}:
\begin{align}\label{n_cancer_cells}
n_{\textrm{C}}(t)=n_{\textrm{C},0}\exp{(\frac{k_\textrm{Gr}}{k_\textrm{Dec}}(1-\exp{(-k_\textrm{Dec}t)}))},
\end{align}
where $k_\textrm{Dec}$ and $k_\textrm{Gr}$ are the fractional decay and growth rates of the cancerous cells, respectively, and $n_{\textrm{C},0}$ is the number of cancerous cells at time $t=0$. For small values of $k_\textrm{Dec}$, i.e., $k_\textrm{Dec} \rightarrow 0$ (which is the case in the early stage of cancer), the number of cancerous cells can be simplified to the following mono-exponential growth function:
\begin{align}
n_{\textrm{C}}(t)=n_{\textrm{C},0}\exp{(k_\textrm{Gr}t)}.
\end{align}

Hence, we have
\begin{align}\label{abnormality_rates}
x_\textrm{C}(t)&=f_\textrm{H}r_\textrm{H}n_{\textrm{C},0}\exp{(k_\textrm{Gr}t)}.
\end{align}
}

\item{ \textit{Medium effects}: abnormality may appear as the change in the medium parameters such as flow velocity \cite{fouron2003changes}, pressure \cite{zhu2017gas}, temperature \cite{jafari2020cell}, and geometry \cite{liu2020detection}. For example, a leakage leads to a pressure drop in oil pipelines \cite{zhu2017gas}; a tumor may change the temperature of the nearby environment \cite{jafari2020cell}, and Atherosclerotic Lesion decreases the cross-section of the blood vessel \cite{liu2020detection}. 
The sensors may directly measure the environment parameters and detect the changes.
Also, a communication system could be influenced by the medium change and indirectly detect the abnormality \cite{ghavami2017abnormality}.}

\item{\textit{Molecule absorption}: the abnormal entity may absorb certain molecule types in the environment.
For example, an eavesdropper (considered as an abnormality) may absorb information or signaling molecules to interfere with a natural or synthetic molecular communication link \cite{guo2016eavesdropper}.
Moreover, cancerous cells may act as competitor cells for other cells inside the body and absorb their released molecules (such as Oxygen) \cite{ghavami2017abnormality}.
The concentration change of the absorbed molecules can be used to detect this abnormal entity.}
\end{itemize}

The sensors may use one (or more than one) of these abnormality recognition methods to obtain information about the abnormality in the environment.

\begin{figure}[t]
\centering
\includegraphics[scale=0.52]{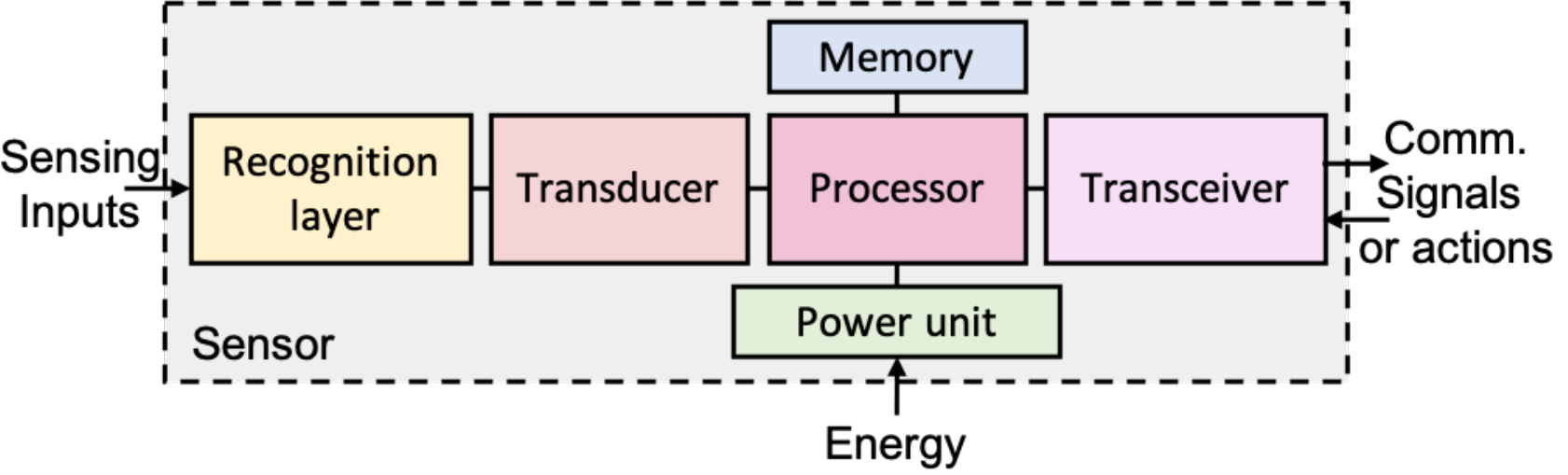}
\caption{Functional units of the sensors.}
\label{fig_sensor_functional_units}
\vspace{-1em}
\end{figure}

\begin{figure*}[t]
\centering
\includegraphics[scale=0.6]{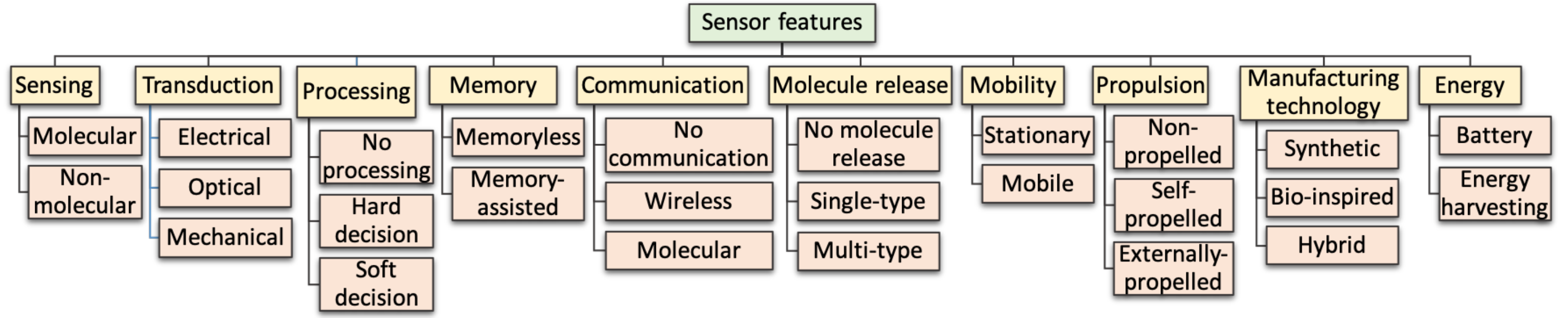}
\caption{Sensor features.}
\label{fig_sensor_features}
\vspace{-1em}
\end{figure*}

\subsection{Sensors}\label{sec:sensors}
The functional units of sensors are shown in Fig. \ref{fig_sensor_functional_units} \cite{agrawal2011designing}.  A sensor generally consists of a recognition layer, where the abnormal events are recognized, a transducer unit, where the perceived abnormal events are converted into another type of signal to be processed further, a processing unit, where the signal is processed, a power unit, {which} provides the energy for the processor, a memory unit to store information, and a transceiver to communicate with other nodes or to perform supplementary interventions at the sensor (e.g., releasing drug molecules) \cite{agrawal2011designing}. 
The communication signal inputs may contain controlling signals for the sensor, e.g., controlling the motion of the sensor. 
The sensing input signals may also cause the moving of sensors towards the abnormality. 
In the following, we categorize the features of the sensors used for ADL (see Fig.~\ref{fig_sensor_features}):

\begin{itemize}[leftmargin=11pt]
\item{\textit{Sensing}: As the first mission, sensors may sense environmental parameters like temperature \cite{chaudhary2020twin} and pressure \cite{xu2018recent} (non-molecular sensing), or sense and measure the concentration of certain molecules (biomarkers) in the environment (molecular sensing), such as different gases and chemicals \cite{mishra2021design, chaudhary2020topas}. For molecular sensing, the receiving process of the sensors can be modeled by transparent, absorbing (fully or partially), {and} reactive receivers \cite{jamali2019channel} (see Table~\ref{tab-features} for different abnormality detection schemes assuming these models).}

\item{
\textit{Transduction}: The converted signal in the transducer may be electrical, optical, and mechanical \cite{kuscu2016physical}. The optical and mechanical sensors need macro-scale elements to detect the transducer's output to be ready for processing, while the electrical sensors can process the inputs in a self-operated manner at the nano-scale.
Electrochemical biosensors are examples of electrical sensors \cite{grieshaber2008electrochemical} and 
biosensors based on nano-mechanical systems
are examples of mechanical sensors \cite{tamayo2013biosensors}. Surface plasmon resonance biosensors \cite{chaudhary2021gold} and Terahertz refractive index sensors \cite{mishra2021terahertz} are examples of optical sensors.}

\item{
\textit{Processing}: The sensors may be able to process their measurements and make decisions on the existence of the abnormality. The decisions made by the individual sensors can be \emph{hard}, representing the absence or presence of the abnormality, or \emph{soft}, interpreting a probability of abnormality existence \cite{solak2020neural}.
}

\item{
\textit{Memory}: The sensors may use memory for information processing. For instance, aggregate sensors in \cite{khaloopour2021theoretical} keep their measurements in different time instances and aggregate them for the final decision. A higher memory enables more complicated information processing algorithms leading to more efficient ADL.}

\item{
\textit{Communication}: 
Besides sensing, the sensors may communicate with other agents in the system (e.g., sensors {and} FC) to report their measurements and/or decisions for further actions regarding the abnormality \cite{khaloopour2021theoretical}. Depending on the application, the sensors may be designed to communicate, which can be accomplished using wireless (e.g., RF \cite{ganesh2021rf, lemic2021survey}, optical \cite{guo2015intra}, ultrasonic \cite{santagati2014medium},) and molecular signals \cite{felicetti2014molecular}. In \cite{lemic2021survey}, nano-sensors with RF-based communication, operate in Terahertz (THz) band. MC is a promising approach for communication in nano-scale in-body applications since it is bio-compatible and energy-efficient \cite{dressler2015connecting}.}

\item{
\textit{Molecule release}: The sensors may be able to release molecules (single type or multiple types) for various purposes, for instance, to communicate with other sensors and FC  \cite{khaloopour2021theoretical}, {and} to deliver drug molecules at the target site \cite{ghavami2017abnormality}. The molecules may be stored in the sensor storage or produced through chemical processes by the sensor itself. The release of molecules should be controlled by the sensor, which can be realized using ion channels \cite{arjmandi2016ionchannel} and ion pumps \cite{arjmandi2016ionpump}.}

\item{
\textit{Mobility}: Sensors may be \emph{mobile} or \emph{stationary}. The locations of the stationary sensors are fixed, e.g., skin-integrated and implantable sensors \cite{rodrigues2020skin, koo2020deep}, tissue-integrated sensors \cite{wisniewski2017tissue}, and virus or air pollutant sensors in a room \cite{amin2021viral}.
On the other hand, the sensors may be able to move passively \cite{varshney2018abnormality} or actively (autonomously) \cite{okaie2014cooperative, soto2020medical} in the environment.}

\item{
\textit{Propulsion}: Mobile sensors move using different propulsion methods. The sensors  
 may move passively in the fluid medium due to the diffusion and/or flow \cite{varshney2018abnormality}, which is called non-propulsion. Sensors may move using their own means for active (autonomous) displacement, which {is referred} to as self-propulsion. For example, flagellated bacteria autonomously move towards the higher gradient of attractant molecules to find food \cite{okaie2014cooperative}. Moreover, the sensors may be externally-propelled for active (autonomous) displacement, i.e., external electric or magnetic fields can control the sensors' motion \cite{soto2020medical, sun2020cooperative}.}

\item{
\textit{Manufacturing technology}: The manufacturing technology of the sensors at nano- and micro-scales includes synthetic (based on micro-electromechanical systems (MEMS)), bio-inspired (engineered biological elements), and hybrid \cite{stelzner2016precise}.
}

\item{
\textit{Energy}: Based on the manufacturing technology, the sensors need a source of energy for their processing that can be harvested or stored in advance in their energy storage (battery). 
The sensors may harvest energy using different techniques including vibration energy harvesting, thermal energy harvesting, biochemical energy harvesting, {and} wireless energy harvesting \cite{katic2015efficient}. 
Biological sensors may use the foods in their environment to harvest energy (biochemical energy harvesting) \cite{katic2015efficient}. 
The sensors may use thermal energy (temperature difference) to harvest energy \cite{katic2015efficient, lu2010thermal}.
They may also harvest energy from the kinetic energy (such as flow) in their environment (vibration energy harvesting) \cite{katic2015efficient, wei2017comprehensive, park2020nanofluidic}.
Further, they may harvest energy from wireless energy sources  \cite{katic2015efficient, phillips2021energy}.}

\end{itemize}

\subsection{FC and GW (Interface)}\label{sec:interfaces}
To provide more processing capabilities in the MC-based ADL layout, the sensors may send their collected information to external computing units through FC and GW, which is referred to as out-messaging. External units can conversely impose controlling commands to the sensors in the reverse direction referred to as in-messaging.
 GW connects the internal communication network (e.g., the nano-network inside the body, called the in-body network) to the external communication network (e.g., the macro-network outside the body, called the off-body network).
 
In healthcare applications, the GW can be placed inside or outside the body \cite{yang2020comprehensive}. Hence, either the FC or the GW needs a nano-macro interface to convert the data from the in-body network to the appropriate form for the off-body network.
The nano-network inside the body can use THz or molecular signals, while for the off-body networks, EM signals are mostly used for communication. For example, \cite{islam2016catch, canovas2018nanoscale} use THz signals and \cite{nakano2014externally} uses molecular signals for communication in the nano-network. 
In THz-based nano-networks, the nano-devices inside the body are non-biological devices and use EM waves for communication, which may be invasive to the human body, while MC-based nano-networks are more bio-compatible and less invasive, due to using molecules and biological elements for communication, and hence they are of most interest \cite{nakano2013molecular, AtakanBook, felicetti2016applications}. 

Based on the communication technologies used in the S2S, S2FC, S2GW, and  FC2GW links, the interfaces   can be categorized into two main types:
 
  \begin{itemize}[leftmargin=11pt]
  	\item{
  		\textit{Wireless to wireless interface}: For sensors that use THz signals for  communication, an interface is needed to convert the input THz signal  to the output EM signal and vice versa. Since both signals are wireless, the interface is an electrical device with a dual transceiver enabling cross-tier communications.
  		For example, in \cite{islam2016catch}, the nano-devices are located inside coronary arteries, which communicate in THz band with a GW. The GW is a wireless to wireless interface located in the intercostal space of the rib cage of the patient which  communicates with the off-body networks using EM signal. 
  		In \cite{canovas2018nanoscale}, multiple nano-devices are inserted into the circulatory system and communicate in THz band with a nano-router playing the role of an FC.
  		The FC is a  wireless to wireless interface
  		 implanted in the hand skin between the dermis and epidermis. The FC communicates in THz band with an external portable device as a GW, which is placed on the back of the hand.
  	}
  	
  	\item{
  		\textit{Molecular to wireless interface}: 
  		 For MC networks, an interface is needed to convert the input molecular signal (originating from the internal network) to the output EM signal (appropriate for the external network) and vice versa. Different interfaces for in-messaging and out-messaging 
 can be implemented using genetically engineered cells \cite{nakano2014externally, wu2009periodic, ozawa2013advances}, artificially synthesized materials (ARTs) (such as photosensitive, temperature-sensitive, magnetic, and luminescent materials) \cite{nakano2014externally}, CNTs \cite{yang2020comprehensive, roman2004single}, and GNRs \cite{yang2020comprehensive, lazar2013adsorption}. Implementing different types of signal conversion is complex using genetically engineered cells, while ARTs provide a wide range of options for signal conversion in interfaces, e.g., converting electrical, optical, or magnetic signals to molecular signals and vice versa. ARTs can be {generated} using Polystyrene bead and pHrodo molecules \cite{nakano2014externally}. 
  		pHrodo molecules are pH-sensitive fluorogenic dyes that are only fluorescent under acidic environments (low pH) and almost non-fluorescent under neutral environments. Hence, the intensity of the fluorescence can be measured using fluorescence microscopy to get information about the acidity and neutrality of the cells (out-messaging). Further, polystyrene bead can be used to implement in-messaging, e.g., it can be designed to release or absorb molecules upon receiving signals from external devices. However, the need for  a fluorescence microscope makes it complicated for ordinary use. 
  		CNTs and GNRs are more recent approaches to construct the interfaces  \cite{yang2020comprehensive, roman2004single}. In these mechanisms, chemical nano-sensors are embedded in CNT or GNR. These nano-sensors absorb certain molecules that generate an electrical signal in the CNT or GNR by changing the number of electrons that move through the carbon lattice.
  	}
  \end{itemize}
 
\subsection{Unified Channel Model and Signal Reception} \label{eq}
In this section, we provide the general channel model for the system considered in Fig.~\ref{fig_network}. The link from the abnormality to the sensors is referred to as the sensing channel. Other links, e.g., S2FC, FC2GW, and S2GW
 links are all referred to as communication channels. 
We model the sensing and communication channels using a generalized advection-diffusion-wave equation in viscoelastic\footnote{Viscoelasticity of a body identifies its capability to simultaneously dissipate and store mechanical energy in a generic continuous media exhibiting a combination of viscosity (i.e., fluid-like materials supporting types of anomalous diffusion processes) and elasticity (i.e., solid-like materials exhibiting wave propagation) \cite{luchko2013fractional, mainardi2010fractional}.} environments, which characterizes the Brownian motion of particles as well as the wave propagation. The generalized space-time fractional advection-diffusion-wave equation of order $(\alpha, \beta)$ for spatio-temporal field $\mathcal{C}(\vect{r},t)$ is described as follows \cite{luchko2013fractional, mainardi2010fractional}:
\begin{align}\label{p1}
\nonumber
\mathcal{D}^{\beta}_t \big(\mathcal{C}(\vect{r},t)\big)&= \nabla^{\alpha} \big(a_{{\vect{r},t}}\mathcal{C}({\vect{r},t})\big)-u_{\beta}\nabla.\big(b_{{\vect{r},t}}\mathcal{C}(\vect{r},t)\big)\\
&\quad -u_{\beta}\mathcal{F}(k_d,\mathcal{C}(\vect{r},t)) +\mathcal{P}\big(S(\vect{r},t)\big),
\end{align}
where $u_{\beta}=\mathbf{1}_{\beta\neq 2}$ {and $u_{2}=0$}. Also, $\nabla^{\alpha}(\cdot)$ and $\mathcal{D}^{\beta}_t(\cdot)$ denote the Riesz space-fractional derivative of order $\alpha$ and the Caputo time-fractional derivative of order $\beta$ \cite{luchko2013fractional, mainardi2010fractional}, respectively. For $\beta\neq 2$, \eqref{p1} describes the space-time fractional diffusion equation with $a_{\mathrm{\vect{r},t}}$ and $b_{\mathrm{\vect{r},t}}$ denoting the spatio-temporal diffusion and drift parts, respectively. Moreover, $\mathcal{F}(\cdot)$ and $\mathcal{P}(\cdot)$ describe the degradation reaction rate function with degradation rate $k_d$ and the source equation, respectively. In the case of diffusion processes, \eqref{p1} models a wide range of diffusion processes from normal to anomalous diffusion.
For normal diffusion, i.e., $(\alpha,\beta)=(2,1)$, the Fokker-Planck equation in \eqref{p1} is given by $\mathcal{P}\big(S(\mathrm{\vect{r},t})\big)=S(\mathrm{\vect{r},t})$. Also, $\beta=2$ characterizes the time-fractional wave equation with  $a_{\mathrm{\vect{r},t}}=c^2$ where $c$ denotes the {corresponding wave} speed.
In the sequel, the provided model in \eqref{p1} is clarified for special cases of reaction-coupled diffusion and wave-propagation channels.

\subsubsection{Reaction-Coupled Diffusion Channels}
In a real biological microenvironment, there may be multiple transceivers, obstacles, and boundaries in the presence of flow-induced diffusion, coupled with arbitrary homogeneous boundary conditions. 
The molecules diffused in the environment may react with  the receptors on the boundaries or obstacles in the environment. The reactions at the boundaries may lead to complicated and non-linear boundary conditions, which can be simplified to linear reaction chains. 
A general homogeneous boundary condition that simplifies modeling of a wide range of reactions at the boundaries is given as follows \cite{arjmandi2020mathematical},
\begin{equation}\label{GHB}
\textbf{D}\nabla \mathcal{C}(\vect{r},t)\cdot\hat{n}=
\textbf{L}\mathcal{C}(\vect{r},t),~~~\vect{r}\in\partial \mathcal{D},
\end{equation} 
where $\nabla$ denotes the spatial gradient operator, $\hat n$ describes the surface normal vector pointing towards the exterior of the diffusion environment at $\vect r \in \partial \mathcal{D}$, where $\partial \mathcal{D}$ denotes the set of all points over the boundary of the environment, and \textbf{D} and \textbf{L} are some time-domain differential operators which are characterized based on the reactions involving information molecules across the boundary.

In the case of  a normal diffusive MC system in a viscous medium with constant diffusion coefficient $a_{\mathrm{\vect{r},t}}={D}$ and uniform bulk flow $b_{\mathrm{\vect{r},t}}={v}$, 
\eqref{p1} is simplified to:
\begin{align}\label{t1}\nonumber
\mathcal{D}_t \big(\mathcal{C}(\mathrm{\vect{r},t})\big)&= D\nabla^{2} \big(\mathcal{C}(\mathrm{\vect{r},t})\big)\\
&\quad -k_d\mathcal{C}(\mathrm{\vect{r},t}) 
-v\nabla.\big(\mathcal{C}(\mathrm{\vect{r},t})\big)+S(\mathrm{\vect{r},t}).
\end{align}

Note that \eqref{GHB} models a wide range of boundary conditions including partially absorbing boundaries and active transport mechanisms. 
\eqref{GHB} can potentially describe an oversimplified model for Transcytosis across membrane announcing physiochemical reactive properties using a second-order reaction-diffusion system \cite{arjmandi2020mathematical}.

\begin{exmp}Simplified adevction-reaction-diffusion channel in an unbounded environment:
 Given the impulsive instantaneous release source $S(\mathrm{\vect{r},t})=\delta(\mathrm{\vect{r}}-\mathrm{\vect{r}_0})\delta(t-\tau_0)$, where $\delta(\cdot)$ denotes the Dirac's delta function, and boundary condition $\mathcal{C}(\mathrm{|\vect{r}|\rightarrow \infty,t})=0$, diffusion coefficient $a_{\mathrm{\vect{r},t}}={D}$, uniform bulk flow $b_{\mathrm{\vect{r},t}}={v}$, and the first-order  degradation of rate $k_d$, i.e.,  $\mathcal{F}(k_d,\mathcal{C}(\mathrm{\vect{r},t}))=k_d\mathcal{C}(\mathrm{\vect{r},t})$, the solution of \eqref{p1} in an unbounded environment is obtained as \cite{farsad2016comprehensive, jamali2019channel}:
	\begin{align}\label{p11}
	\mathcal{C}(\mathrm{\vect{r},t})&=\frac{1}{(4\pi D (t-\tau_0))^{3/2}}\\\nonumber
	&\quad \times \exp\big(-k_d(t-\tau_0)-\frac{|\mathrm{\vect{r}}-(t-\tau_0){v}-\mathrm{\vect{r}_0}|^2}{4D (t-\tau_0)}\big),
	\end{align}
	for $t\geq \tau_0$.
\end{exmp}

\subsubsection{Wave-propagation Channels}
In typical wireless communications, various technologies, e.g., electromagnetic, ultrasonic, and RF, are employed for information transmission. The S2S, S2FC, FC2GW, and {S2GW} 
links may be modeled using \eqref{p1} with $\beta=2$,  $a_{\mathrm{\vect{r},t}}=c^2$, and $\mathcal{P}\big(S(\mathrm{\vect{r},t})\big)=\Gamma \frac{\partial }{\partial t}\mathcal{S}(\mathrm{\vect{r},t})$, in which $\Gamma$ is the  Gr\"uneisen coefficient which depends on the medium characteristic for the propagating wave \cite{santagati2014medium}. 

\begin{exmp}Opto-ultrasonic channel model:
One of the interesting channels in WBAN is the opto-ultrasonic channel described by \eqref{p1} \cite{santagati2014medium, santagati2013opto}. 
This channel is based on ultrasonic waves (i.e., high-frequency\footnote{For instance, wideband ultrasonic signals (in the range of hundreds of MHz for both bandwidth and central frequency) are generated in response to nano-seconds-long optical pulses.} acoustic waves) generated in response to irradiation of optical signals to the medium. Joint deployment of light beams and ultrasonic waves allows for external interventions at the nano-scale, which is critical for connecting in-body nano-networks to off-body WBANs. Assuming Gaussian spatio-temporal pulse profile, i.e., {$S(\mathrm{\vect{r},t})=G(\mathrm{\vect{r}})\delta(t)$}, the received pressure at position $\vect{r}$ and time $t$ under the far-field assumption is described as follows:
\begin{equation} \label{22}
\begin{aligned}
\mathcal{C}(\mathrm{\vect{r},t})=-\frac{\Gamma}{2(2\pi)^{3/2}c^2\vect{r}} \frac{t-\tau}{\tau_e^3}\exp{\bigg(-\frac{1}{2}(\frac{t-\tau}{\tau_e})^2\bigg)},
\end{aligned}
\end{equation}
where $\tau_e$ depends on both the width of the optical beam and optical pulse duration.  
Also, $\tau=\frac{\vect{r}}{c}$ and the source ${S}(\mathrm{\vect{r},t})$ is the normalized absorbed heat by tissue representing the spatio-temporal source of pressure \cite{santagati2013opto}.
\end{exmp}

We note that the concentration signal and corresponding signal received by the individual sensors, FC, or GW are random variables whose mean satisfies the obtained average field given by \eqref{p1}. In most scenarios, the received signal is well approximated by a Gaussian distribution, i.e., 
$y(t)\sim \mathcal{N}(\mu(\mathrm{\vect{r},t}), \sigma^2(\mathrm{\vect{r},t}))$ where $\mu(\mathrm{\vect{r},t})$ and $\sigma^2(\mathrm{\vect{r},t})$ denote the mean  and the variance of the received signal, respectively. The variance $\sigma^2(\mathrm{\vect{r},t})$ can be either signal-dependent (diffusion-based links) or {signal-independent}  (wave-propagation-based links).

\renewcommand{\arraystretch}{0.7}
\renewcommand{\tabcolsep}{2pt}
\newcolumntype{C}[1]{>{\centering\let\newline\\\arraybackslash\hspace{0pt}}m{#1}}
\begin{table*}[h]
	\centering
	\rowcolors{6}{light-gray}{}
	\caption{Summary of the {ADL} methods.}
	\begin{tabular}{|*{11}{c|}}  % repeats {c|} 18 times
		\hline
		\multirow{8}{*}{\normalsize Ref.}  
		&\multicolumn{4}{C{6.25cm}|}{\normalsize Abnormality Detection} 
		& \multicolumn{6}{c|}{\normalsize Abnormality Localization} \\\cline{2-11}
		&\multicolumn{2}{C{3.35cm}|}{Stationary sensors} 
		&\multicolumn{2}{C{2.75cm}|}{Mobile sensors (non-propelled)} 
		& \multicolumn{3}{c|}{Stationary sensors (passive)} 
		& \multicolumn{3}{c|}{Mobile sensors} \\\cline{2-11}
		& \multicolumn{1}{C{1.5cm}|}{\scriptsize Non-cooperative detection (single sensor)} 
		& \multicolumn{1}{C{1.7cm}|}{\scriptsize Cooperative detection (multiple sensors and an FC)} 
		&\multicolumn{1}{C{1.3cm}|}{\scriptsize Non-cooperative sensing/ activation} 
		&\multicolumn{1}{C{1.3cm}|}{\scriptsize Cooperative sensing/ activation} 
		& \multicolumn{1}{C{1.4cm}|}{\scriptsize Classification}
		& \multicolumn{1}{C{1.2cm}|}{\scriptsize Discrete-model}  
		& \multicolumn{1}{C{1.3cm}|}{\scriptsize Continuous-model} 
		& \multicolumn{1}{C{1.3cm}|}{\scriptsize Non-propelled (passive)} 
		& \multicolumn{1}{C{1.3cm}|}{\scriptsize Self-propelled (active)}
		& \multicolumn{1}{C{1.3cm}|}{\scriptsize Externally-propelled (active)}
		\\ \hline 
		\multicolumn{1}{|C{2.5cm}|}{\cite{khalid2018system, khalid2020modeling, amin2021viral}} &  $\checkmark $&  &&&&&&  &   &  \\ \hline	
		\multicolumn{1}{|C{2.5cm}|}{\cite{ghavami2017abnormality,ghavami2020anomaly,mosayebi2017cooperative,mosayebi2018advanced,ghoroghchian2019abnormality,solak2020neural, solak2020rnn, solak2020sequential,mai2017event}} &  & $\checkmark $&  & & &&& &  & \\ \hline
		\multicolumn{1}{|C{2.5cm}|}{\cite{mosayebi2018early, varshney2018abnormality, rogers2016parallel, stelzner2016precise, liu2020detection, gomez2021machine}} &  & &$\checkmark $  &&& && &  &   \\ \hline	
		\multicolumn{1}{|C{2.5cm}|}{\cite{felicetti2014molecular}} &  &  &$\checkmark $&&&&& &$\checkmark $&   \\ \hline
		\multicolumn{1}{|C{2.5cm}|}{\cite{khaloopour2021theoretical}} &  & &  & $\checkmark $ &  &&& $\checkmark $&   & \\ \hline
		\multicolumn{1}{|C{2.5cm}|}{\cite{
				kumar2020nanomachine,
				baidoo2020channel,
				gulec2020localization,
				qiu2015long,
				yetimoglu2021multiple,
				matthes2005source
				}} &  &  & & &  &&$\checkmark $& &  &   \\ \hline
		\multicolumn{1}{|C{2.5cm}|}{\cite{
						kim2019source,
						vijayakumaran2007maximum,
						murray2015estimating,
						vergassola2007infotaxis,
						li2020inference
					}} &  &  & & & $\checkmark $ &$\checkmark $&& &  &   \\ \hline
		\multicolumn{1}{|C{2.5cm}|}{\cite{soto2020medical,yang2020recent}}  &  &  & & &  & &&&$\checkmark $&$\checkmark $  \\ \hline
		\multicolumn{1}{|C{2.5cm}|}{\cite{fabbiano2016distributed}}   &  &  & && &&&& & $\checkmark $  \\ \hline
		\multicolumn{1}{|C{2.5cm}|}{\cite{giaretta2015security}}  &  &  & &&&&&&  $\checkmark $&  \\ \hline
		\multicolumn{1}{|C{2.5cm}|}{\cite{okaie2014cooperative,nakano2016performance}}  &  & &&&& & & &$\checkmark $ &   \\ \hline
		\multicolumn{1}{|C{2.5cm}|}{\cite{tran2014localization, zhou2017pulse, el2018high, lemic2021toward}} &  &&&  &&& &  & &$\checkmark $   \\ \hline
		\multicolumn{1}{|C{2.5cm}|}{\cite{shi2020nanorobots,rady2020biosensors,raz2015bioinspired}} & & &  &&$\checkmark $&& &  & &$\checkmark $   \\ \hline
		\multicolumn{1}{|C{2.5cm}|}{\cite{odysseos2021bionanomachine}} &  &  &&&&&&& $\checkmark $  & $\checkmark $  \\ \hline
		\multicolumn{1}{|C{2.5cm}|}{\cite{ishiyama2020cooperative, okaie2018leader}} &&  & && &&  &   &$\checkmark $ & $\checkmark $ \\ \hline
		\multicolumn{1}{|C{2.5cm}|}{\cite{yang2020comprehensive, chen2015touch}} &  &  &&& &&  & $\checkmark $ & &$\checkmark $   \\ \hline
	\end{tabular}
	\label{tab-seg}
		\vspace{-0.5em}

\end{table*}%

\section{Abnormality Detection Schemes} \label{sec:detection}
MC-based abnormality detection schemes 
can be categorized based on using stationary or mobile sensors. 
In both cases, the abnormality detection can be cooperative or non-cooperative.
In addition to the detection scheme, the sensors' sensing/activation can be achieved in either a cooperative or a non-cooperative manner. 
See Table~\ref{tab-seg} for the list of abnormality detection schemes, which will be discussed in this section. 
Further, we discuss the sensing and communication features in abnormality detection schemes, including abnormality recognition methods, sensor mobility, receiver type of the sensors and the FC, and sensing and S2FC/GW communication links, summarized in Table \ref{tab-features}.

We first formulate the abnormality detection problem using the hypothesis testing approach in Subsection~\ref{subsec_det_formulation}. Then, in Subsection~\ref{subsec_det_st}, we describe the abnormality detection schemes using stationary sensors, and in Subsection~\ref{subsec_det_mb}, we describe the schemes using mobile sensors.

\subsection{Abnormality detection problem formulation} \label{subsec_det_formulation}
The abnormality detection problem is modeled as a binary hypothesis testing problem;
$H_0$, the absence of abnormality, and $H_1$, the existence of abnormality, based on the observation vector $\vect{y}=[y_1,y_2,...y_n]$.
The decision is made using a test statistic $T(\vect{y})$ and a set $A$ as follows:
\begin{align}
\begin{cases}
H_1, ~ T(\vect{y}) \in A,\\
H_0, ~ T(\vect{y}) \in A^\textrm{c}.
\end{cases}
\end{align}
Assume $f_{H_0}(\cdot)$ and $f_{H_1}(\cdot)$ are the probability density functions of $T(\vect{y})$ conditioned on hypotheses $H_0$ and $H_1$, respectively. The probability of detection in binary hypothesis testing problem, i.e., the probability of correctly deciding $H_1$, is
\begin{align}
P_\textrm{d}=\mathbb{P}(T(\vect{y}) \in A|H_1)=\int_{l \in A}f_{H_1}(l)dl,
\end{align}
and the probability of false alarm, i.e., the probability of falsely deciding $H_1$, is
\begin{align}
P_\textrm{fa}=\mathbb{P}(T(\vect{y}) \in A|H_0)=\int_{l \in A}f_{H_0}(l)dl.
\end{align}
Further, the miss-detection probability, i.e., the probability that the presence of $H_1$ is missed, is defined as 
\begin{align}
P_\textrm{m}=\mathbb{P}(T(\vect{y}) \in A^\textrm{c} |H_1)=1-P_\textrm{d}.
\end{align}
There is a trade-off between the false alarm and miss-detection probabilities, i.e., reducing one of them increases the other one. 
The goal in this problem is to find a good test statistic $T$ and the set $A$, leading to a low miss-detection probability $P_\textrm{m}$ as well as a low false alarm probability $P_\textrm{fa}$. 

We first describe a binary hypothesis test with a fixed sample size $n$ and then {we extend it} to a sequential test with a variable sample size. For the case of the binary hypothesis test, the Neyman-Pearson lemma gives the optimal test, where the
Neyman-Pearson criterion deals with the trade-off between $P_\textrm{fa}$ and $P_\textrm{m}$ by solving the following optimization problem \cite{poor1994elements}:
\begin{align}
\nonumber
&\min_{T,A} P_\textrm{m}\\
&\textrm{subject to}~P_\textrm{fa}\leq \gamma.
\end{align}
According to the Neyman-Pearson lemma, given $f_{H_0}$ and $f_{H_1}$, the optimal decision rule that minimizes $P_\textrm{m}$ (or equivalently maximizes $P_\textrm{d}$), for a fixed sample size $n$ and any $P_\textrm{fa}$, is the likelihood ratio test (LRT) as follows \cite{poor1994elements}:
\begin{align}
\lambda_\textrm{LR}(\vect{y}) \underset{H_0}{\overset{H_1}{\gtrless}} \mathcal{T},
\end{align}
where $\mathcal{T}$ is the threshold that satisfies $P_\textrm{fa}=\int_{\mathcal{T}}^{\infty}f_{H_0}(l)dl$ and $\lambda_\textrm{LR}(\vect{y})$ is the likelihood ratio as follows:
\begin{align}
\lambda_\textrm{LR}(\vect{y})=\frac{\mathbb{P}(\vect{y}|H_0)}{\mathbb{P}(\vect{y}|H_1)}.
\end{align}
The optimal decision rule can also be written using log-likelihood ratio (LLR), i.e., $\lambda_\textrm{LLR}(\vect{y})=\log(\lambda_\textrm{LR}(\vect{y}))$, as
\begin{align}
\lambda_\textrm{LLR}(\vect{y}) \underset{H_0}{\overset{H_1}{\gtrless}} \log(\mathcal{T}).
\end{align}

An alternative approach to deal with the trade-off between $P_\textrm{fa}$ and $P_\textrm{m}$, introduced by Wald \cite{wald2004sequential}, is to assume a sequential test with variable sample size $n$ and to minimize $n$ (i.e., to minimize the decision delay) for predetermined values of $P_\textrm{fa}$ and $P_\textrm{m}$. This criterion yields  
Wald's sequential probability ratio test (SPRT) \cite{young1998sequential}, which uses the observation sequence $(y_1,y_2,...)$ and obtains the likelihood ratio sequence $(\lambda_1,\lambda_2,...)$, where
\begin{align}
\lambda_n=\lambda_\textrm{LR}(y_1,y_2,...,y_n)=\frac{\mathbb{P}(y_1,...y_n|H_0)}{\mathbb{P}(y_1,...y_n|H_1)},
\end{align}
for $n \in \mathbb{Z}^+$. The decoder continues sampling as long as $\mathcal{T}_1 < \lambda_n < \mathcal{T}_2$ and stops collecting the samples after the $\tilde{n}$-th observation and decides {on a} hypothesis if $\lambda_{\tilde{n}} \leq \mathcal{T}_1$ (decides $H_0$) or $\lambda_{\tilde{n}} \geq \mathcal{T}_2$ (decides $H_1$). Hence, Wald's SPRT can be characterized as
\begin{align}
\tilde{n}&=\inf\{n>0:  \lambda_n \notin(\mathcal{T}_1, \mathcal{T}_2)\},\\\label{sprt}
&\qquad\begin{cases}
H_0:~\lambda_{\tilde{n}} \leq \mathcal{T}_1,\\
H_1:~\lambda_{\tilde{n}} \geq \mathcal{T}_2,
\end{cases}
\end{align}
where $\mathcal{T}_1$ and $\mathcal{T}_2$ are obtained according to the predefined values of $P_\textrm{fa}$ and $P_\textrm{m}$ as
 ${P_\textrm{fa}=\int_{l \geq \mathcal{T}_2}f_{H_0}(l)dl}$ and ${P_\textrm{m}=\int_{l \leq \mathcal{T}_1}f_{H_1}(l)dl}$. This approach is called sequential hypothesis testing which has been used for MC-based abnormality detection in \cite{solak2020sequential, ghoroghchian2019abnormality}.

\renewcommand{\tabcolsep}{2pt}
\newcolumntype{C}[1]{>{\centering\let\newline\\\arraybackslash\hspace{0pt}}m{#1}}

 \begin{table*}[h]
	\centering
		\rowcolors{2}{}{light-gray}
	\caption{Sensing and reporting features in abnormality detection schemes.}
	\begin{tabular}{|*{12}{c|}}  % repeats {c|} 18 times
		\hline
		\multicolumn{1}{|C{2.2cm}|}{\normalsize Ref.}  
		&\multicolumn{1}{C{2.2cm}|}{\normalsize Abnormality recognition} 
		& \multicolumn{1}{C{3.1cm}|}{\normalsize Sensor and FC mobility}
		& \multicolumn{1}{C{3cm}|}{\normalsize Receiver type at the sensors and FC} 
		& \multicolumn{1}{C{1.8cm}|}{\normalsize Sensing link}
		& \multicolumn{1}{C{2.5cm}|}{\normalsize {S2FC/GW} communication link}
		& \multicolumn{1}{C{1.8cm}|}{\normalsize Sensor decision}\\ \hline
		\cite{khalid2020modeling, amin2021viral} & \scriptsize Molecule release & \scriptsize Stationary sensor& \scriptsize Reactive sensor & \scriptsize Molecular & --- & \scriptsize Hard\\ \hline	
		\multicolumn{1}{|C{2.2cm}|}{
		\cite{ghavami2017abnormality}} & \scriptsize General &\multicolumn{1}{C{3.1cm}|}{ \scriptsize Stationary sensors and FC}& \scriptsize Reactive sensors & \scriptsize Molecular & \scriptsize General & \scriptsize Hard \\ \hline
		\multicolumn{1}{|C{2.2cm}|}{\cite{mosayebi2017cooperative}} & \scriptsize General & \multicolumn{1}{C{3.1cm}|}{\scriptsize Stationary sensors and FC}& \scriptsize Absorbing FC & \scriptsize General & \scriptsize Molecular & \scriptsize Soft\\ \hline
		\multicolumn{1}{|C{2.2cm}|}{\cite{ghavami2020anomaly}} & \scriptsize Molecule release & \multicolumn{1}{C{3.1cm}|}{\scriptsize Stationary sensors and FC} & \scriptsize Absorbing  & \scriptsize Molecular & \scriptsize Molecular & \scriptsize Hard\\ \hline		
		\multicolumn{1}{|C{2.2cm}|}{\cite{mosayebi2018advanced}} & \scriptsize Molecule release & \multicolumn{1}{C{3.1cm}|}{\scriptsize Stationary sensors and FC} & \scriptsize Reactive sensors and FC & \scriptsize Molecular & \scriptsize Molecular & \scriptsize Hard \\ \hline
		\multicolumn{1}{|C{2.2cm}|}{\cite{solak2020neural,
		solak2020sequential}} & \scriptsize General & \multicolumn{1}{C{3.1cm}|}{\scriptsize Stationary sensors and FC} & \scriptsize Absorbing FC& \scriptsize General & \scriptsize Molecular & \scriptsize Soft \\ \hline
		\multicolumn{1}{|C{2.2cm}|}{\cite{mai2017event, ghoroghchian2019abnormality}} & \scriptsize General & \scriptsize Stationary sensors and FC& \scriptsize Absorbing FC & \scriptsize General & \scriptsize Molecular & \scriptsize Hard \\ \hline
		\multicolumn{1}{|C{2.2cm}|}{\cite{mosayebi2018early}} & \scriptsize Molecule release & \multicolumn{1}{C{3.1cm}|}{\scriptsize Mobile (non-propelled) sensors and stationary FC} & \scriptsize Transparent  & \scriptsize Molecular & \scriptsize Molecular & \scriptsize Soft \\ \hline	
		\multicolumn{1}{|C{2.2cm}|}{\cite{varshney2018abnormality}} & \scriptsize General & \multicolumn{1}{C{3.1cm}|}{\scriptsize Mobile (non-propelled) sensors and FC} & \scriptsize Absorbing FC & \scriptsize General & \scriptsize Molecular & \scriptsize Hard \\ \hline	
		\multicolumn{1}{|C{2.2cm}|}{\cite{rogers2016parallel}} & \scriptsize Molecule release & \multicolumn{1}{C{3.1cm}|}{\scriptsize Mobile (non-propelled) sensors and stationary FC} & \scriptsize Transparent FC & \scriptsize General & \scriptsize Molecular & \scriptsize Hard \\ \hline	\multicolumn{1}{|C{2.2cm}|}{\cite{gomez2021machine}} & \scriptsize Molecule release & \multicolumn{1}{C{3.1cm}|}{\scriptsize Mobile (non-propelled) sensors and stationary GW}& \scriptsize General  & \scriptsize Molecular & \scriptsize {Wireless} & \scriptsize Hard \\ \hline 
		\multicolumn{1}{|C{2.2cm}|}{\cite{khaloopour2021theoretical}} & \scriptsize General &\multicolumn{1}{C{3.1cm}|}{\scriptsize Mobile (non-propelled) sensors ans stationary FCs}&  \scriptsize Transparent FC & \scriptsize General & \scriptsize Molecular & \scriptsize Hard   \\ \hline
	\end{tabular}
	\label{tab-features}
		\vspace{-0.5em}

\end{table*}%

Alternatively, the decision-maker may choose from a subset of actions to control the information 
extracted from the observation data and enhance the decision-making fidelity \cite{naghshvar2013active, kartik2019active}. This approach, which is called active hypothesis testing, was first introduced by Chernoff \cite{chernoff1959sequential} for the sequential case. However, it can be used in both fixed sample size and sequential cases \cite{nitinawarat2012controlled}. Active hypothesis testing has been used for abnormality detection in \cite{kartik2020testing, cohen2015active}. 
However, there are no works in the {MC literature} that use this approach.

\subsection{Abnormality Detection Schemes using Stationary Sensors}\label{subsec_det_st}
In this subsection, we review the abnormality detection schemes that use stationary sensors. The schemes are divided into non-cooperative detection schemes (single sensor) and cooperative detection schemes (multiple sensors and an FC). In addition to the detection scheme, 
in principle, the sensors can be cooperative or non-cooperative in their sensing/activation. However, there are no schemes in the literature of MC that use stationary sensors with cooperative {sensing/}activation.

\subsubsection{Non-cooperative detection (single sensor)}
Motivated by to the COVID-19 pandemic, \cite{khalid2020modeling, amin2021viral} use a single electronic-based bio-sensor (Silicon NanoWire (Si-NW) field-effect transistor (FET)) in the environment to detect virus-laden aerosols from the exhaled breath of an infected human to viruses (such as influenza or COVID-19). 
To detect viruses using Si-NW FET, the antibody receptors of the virus are placed on Si-NW between the transistor's source and drain. The antigens on the surface of the viruses bind to these receptors, and as a result, a current change occurs across the source-drain channel.

In \cite{khalid2020modeling}, the virus transmission using aerosols is modeled as an MC system. The aerosol channels with continuous sources (associated with breathing) and impulsive (jet) sources (associated with coughing and sneezing) are analyzed for a large room in the presence of airflow.
The steady-state and transient channel responses are obtained for certain initial and boundary conditions. The frequency response of the system is also derived for choosing the appropriate sampling frequency.
Further, a bio-sensor detector along with an air-sampler is proposed to detect the viruses using a predefined threshold. The air-sampler consists of an ionizer, which induces the aerosols with the negative charge, and a charged electrode, which is induced with the positive charge and hence attracts the negatively charged aerosols. {The} performance of the system is evaluated by obtaining the miss-detection probability using simulation, and the effects of different parameters such as distance, virus flow rate, and air flow velocity are analyzed on the miss-detection probability. It is shown that these parameters have a significant effect on the system performance.

In \cite{amin2021viral}, the viral aerosol detection problem is considered in a bounded environment with quiescent air, and partial/full absorption and reflection from the walls. The breathing, coughing, and sneezing sources are modeled as clouds with a relatively symmetrical conical shape and an initial velocity that vanishes eventually through space and time. Hence, there is a zero-velocity zone where the aerosols move only due to diffusion. The authors in \cite{amin2021viral} focus only on the zero-velocity zone and obtain the concentration of the virus-laden aerosols in this zone using an impulsive point source and continuous circular source. The maximum-likelihood threshold for the virus detector and the probability of miss-detection are obtained and the effect of the channel parameters (such as detector location and sampling time) is investigated using simulation. The authors show that the reflecting and absorbing boundaries have a significant effect on the concentration profile of the aerosols and the virus detector performance.

\subsubsection{Cooperative detection (multiple sensors and an FC)}
In \cite{ghavami2017abnormality, %ghavami2012,
 mosayebi2017cooperative, ghavami2020anomaly, mosayebi2018advanced, solak2020neural, %, solak2020rnn, 
solak2020sequential, mai2017event, ghoroghchian2019abnormality}, cooperative abnormality detection is considered with a two-tier network consisting of multiple stationary sensors with non-cooperative sensing/activation and a stationary FC.

In \cite{ghavami2017abnormality}, two scenarios are studied which consider different abnormality recognition methods, categorized as molecule release, medium effects, and molecule absorption. In the first scenario, specific molecules are released from abnormal entities (molecule release). In the second scenario, it is assumed that the abnormal entity changes the medium parameters or absorbs the molecules released from normal entities. For example, it may change the absorbing rate of the molecules at the sensors, devitalize the receptors or change the temperature of the receptors (medium effects), or it may be a competitor cell for the sensors that absorbs the molecules (molecule absorption). 
The molecule sources in both scenarios are modeled by transmitters. In the first scenario, the molecule source is the abnormal entity that releases some molecules. In the second scenario, the molecule sources are other normal entities that release some molecules both in the presence and absence of the abnormality and the abnormality changes the concentration of these molecules by absorbing the molecules or changing the channel parameters.
Multiple sensors that can react with these molecules, acting as molecular receivers, are injected into the medium and the presence of the abnormality is detected by noticing the change in the received molecular concentration.
The sensors communicate their hard decisions (activation flags) to the FC through a micro-communication channel, which can be molecular or non-molecular
{and} is modeled by an additive white Gaussian noise channel. 
The final decision is made by applying a threshold on the received signal at the FC, where the optimum threshold is obtained using the maximum a-posteriori (MAP) rule. This is equivalent to the sub-optimal OR rule, i.e., the FC decides on the existence of the abnormality if at least one of the sensors reports its existence. 
Further, the 
detection and false alarm probabilities of the two-tier network are analyzed, and the number of sensors is optimized for the desired performance.
 Moreover, \cite{ghavami2017abnormality} considers the {impact} of temporal and spatial correlation of the Gaussian observations {which may lead to noticeably inaccurate results if they are not taken into account}.

In \cite{mosayebi2017cooperative}, the sensors measure some input variables which are affected by the abnormality (such as the concentration of biomarker molecules, temperature,  pH value). The sensors make soft decisions by normalizing and quantizing the measured variables. These decisions are reported to an absorbing FC by releasing the same types of molecules (STM) or different types of molecule types (DTM)
For the STM scenario, the optimal detector is obtained at the FC using LLR, leading to a simple decision rule. For the DTM scenario, since the optimal rule has a complex form, sub-optimal decisions are made based on the max-log, maximum ratio combining, and Chair-Varshney approximations of LLR. Further, a two-stage sub-optimal detector is proposed for the DTM scenario: In the first stage, the FC makes local decisions for each sensor using their soft observations; in the second stage, the FC makes the final decision using the local decisions and applying a threshold on the sum of the decisions. 
The overall detection and false alarm probabilities are investigated for both scenarios. A numerical method is proposed for the performance evaluation of LLR-based detectors in terms of their false alarm and miss-detection probabilities. Closed-form expressions are derived for the performance of optimal detector for STM and sub-optimal maximum ratio combining and two-stage detectors for DTM. Further, the asymptotic performance for the large number of sensors is evaluated for the STM and DTM, providing upper bounds for the schemes. It has been shown that for small background noise, the optimal detector for the DTM scenario has a better performance compared to the STM, while for large noise, STM outperforms DTM.

In \cite{ghavami2020anomaly}, the sensors are absorbing artificial cells and the FC is a fixed smart probe, located along a blood vessel, which sends its information to a bio-cyber interface (GW). The abnormal entities are diseased cells that release biomarkers in the medium and the range of the biomarker concentration value distinguishes the healthy and diseased cells. The biomarker release waveform is modeled by the Weibull function, and the sensors detect the abnormality by capturing the variations in the biomarker concentration based on its steady-state
 value and the generalized LRT for independent Poisson observations.
The sensors send a molecular message to inform their hard decision to the FC. If a sensor detects the abnormality, it releases a concentration waveform over time, which is considered the Weibull function; otherwise, it stays silent. The generalized LRT yields
a complex decision rule at the sensors. Hence, a simplified decision rule and a lower bound on its detection probability for a limited false alarm probability are obtained by bounding the likelihood ratio using the maximum likelihood (ML) estimate of the average number of received molecules at the sensors. Further, assuming a symmetric topology for the network, the decision rule at the FC, a lower bound on the total detection probability, and an upper bound on the total false alarm probability are derived. Moreover, the optimum thresholds at the sensors and the FC, which minimize the total detection probability for a limited total false alarm probability, are obtained numerically, and the effect of different parameters on the system performance is numerically analyzed.

The authors in \cite{mosayebi2018advanced} investigate the early detection of the abnormality (such as a tumor) inside the body. The sensors and FC, equipped with reactive receivers, are located near a suspicious area to detect the abnormal entity releasing biomarkers in the medium.
The target location and the biomarker release rate, assumed to be constant, are unknown at the sensors and the FC, while the locations of the sensors are known at the FC.
The average received signal, i.e., the average number of activated receptors at each sensor, and its steady-state probability mass function are derived, which is used to obtain the optimal decision rule at the sensors. The presence of the target is detected at the sensors using the optimal detector with unknown parameters which turns out to be uniformly most powerful (UMP) test, having a performance equal to the Neyman-Pearson detector with known parameters. The UMP test compares the received signal with a threshold instead of using the LLR in the Neyman-Pearson detector. 
Each sensor makes its hard decision using the optimal decision rule and sends it to the FC using different molecule types (different from the biomarker molecules). The final decision is made at the FC using sub-optimal detectors including generalized LRT and generalized locally optimum detector (the detector proposed in \cite{davies1987hypothesis}). Further, a genie-aided detector (aware of the location of the target and its biomarker release rate) is developed at the FC leading to an upper bound on the performance of the proposed detectors.
 
In \cite{solak2020neural, solak2020sequential}, the authors consider 
MC-based abnormality detection for a variety of medical and environmental applications. The sensors sense some variables in the medium affected by the abnormality (molecule release/absorption or medium effects) and make general soft decisions based on their measurements, i.e., the output of the sensors are quantized values between $0$ and $1$ (this is equivalent to the hard decision scenario by setting two quantization levels). The soft decision of the sensors is then sent to an absorbing FC using the same molecule type for all sensors and the final decision at the FC is made using neural network detectors in \cite{solak2020neural} and sequential tests in \cite{solak2020sequential}. In \cite{solak2020neural}, feed-forward neural network (FF-NN) and recurrent neural network (RNN) structures are used for abnormality detection based on the training data obtained from the existing statistical models (Poisson channel model). The training data is a set of received signal sequences labeled with the corresponding hypothesis.
The neural network detector, trained by this set, decides on the hypothesis of a given sequence of the received signal (test). It is shown that the neural network-based detectors perform better than the LLR detector in \cite{mosayebi2017cooperative} for unknown channel parameters, since they provide a better approximation to the unknown parameters. 
In \cite{solak2020sequential}, a decision fusion approach is proposed based on Wald's SPRT in \eqref{sprt}, which uses a variable observation window size and minimizes the average sample size required for the decision. The authors in \cite{solak2020sequential} propose to use the average LLR (taking the average from the likelihood function over the sum of sensor outputs) for both fixed sample size and sequential detectors. It is shown that the sequential detector needs a lower sample size on average and hence a shorter decision delay compared to the fixed sample size approaches for a given detection performance.

In \cite{mai2017event}, the event detection problem is considered in mediums with anomalous diffusion such as turbulent flow, which is usually modeled by L\'evy walk, instead of the normal diffusion where molecules diffuse according to Brownian motion. Multiple sensor nano-machines monitor the environment for certain events which may occur in the medium. The sensors may detect the event  in each time slot and release one molecule of the same type upon detecting the event at the beginning of the next time slot. The released molecule is assumed to diffuse according to L\'evy walk to reach an absorbing FC. The inter-symbol interference caused by the released molecules from the previous time slots is also considered.
Thereby, the number of received molecules from the sensors for each hypothesis, corresponding to the presence and absence of the event, is modeled as a Poisson-Binomial random variable which is approximated to have a Poisson distribution for a large number of time slots. Based on the Poisson observations, the LLR detector is obtained at the FC to detect the abnormality. Further, an online algorithm based on reinforcement learning is proposed to optimize the time slot duration, and thereby the network throughput.
The performance of the proposed algorithm, in terms of the network throughput, and the complexity of the algorithm, in terms of the training phase duration, are investigated using numerical simulations. It is shown that there is a tradeoff between performance and complexity. But, the proposed algorithm has a good performance for a short training phase duration.

In \cite{ghoroghchian2019abnormality}, a moving abnormality is assumed in the medium starting from a random location and time. A sensor network is considered to monitor the environment. The sensors are placed at known locations and partition the environment into different sections. Each sensor senses the environment for the abnormality with certain probabilities of detection and false alarm and if it detects the abnormality, it releases some molecules (which are of different types for different sensors) to inform the FC with an absorbing receiver. Two scenarios are considered at the FC, referred to as stopping time and monitoring scenarios, in which the initiation time of the abnormality and how it spreads in the environment are investigated, respectively. For the stopping time scenario, binary (two hypotheses) and non-binary (more than two hypotheses) cases are considered regarding the abnormality.
In each case, to detect the abnormality initiation time, the quickest change detection and quickest detection and identification of change problems are investigated. For these problems, the stopping time and its corresponding hypothesis that minimize the decision delay are obtained, with constraints on the false alarm and/or miss-detection probabilities. The optimal solutions for these problems are obtained using a partially observable Markov decision process framework, which is complex, and hence a sub-optimal solution is obtained using myopic policy, which is shown to have an acceptable performance.

 \subsection{Abnormality Detection Schemes using Mobile Sensors}\label{subsec_det_mb}
Here, we describe the abnormality detection schemes using mobile sensors. In most of these schemes, the sensors move due to the diffusion and/or the medium flow, i.e., they are non-propelled mobile sensors. Similar to the schemes using stationary sensors, the schemes using mobile sensors can have cooperative detection (multiple sensors and an FC) or non-cooperative detection (a single sensor). In addition, the sensors can be cooperative or non-cooperative in their sensing/activation. However, there are no schemes with a single mobile sensor in the literature and all existing schemes with mobile sensors use cooperative detection. Hence, we divide the scheme into cooperative and non-cooperative sensing/activation.

\subsubsection{Non-cooperative sensing/activation}
In \cite{mosayebi2018early, varshney2018abnormality, felicetti2014molecular, rogers2016parallel, stelzner2016precise, liu2020detection}, multiple mobile sensors with non-cooperative sensing/activation and a stationary/mobile FC is used for cooperative detection of the abnormality in the medium. 
In these schemes, the sensors may be absorbed by the FC to report their measurements and data. More accurately the sensors may move towards the FC and the FC absorbs them for reading their activation levels \cite{mosayebi2018early, khaloopour2021theoretical}.

 In \cite{mosayebi2018early}, early cancer detection using multiple mobile nano-sensors in blood vessels is studied. The cancer cells emit biomarkers in the blood vessels, which propagate by the diffusion and flow, and may be detected by the sensors with transparent receivers. A simplified 2-D model is considered for the cardiovascular system, and the production of the biomarkers from the cancerous and healthy cells are assumed as in \eqref{n_cancer_cells}.
   The sensors move in different paths according to the topology of the blood vessel network and blood flow direction and are eventually collected by a stationary FC (see Fig.~\ref{fig_mosayebi}), where the final decision is made by reading the activation level of the sensors. The optimal LRT is obtained at the FC, which 
   depends on the knowledge of the cancerous cell location and network topology, and hence it is not suitable for practical applications. Therefore, a sub-optimal sum detector is proposed without requiring the network topology. The system model assumptions are validated using particle-based simulations, and the miss-detection probability is obtained using Monte Carlo simulations {for evaluating the performance of the proposed scheme. It is shown that the proposed scheme substantially outperforms the benchmark scheme,
which uses stationary sensors fixed at the location of the FC (equivalent to the conventional blood test), and applies equal gain combining on the sensor observations
(see Fig.~\ref{fig_mosayebi}). 

\begin{figure}[t]
\centering
\includegraphics[scale=0.36]{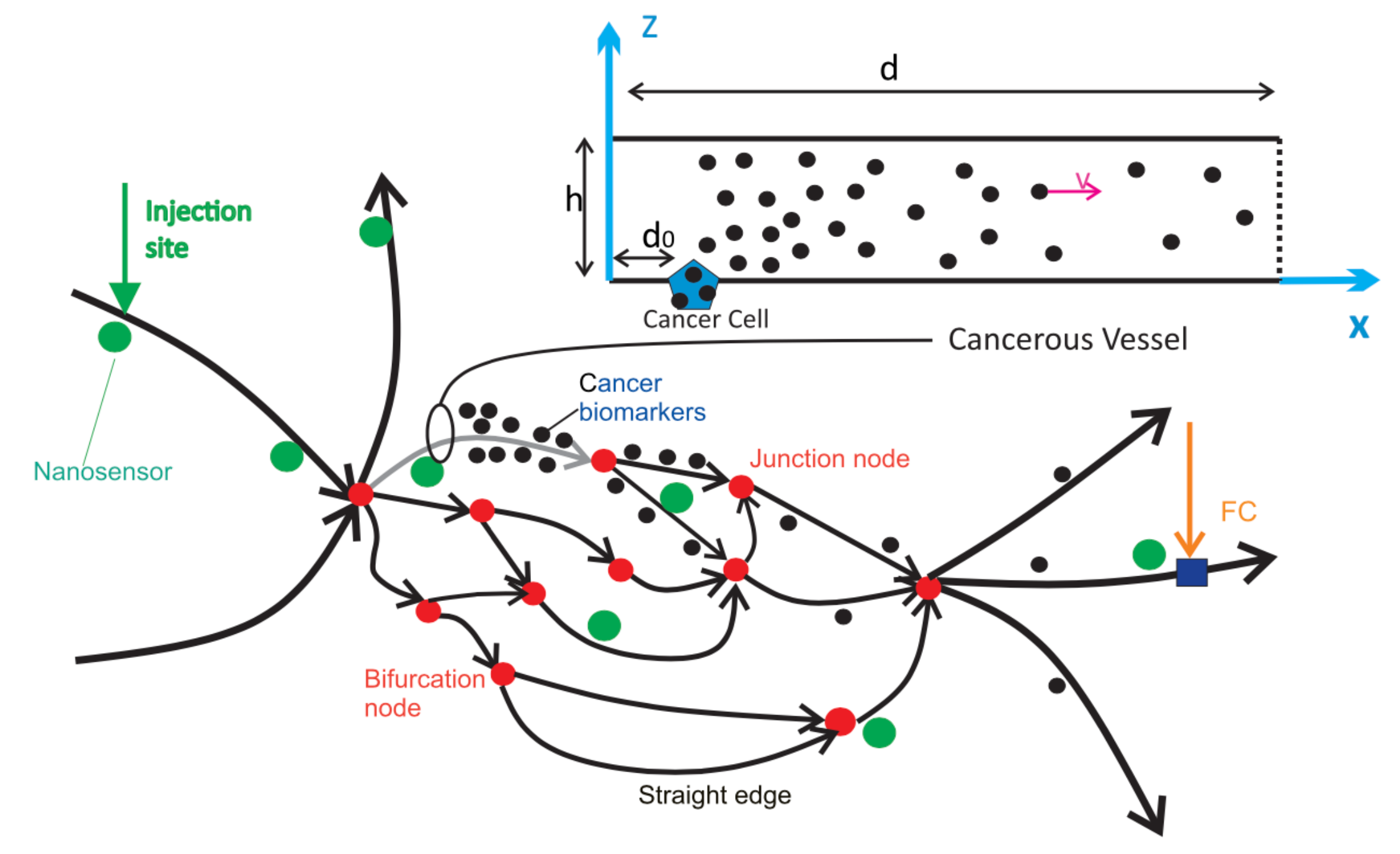}
\caption{Abnormality detection with mobile cells in the cardiovascular system \cite{mosayebi2018early}.}
\label{fig_mosayebi}
\vspace{-1em}
\end{figure}

\begin{figure}[t]
\centering
\includegraphics[scale=0.26]{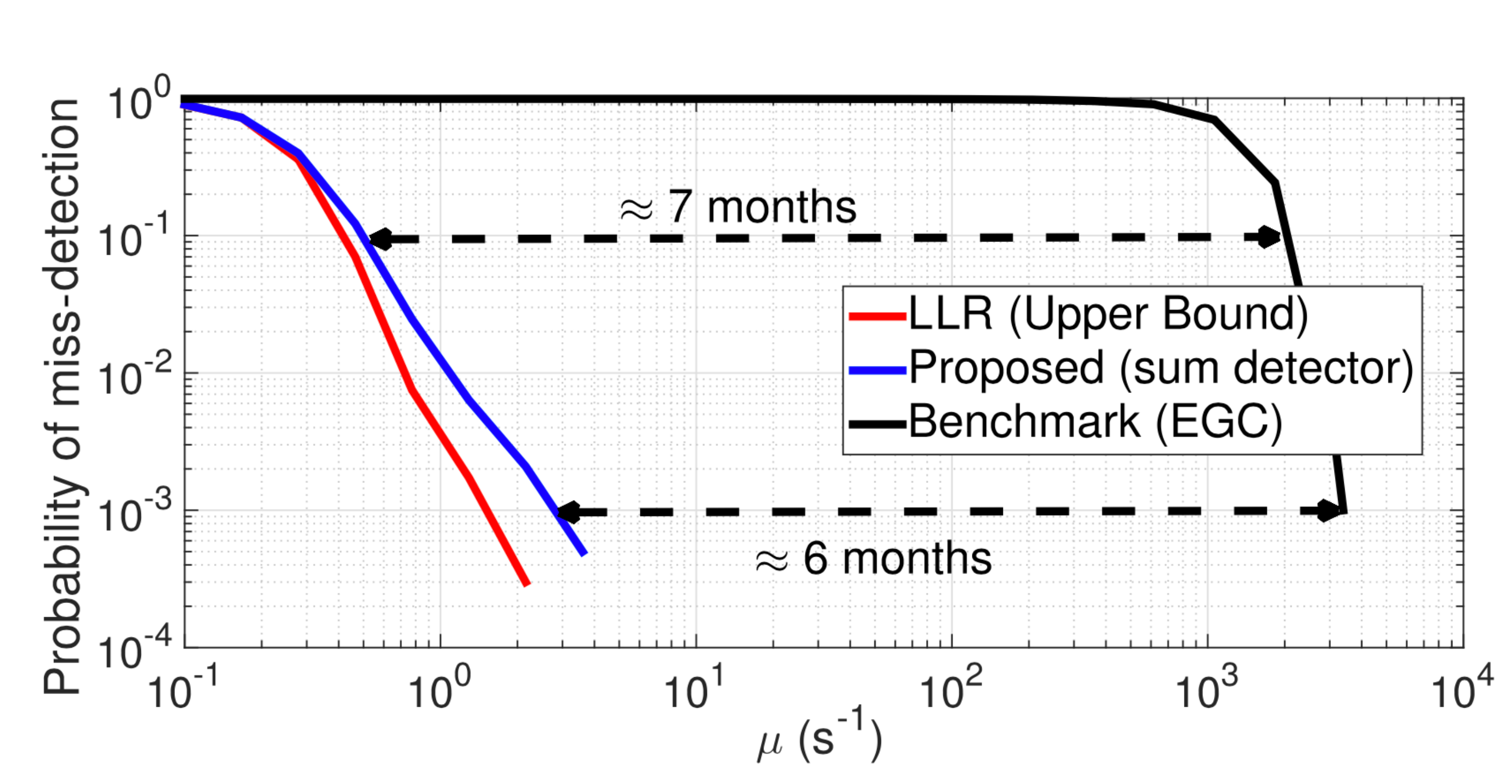}
\caption{Probability of miss-detection versus biomarker release rate ($\mu$) \cite{mosayebi2018early}. During the 7 month time interval, that $\mu$ changes from $0.5~\textrm{s}^{-1}$ to $2000~\textrm{s}^{-1}$, the miss-detection probability of the benchmark scheme (conventional blood test) is more than $10^{-1}$. Further, using the proposed sum detector, for the example considered, the cancer is detected at least 6 months earlier than the conventional blood test.}
\label{fig_mosayebi_comp}
\vspace{-1em}
\end{figure}

In \cite{varshney2018abnormality}, multiple mobile nano-sensors and a mobile FC with absorbing receivers are used inside a single blood vessel to detect an abnormality. The sensors and the FC move along the vessels due to the blood flow. The sensors sense the abnormality with certain detection and false alarm probabilities and send their local decisions to the FC using different molecule types and ON-OFF keying modulation. The FC 
receives the sensors' transmission signals and uses a sub-optimal detector that is based on the approximation of LLR. The FC makes the final decision about the presence of the abnormality using AND/OR rules. For the AND rule, the FC decides on the existence of the abnormality if all of the sensors report its existence.

In \cite{rogers2016parallel}, the \emph{in-vivo} detection of the abnormality is considered using an MC system consisting of
multiple mobile nano-sensors and a stationary FC in a fluid medium with drift (such as blood vessels). 
 The abnormal entity releases biomarkers at a specific rate to the blood vessels detectable by the sensors. The nano-sensors are assumed transparent receivers randomly distributed in the environment with limited movability, i.e., they can hold their position to sense and sample the biomarkers to detect the presence of abnormality. Each nano-sensor decides individually on the presence of the abnormal entity based on its observation, with some probabilities of detection and false alarm, and informs its decision to the FC using two different molecule types (type 1 for the presence and type 2 for the absence of abnormality). The final decision is made at the FC using the optimal Chair-Varshney fusion rule and sub-optimal counting rule, also called $K$ out of $N$ fusion rule, i.e., the FC decides on the presence of abnormality if $K$ out of $N$ sensors report its existence. The false alarm and detection probabilities are obtained for the optimal and sub-optimal fusion rules and it is shown that the sub-optimal counting rule gives an acceptable performance.

In \cite{felicetti2014molecular}, an MC system is proposed for tumor detection using mobile sensors inside the blood vessels and a stationary smart probe that communicates the detection of the cancer cells to the outside of the body. The smart probe can be used to detect the tumor cells originating from the tumor site and circulating in the cardiovascular system. Due to the number of circulating cells in the early stage of cancer, another approach is to use small mobile sensors circulating in the blood vessels to detect circulating cells and release a burst of molecules to inform their decisions to the smart probe. In another scenario, small mobile sensors with slow speed (bacteriobots with self-propulsion mechanism) move towards the tumor cells to detect the tumor site. They inform their decisions to several large mobile sensors that move faster in the blood vessels and eventually reach the smart probe, where the final decision is made and sent to the outside of the body.

\begin{figure*}
\centering
\includegraphics[scale=0.42]{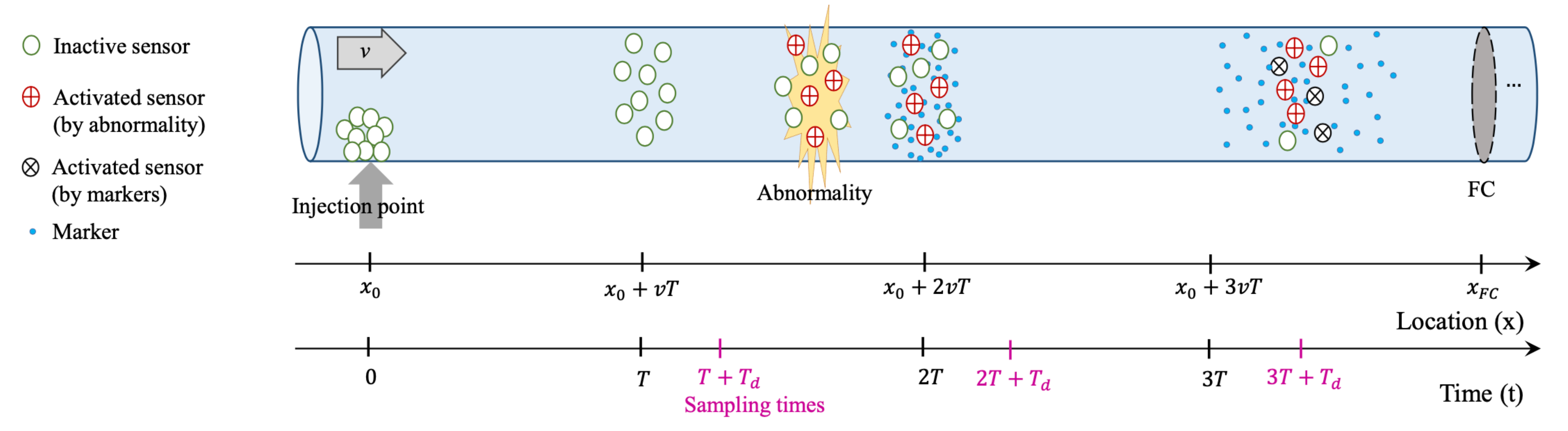}
\caption{Abnormality detection with cooperative sensors in pipelines \cite{khaloopour2021theoretical}. }
\label{fig_khaloopour}
\vspace{-1em}
\end{figure*}

\begin{figure}
\centering
\includegraphics[scale=0.36]{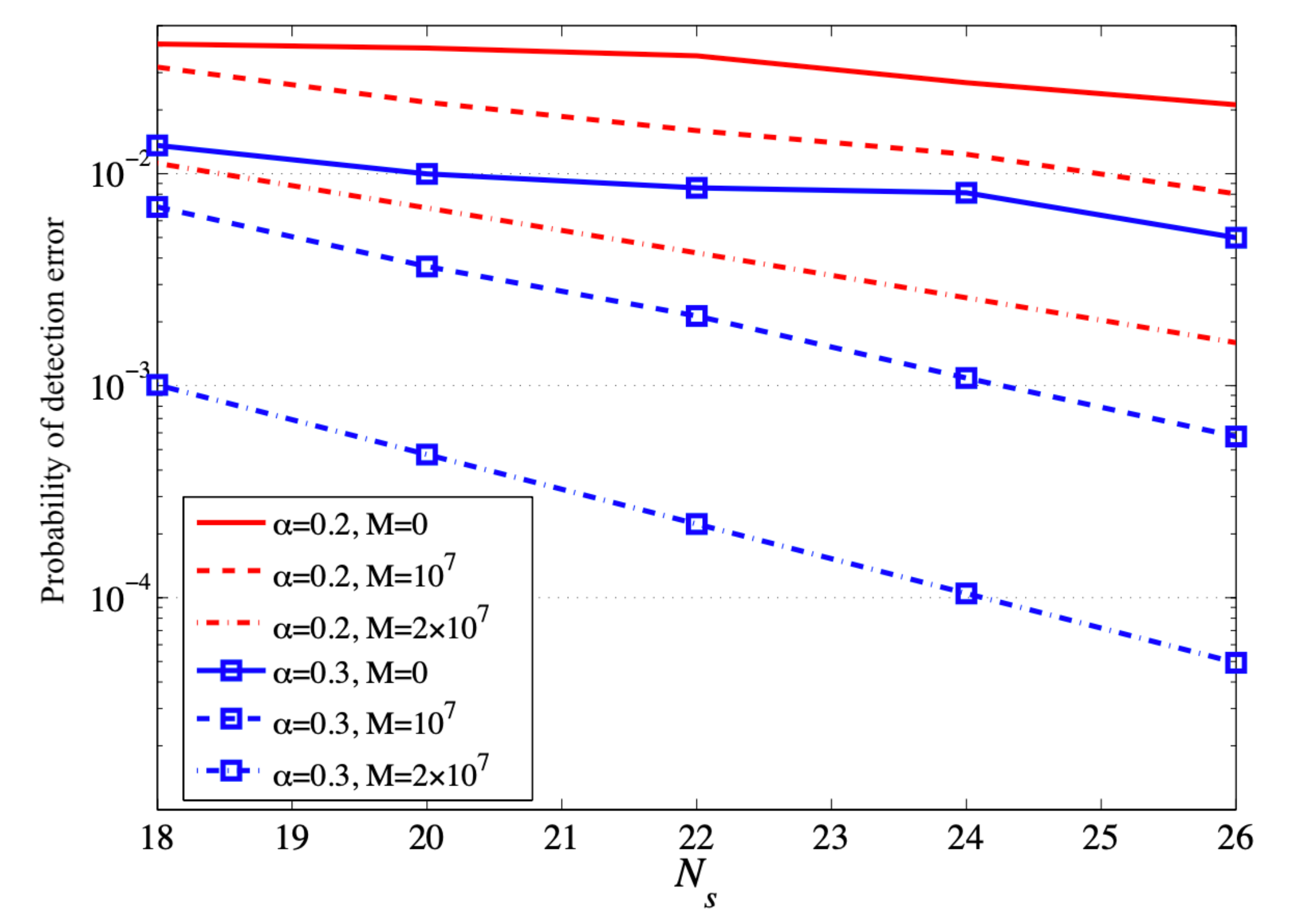}
\caption{Probability of detection error versus the number of sensors ($N_\textrm{s}$) using cooperative sensors in pipelines \cite{khaloopour2021theoretical}. $\alpha$ is the detection probability of the sensors at the abnormality and $M$ is the number of stored molecules at the sensors ($M=0$ corresponds to the non-cooperative activation for the sensors). }
\label{fig_khaloopour_comp}
\vspace{-1em}
\end{figure}

In \cite{gomez2021machine}, early infection detection is considered using mobile nano-sensors inside blood vessels, which is faster than conventional blood tests that take 48-72 hours. The bacterial infections release quorum-sensing molecules for communication among each other. These molecules also reach the blood vessels and may be detected by the nano-sensors. The sensors circulate inside blood vessels and may reach the sensing region (where the concentration of the quorum-sensing molecules is above a detection threshold), which is obtained using COMSOL simulations. If a sensor enters the sensing region and detects the infection, it reports its sensing data to a GW, which is placed out of the body, using ultrasonic or THz signals (though the channel is considered to be ideal). The GW applies the counting rule on the sensor observations to make the final decision. To evaluate the performance of this system, the probability of a sensor entering the sensing region is obtained using machine learning.
To this end, the traveling behavior of the sensors inside the human circulatory system is modeled using BloodVoyagerS \cite{geyer2018bloodvoyagers} (a simulation framework for nano-networks). Further, the traveling path of the sensors in the circulatory system (e.g., the vessels) is modeled as a Markov chain and the transition probabilities are obtained using ML methods. Then, using the transition probabilities, the stationary probability of a sensor located in a vessel corresponding to the sensing region is obtained for each organ inside the body.

Similarly, the authors in \cite{stelzner2016precise} propose an MC network to detect and treat early-stage infections in the human body. Several nano-machines inside the body sense their environment to detect infections and are able to send, receive, and forward messages forming an in-body network, which may {communicate using} molecular or wireless signals. The nano-machines communicate with a control station out of the body (off-body network) using GWs which are implants or on-body micro-organisms. Since the bio-nano-machines operate inside the body, they are mobile and move in the vessels due to the blood flow.

\subsubsection{Cooperative sensing/activation}
In \cite{khaloopour2021theoretical, ishiyama2018epidemic}, the sensors share their decisions with other sensors to cooperate effectively in sensing/activation.

In \cite{khaloopour2021theoretical}, an MC setup is proposed in the cylindrical environment that uses cooperative mobile sensors to detect and localize the abnormality. While most of the ADL methods in MC literature focus on the micro-scale in-body applications, \cite{khaloopour2021theoretical} considers the macro-scale applications and investigates detection and localization of the abnormality (such as leakage) in pipelines. 
For cooperative activation, each sensor after detecting the abnormality and getting activated releases molecules to activate other sensors. There are multiple sensory regions with local FCs at the end of each region (see Fig.~\ref{fig_khaloopour}) that collect the sensors and read their activation levels to decide about the presence of the abnormality using binary hypothesis testing. Two types of sensors are considered in \cite{khaloopour2021theoretical} based on their activation strategies: memoryless and aggregate. The memoryless sensor is activated based on its observation in a single time instance. The aggregate sensor is activated based on the summation of its observations in multiple time instances. The probabilities of false alarm, miss-detection, and detection error are obtained for each sensor type.
As expected, the aggregate sensors outperform the memoryless sensors to detect the abnormality. Further, the authors show that cooperative activation of sensors improves the performance significantly (see Fig.~\ref{fig_khaloopour_comp}).

In \cite{ishiyama2018epidemic}, an MC system using multiple mobile nano-machines with reactive receivers is proposed for information dissemination in the system.
In this system, the nano-machines have two \emph{infected} and \emph{uninfected} states that correspond with awareness and unawareness of the information being disseminated, respectively. The infected nano-machines release information molecules, and the uninfected nano-machines may sense and detect them to change their states and become infected. By repeating this process, the information originating from a source node can be disseminated in the system to reach a destination node. This information dissemination method can be used for target detection by propagating the information about the target in the system, in which the source node is the abnormal entity and the destination node is an FC or a controlling node, and the sensors have cooperation in their sensing/activation.

%%%%%%%%%%%%%%%%%%%%%%%%%%%%
	
%%%   Authors: Dr. Etemadi
\section{Abnormality Localization Schemes}  \label{localization} 
In this section, we review the papers in the area of abnormality 
%\redsout{source} 
localization mainly for medical applications.  
To localize the abnormality, mobile \cite{khaloopour2021theoretical, sinha2018consensus, vergassola2007infotaxis, mesquita2012jump} and stationary \cite{vergassola2007infotaxis, 
	kumar2020nanomachine, 
	matthes2005source, 
	baidoo2020channel,
	guo2016eavesdropper,
	huang2020channel, 
	zhu2021target, 
	bao2021relative, 
	miao2019cooperative, 
	qiu2015under, 
	qiu2015long, 
	li2020inference, 
	farsad2017novel, 
	regonesi2020relative, 
	gulec2020localization, 
	yetimoglu2021multiple, 
	vijayakumaran2007maximum, 
	murray2015estimating} sensors may be used. 
The localization schemes 
are also categorized into passive or active schemes.
In passive localization, the sensors solely determine the location of the abnormality while in the latter, the sensors collectively migrate towards the abnormality using a propulsion mechanism, e.g., in response to an abnormal gradient  or actuating external field; see Table~\ref{tab-seg}.

\subsection{Abnormality Localization Problem Formulation}
Generally, the abnormality {propagation field is} characterized  by \eqref{p1}. 
For the special case of a diffusion field, it is assumed {that} the abnormality
source releases type-$A$ molecules at an unknown rate $Q_{\mathrm{Ab}}$ at an unknown position ${\mathrm{\vect{r}}_{\mathrm{Ab}}}$. The diffusion coefficient for type-$A$ molecules is $D$. Assume {that} the $j$th sensor is located at $\mathrm{\vect{r}}_{\mathrm{s}}(j)$.  
{Considering} an analytical concentration profile $\mathcal{C}(\mathrm{\vect{r}}_{\mathrm{s}}(j),{\mathrm{\vect{r}}_{\mathrm{Ab}}}, Q_{\mathrm{Ab}})$, the standard estimation problem centralized in the FC can be formulated using the following least-square-based method:  
\begin{equation}\label{LS}
\begin{aligned}
& \underset{{\mathrm{\vect{r}}_{\mathrm{Ab}}, Q_{\mathrm{Ab}}}}{\text{minimize}}
& & \sum_{j=1}^{M} \big| \mathcal{C}(\mathrm{\vect{r}}_{\mathrm{s}}(j), {\mathrm{\vect{r}}_{\mathrm{Ab}}}, Q_{\mathrm{Ab}})-y_j\big|^2, \\
\end{aligned}
\end{equation}
where $M$ denotes the number of the sensors in the environment and $y_j$ is the noisy concentration measured by the $j$th sensor. Also, $Q_{\mathrm{Ab}}$ and ${\mathrm{\vect{r}}_{\mathrm{Ab}}}$ are optimization variables of the above least-squares problem.

{The system may} include mobile sensors that should localize and potentially follow the abnormality, as presented in Fig.~\ref{fig_network}. 
The sensors may be classified into leaders or followers.
A leader is responsible for both abnormality detection and localization. Once the abnormality is detected, two options are available to steer the follower agents towards the abnormality for further tasks. The steering procedure depends on the sensor propulsion type. 
In the case of self-propelled sensors in the MC framework, the leader may release type-$B$ molecules to coordinate the follower sensors towards  the abnormality.
For the externally-propelled sensors, a feedback system consisting of sensors, FC, and GW is required to localize the abnormality and modulate the external intervention, accordingly.

%------------------------------------------------
\begin{figure}
	\centering
	\includegraphics[width=7 cm]{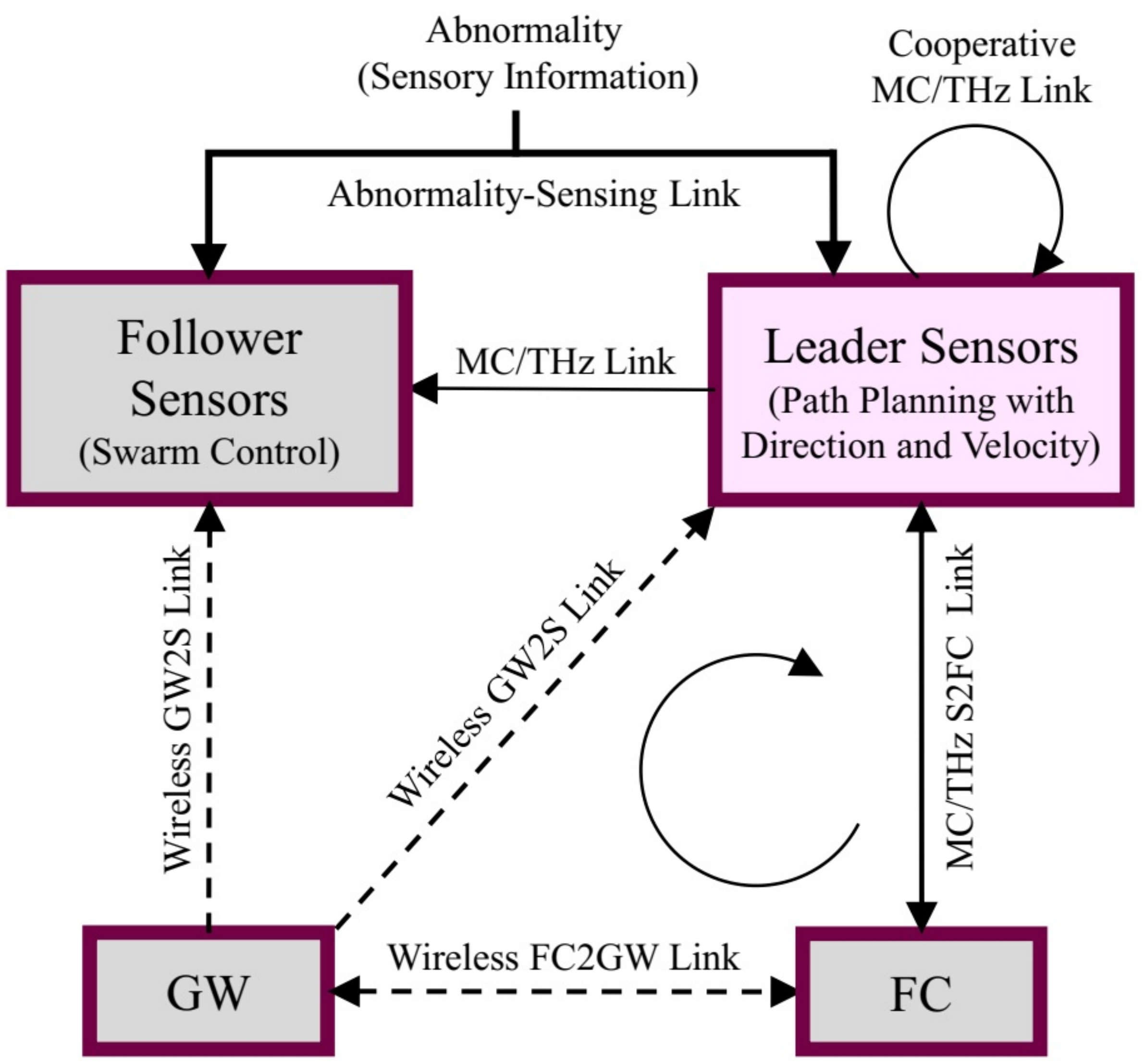}
	\setlength{\abovecaptionskip}{0.2 cm}
	\caption{ Schematic representation of the hybrid localization system. The solid lines refer to links having both MC and wireless (RF, magnetic, and acoustic) technologies. The dashed lines refer to only the wireless technology.  }
	\label{localg2}
	\vspace{-1em}
\end{figure}
%------------------------------------------------

Fig.~\ref{localg2} depicts the hybrid feedback signaling flow among leader sensors (via cooperative MC/THz links), follower sensors, FC, and GW. The solid lines refer to {the} links having both MC and wireless (RF,  magnetic, and acoustic) technologies. The dashed lines refer to only the wireless technology. {Noteworthy}, the links among the same sensor types or between leader sensors and follower sensors are considered to be MC or THz (as a nano-scale RF technology). 
The FC plays the role of a hub that can communicate with either sensor (via S2FC links) or GW (via FC2GW link) to carry the inbound or outbound messaging.
The GW can directly communicate and control the leaders and followers via GW2S inbound links to enforce a dedicated task in addition to the localization. 
{Once leader sensors detect the abnormality, the FC and/or potentially the GW get engaged in the \emph{hybrid localization process}\footnote{{Our aim of using \emph{hybrid localization} is to highlight the role of external interventions (e.g., collaboration of FC and GW using various types of transmission technologies) in addition to the MC-based localization.}}. The hybrid localization enables the  follower sensors to collectively migrate towards the abnormality.}
In the sequel, we explain localization schemes using different sensor types and related communication links in more detail.

\subsection{Abnormality Localization  using Stationary Sensors}
{Typically}, multiple (a network of) sensors are used to localize the abnormality \cite{matthes2005source, kumar2020nanomachine, baidoo2020channel}.
The {MC-based} localization methods using stationary sensors are mainly based on \textit{i}) classification, \textit{ii}) discrete models, {and} \textit{iii}) continuous models \cite{matthes2005source}. 
Classification methods \cite{alpay2000model} simulate the received signals and measurements based on a set of pre-known abnormality positions and compare them with the real measurements. 
Methods based on the discrete models employ a lattice-based structure to discretize the spatial dispersion \cite{murray2015estimating, alpay2000model}. The number of sensors should exceed the number of lattice nodes for better performance which makes these methods costly. 
Methods based on continuous models use analytical dispersion models to formulate the inverse problems and estimate the parameters.
These methods require an analytical dispersion model of the molecule concentration profile in the medium and convergent solutions for corresponding inverse problems. 
The methods using continuous models are the most prevalent ones that we review in Subsection~\ref{subsec_continuous}. We also review the methods using classification or discrete models in Subsection~\ref{subsec_disc_clas}. See Table~\ref{tab-seg} for the list of abnormality localization schemes.

\subsubsection{Classification and Discrete Models}\label{subsec_disc_clas}
The classification models consider multiple \emph{a-priori} known positions for the source of abnormality \cite{kim2019source, li2020inference}. Assuming \emph{a-priori} known position for every source, simulation or  experimental data is generated for all sensors. For classification task, the simulated data is compared with  the {received} signals which is computationally inefficient. On the other hand, discrete models use discretization of the dispersal model. To this end, a lattice-based model for source localization assuming sensors and sources located at lattice points are proposed in the literature \cite{murray2015estimating, alpay2000model}. To attain an acceptable localization accuracy, the number of sensors should be in the same order of the state-space system.

Motivated by classification methods, {the} authors in \cite{kim2019source} propose a recurrent neural network for odor source localization (OSL) problem with stationary sensors. The proposed deep learning approach is based on a combination of classification and discrete-model methods. It is revealed that learning-based approaches in line with probabilistic methods may provide more reliable approaches. Also, feature extraction methods using the support vector machine classification method to learn the environment with stationary sensors should be manipulated to process the measured time series data, gathered from the sensors. It is shown that the model learns the inverse problem and can accurately estimate the leakage point as the source of abnormality.

The papers \cite{matthes2005source,vijayakumaran2007maximum, murray2015estimating} have investigated the macro-scale abnormality OSL problem using multiple sensors.
In \cite{matthes2005source}, the location of a continuously releasing point source is estimated using multiple stationary sensors assuming a turbulent advection-diffusion model. The localization problem in \eqref{LS} is formulated as a quadratically constrained least-squares problem.
Motivated by the discrete-model methods, {the} authors in \cite{vijayakumaran2007maximum} investigate the localization of a diffusive point source provided by a distributed chemical sensor network using the sequential ML estimation algorithm.
In \cite{murray2015estimating}, a sensor network is deployed to localize multiple sources of diffusion at the macro-scale. The sources are classified into instantaneous and non-instantaneous sources. The paper provides simple and exact closed-form inversion formulas based on the dispersal model for both cases of concentration profiles. Using Green's second theorem \cite{zoofaghari2021semi}, a sequence based on Prony's method was developed to estimate the intended source parameter. 
The use of Green's second theorem links the boundary and obstacle surface concentration measurements to the abnormality sources.
The discrete spatio-temporal samples may robustly construct diffusion fields which is critical in multi-focal localization problems. Despite the need for continuous full-field measurements of the environment, the proposed method can be leveraged to recover the source by approximating the generalized measurement model. 
The authors also address the problem in the case of noisy measurements which needs the inversion formula to be adjusted to improve robustness. It is revealed that the proposed method robustly combats noises and model mismatches. The authors also {implemented} an experimental setup and successfully recovered the multi-focal temperature fields.

In \cite{li2020inference}, a machine learning  framework is developed to infer the diffusion channel parameters in a complex turbulent diffusion environment.
A rectangular environment with non-reflecting transparent boundaries is considered with an injector at the origin. The source of abnormality is considered to be the velocity of individual puffs by the injector (as a transmitter) in the environment. It is assumed that the abnormality injects multiple molecular puffs into the environment having different velocities. The goal is to discriminate the velocities using a time difference concentration method based on the support vector machine.
Also, a stepwise maximum variance algorithm is proposed which selects dominant sensing points in a multi-sensor environment (in a lattice as a discrete model approach), leading to the less dimension-related complexities inherent in localization problems with stationary sensors.
Employing different regularization schemes \cite{li2020inference}, the methods based on {machine learning} are approved to accurately estimate the velocities.

In \cite{regonesi2020relative}, the receiver estimates the relative angle of an absorbing interferer, which causes a perturbation in the diffusion environment. A point transmitter is considered in the environment and both the receiver and interferer are assumed to be fully absorbing. 
The results revealed non-trivial behavior of the considered MC system where the variance of the localization error increases with the distance between the transmitter and receiver and their relative look angle. Notably, when the transmitter and receiver are very closely located, the estimation is also unreliable. Moreover, a look angle of $30^{\circ}$ is demonstrated to be  optimal  for accurate interferer localization. The potential application of the proposed framework arises in the self-propulsive motion planning for mobile MC systems.

\subsubsection{Continuous Models}\label{subsec_continuous}
These methods use the analytical dispersal model based on \eqref{p1}, subject to various boundary conditions.
The standard localization problem, formulated in \eqref{LS}, solves inverse problem to find $Q_{\mathrm{Ab}}$ and ${\mathrm{\vect{r}}_{\mathrm{Ab}}}$. However, finding closed-form analytical solution to \eqref{p1} is cumbersome and only semi-analytical or numerical solutions \cite{zoofaghari2021semi} are attainable in case of realistic scenarios.
Assuming the analytical dispersion model in \eqref{LS} and given the location of the sensors in a $d$-dimensional space, at least $d+1$ sensors are theoretically needed to localize the transmitter \cite{matthes2005source}.

In \cite{kumar2020nanomachine}, a multi-sensor framework was proposed to localize a diffusion source of concentration having potential applications in biomedical engineering, industrial applications, and environmental monitoring. Assuming the analytical dispersal model in \eqref{p11} with zero drift, the location of the transmitter as a source of abnormality is estimated and the Hammersley-Chapman-Robinson lower bound (as a generalization of Cram\'er-Rao lower bound) is investigated considering time-dependent noisy measurements and ISI. Also, the condition for a unique recovery of the {abnormality's location are obtained assuming known and unknown positions of the sensors. 
The authors show that the proposed algorithm has a low root-mean-squared (RMS) error.

In \cite{baidoo2020channel} a localization method {using} the peak of the channel impulse response is proposed employing multiple sensing nodes and an FC  with transparent stationary receivers. The sensors report the measurements to the FC for further processing. It is assumed that each sensor picks the largest value for making further decisions. A method based on triangulation (equivalently known as the least-square problem defined in \eqref{LS} or multi-point positioning method), as well as a gradient-descent method, are adopted to minimize a non-convex cost function. The results demonstrate that the gradient-descent approach outperforms the triangulation approach for a wide range of signal-to-noise ratios.  

In \cite{zhu2021target}, the authors consider the source localization problem using a cluster of a limited number of sensing points in a turbulent diffusion channel. A novel back propagation-based neural network, based on Bayesian regularization, is proposed in a bounded environment. A concentration total variation comparison  algorithm was employed to select the most important sensing points. It is revealed that the proposed scheme characterizes the radius and location of the abnormality with {small} error.

In \cite{bao2021relative}, the authors investigate the localization of a silent abnormal entity that absorbs molecules from the medium. The diffusive MC channel in the presence of an absorbing target (abnormality entity) is modeled using the probability equivalent modeling method. 
To localize the abnormality source, the maximum-likelihood problem is formulated and solved based on the Newton-Raphson algorithm. The proposed method is extended to the channel with multiple absorbing abnormalities. 
The results reveal that the localization accuracy depends on the perceived differentiated concentration of molecules at the receiver.

In \cite{guo2016eavesdropper}, the localization problem of an absorbing source (e.g., an eavesdropper) is solved using the finite return probability of the random walk process.  
Also, a 1-D probability density function of the molecules with absorbing and reflecting boundaries, employing mirroring principles, is derived.
It is shown that  abnormalities closer to the transmitter may need more costly computations for reverse estimation,
and for abnormalities closer to the receiver, a simple first-term approximation of probability density function is appropriate with high accuracy for low diffusion coefficient scenarios.
Furthermore, an eavesdropper very close to the transmitter is highly probable to be detected due to the higher rate of molecule absorption.
The proposed method is complicated to localize the eavesdropper in a 3-D environment. 

In a framework similar to \cite{guo2016eavesdropper}, authors in \cite{huang2020channel} characterize the 1-D closed-form analytical channel impulse response in the presence of a point transmitter located between two fully-absorbing receivers. 
The 1-D environment is a good approximation for close inter-cellular regions, e.g., a synaptic cleft.
They analytically derive the fraction of absorbed molecules, the corresponding hitting rate, and the asymptotic fraction of the absorbed molecules as a function of time. The number of molecules absorbed at the receiver is introduced as a metric to localize the eavesdropper abnormality.
It is shown that increasing degradation rate results in a reduction in the mutual influence of two absorbing receivers.

In \cite{miao2019cooperative}, the transmitter positioning is adopted using a single-input multiple-output MC system. 
The cumulative number of molecules hitting a fully-absorbing receiver until a predefined time is considered as the dispersal model.  
The distance from the transmitter and the attenuation factor are estimated using a curve-fitting method in the form of a non-linear least-square problem. This optimization problem is solved using the Levenberg-Marquardt algorithm. After estimating distances, the multi-point positioning method is employed to localize the transmitter.

In \cite{gulec2020localization}, the abnormality localization in the macro-scale is studied using a clustered sensor network in an experimental platform. The network consists of a point transmitter and 24 MQ-3 alcohol sensing nodes placed on a rectangular surface. Also, a Gaussian plume is assumed to model the release of the molecules by evaporation at room temperature to identify the location of the unknown transmitter. 
Assuming the turbulent dispersal model given by \eqref{p11}, joint estimation of location and detection time is considered using numerical algorithms. 
Moreover, a clustering method for estimating wind velocity is proposed based on the estimated detection times. Then, the estimated velocity is employed for estimating the evaporation rate of ethanol in the air.

In \cite{qiu2015long}, the authors investigate the localization of a chemical abnormality (e.g., submarine disaster or airplane crash discovery) using a mobile robot equipped with a transparent receiver.
The background chemical noise is considered to be Gaussian distributed with its variance depending on the receiver's sensitivity defined as the limit of detection (LOD).
The authors assume that due to the motion of the molecules by oceanic waves, the LOD is dominant compared to the counting noise which is typically assumed in the MC literature.
To solve the localization problem, a multi-stage gradient method based on the Rosenbrock gradient ascent algorithm is proposed. The Rosenbrock algorithm, as a numerical optimization algorithm or a particular form of derivative-free search, is well adapted to searching environments with blind methods in the presence of zones with zero gradients. The proposed methods outperform the conventional acoustic methods for underwater source localization.

In \cite{yetimoglu2021multiple}, a  source localization approach for localizing multiple transmitters (i.e.,  diffusing multi-focal abnormality sources) considering a fully absorbing receiver is proposed. 
The scenario considers multiple point transmitters located at different positions which release the same amount of molecules in the environment. A 3-D unbounded environment with zero drift is considered.
The method employs clustering algorithms, i.e., $K$-means, Gaussian mixture models, Dirichlet model, and Bayesian mixture models.
The Gaussian mixture and Dirichlet models are based on the iterative expectation-maximization algorithm. After the clustering stage, the results are employed to provide the direction of arrival by averaging {the} coordinates of the hitting molecules at the surface of the absorbing receiver. Using estimated distance and direction of arrival information, the source can be efficiently localized.

THz-band communications can also be used to establish S2FC links or cooperation among sensors \cite{lemic2021survey, jornet2012joint} to detect abnormal events at the nano-scale. 
However, the wireless nano-sensors have low communication coverage \cite{tran2014localization,zhou2017pulse}. To address this problem, multi-hop localization strategies may be proposed at the cost of error propagation by increasing number of the hops.

\subsection{Abnormality Localization using Mobile sensors}\label{subsec:mobile}
This section introduces the localization using mobile sensors, referred to as mobile microrobots in the literature \cite{yang2020recent}.
As Table~\ref{tab-seg} demonstrates, mobile sensors can be classified into three categories including non-propelled sensors \cite{khaloopour2021theoretical}, self-propelled sensors \cite{vergassola2007infotaxis, mesquita2012jump}, and externally-propelled sensors \cite{soto2020medical,yang2020recent}. The non-propelled sensors do not have a propulsion source, e.g., motor or other types of responses to the concentration changes in the environment. They may be freely distributed in the environment and move due to the inherent forces of the environment (e.g., the blood flow in the body, other forms of flows like in the lymphatic system, {and} the laminar or bulk flow in the industrial liquid ducts) to monitor and localize the abnormality. For self-propelled sensors, the source of propulsion can be either a specific response of the agent to the changes in the environment \cite{tsang2020roads}, e.g., taxis-based approaches \cite{berg2000motile, bazylinski2004magnetosome, dai2016programmable, zhang2016human, liebchen2018viscotaxis}, or other cooperative control-driven approaches \cite{sinha2018consensus, fabbiano2016distributed}. 
Understanding the motion control mechanisms inspires designing multi-functional systems incorporating micro-agents \cite{yang2020recent}. 
In two latter types of motion, the agents are steered toward the target site using self-propulsion mechanisms or an external field force. 
In the sequel, we review the localization methods in these three categories in more detail.

\subsubsection{Non-propelled sensors}\label{non}
While these sensors do not move intentionally, the inherent forces of the environment may move them.

In \cite{khaloopour2021theoretical}, abnormality localization is conducted using non-propelled sensors, non-cooperatively. An FC is required to integrate the local sensing information to use a multi-hypothesis testing problem. Two types of FCs (namely class-A and class-B FCs) are considered with/without requiring sampling of the sensors' marker signal. These markers can be further used by the FC to find the location of the abnormality.
The class-A FCs use markers for localization (activated sensors are known but the corresponding sub-regions where the sensors were activated are unknown) while the class-B FCs know both the flag value (representing the activation statue of individual sensors) and the storage level of each sensor.
It is revealed that employing multiple sensors, markers, and increasing resolution of the FC sensors storage levels (an indicator of the location of the released molecules at the sensor) improves the performance of the localization.
Also, the performance of the localization error in perfect and imperfect sensing regimes for both type-A and type-B FCs is obtained.
For class-A FCs in the perfect sensing regime, by increasing the number of sensors, the probability of localization error asymptotically vanishes. Also, class-B FCs in the perfect sensing regime can accurately (with no error) localize the abnormality.

In \cite{giaretta2015security}, the security challenges of MC systems employing two types of Blackhole and Sentry attacks {are} introduced. The  sensors are classified into two types: malicious and legitimate. In the Blackhole attack, the malicious agent emits attractant molecules to disrupt the legitimate sensors from normal searching for the target site. In the Sentry attacks, the malicious sensors emit repellent chemicals to disperse the legitimate sensors from the target site. A combination of the Bayes' rule and a threshold-based decision-making approach was conducted to induce robustness against such attacks. To this end, a cooperative form of the engineered bacterial swarms is investigated. Once any probable abnormality is detected by any individual legitimate sensors, it switches from emitting repellents to emitting attractants to drag the other legitimate sensors towards itself.

\subsubsection{Self-propelled sensors}
Establishing cooperation among sensors in MC is challenging due to the inherent interference, time variability, and the need for synchronization.
However, swarm behaviors of biological organisms may be used to address this challenge.   
Self-assembly, self-replication, self-feeding, locomotion, and communication are operated mainly in swarm-like collections \cite{pane2019imaging}.
This enables the swarm of sensors with limited intelligence (i.e., having limited storage and computational capabilities)  to realize the localization \cite{berg2000motile, bazylinski2004magnetosome, dai2016programmable, zhang2016human, liebchen2018viscotaxis}.
The self-propelled agents are considered for their potential applications in water decontamination, DNA sensing and detection, water remediation and cargo transport, wastewater treatment, biocompatible engine, biomedical applications, {and} propagation of pollutants \cite{tsang2020roads}. 

The self-propulsion mechanisms include bubble-based, self-electrophoresis, and taxis-based propulsion mechanisms. 
The taxis-based propulsion mechanisms include chemotaxis \cite{berg2000motile}, magnetotaxis \cite{bazylinski2004magnetosome}, phototaxis \cite{dai2016programmable}, rheotaxis \cite{zhang2016human}, and viscotaxis \cite{liebchen2018viscotaxis}.
For instance, in chemotaxis, self-propelled micro-agents are navigated {due to} chemoattractants and chemo-repellents inducing direct migration of the agent towards and away from a diffusion source, respectively \cite{yang2020recent}.
Generally, the self-propulsion mechanism is generated using various types of micro-motors, e.g., droplet swimmer as a chemotactic-based method \cite{francis2015self}. 
Also, some of the taxis-based approaches (e.g., magnetotaxis) are classified as externally-propelled approaches which {will be} discussed in the next subsection.

In \cite{okaie2014cooperative}, the authors design and model a mobile MC system based on the Bacterium-inspired quorum-sensing method, in which repellent and attractant molecules are used to search for and localize the abnormality.
 {To} widely spread in the environment, the sensors release repellent molecules.    
Once an abnormality is detected, the activated sensors start to release attractant molecules to gather other sensors to the target site. To evaluate the {efficiency} of the localization system, the density of the sensors gathered around the target site is considered as a performance metric using a simulation-based environment. 
Employing an FC (i.e., a millimeter-sized implantable device communicating with GW via FC2GW link) in the monitoring environment may help the sensors efficiently coordinate their movements towards the abnormality. A mean-square displacement measure is used to account for the efficiency of the network expansion in response to the repellent molecules released by sensors. 
Similarly, without attractive sensors, the abnormality localization becomes opportunistic and time-consuming.
For feasibility analysis, a wet-lab experiment is also conducted to show the {efficiency} of the localization framework.

In the more recent work \cite{nakano2016performance}, a non-diffusion-based framework for {a} targeted drug delivery system is proposed which uses adhesive molecules instead of diffusive molecules for navigation. The sensors (or nano-machines) are divided into two main groups of leaders and followers. The leaders are spatially distributed and responsible for the exploration of the environment while the followers carry the drug particles.  
The mentioned non-diffusion-based MC system is based on the abundant adhesive molecules (e.g., fibronectin, laminin, elastin, and collagen) in the extra-cellular matrix (ECM) \cite{kim2011extracellular}, which enable a probable binding mechanism for the cells. 
A wet-laboratory experiment relying on mathematical techniques is adopted. 
This is a form of chemotaxis where endothelial cells move up the fibronectin gradient. For experimental setup, calf pulmonary artery endothelial  cells were cultured  at $37^{\circ}$C. Maximum-likelihood estimation was adopted to estimate the parameters of the Langevin equation for both leaders and followers. 
Fig.~\ref{nakano} depicts a sample realization of the leader-follower-based model in \cite{nakano2016performance}. 
Abnormality is depicted on the top left side of Fig.~\ref{nakano} as a circular area in red color. Once every leader sensor touches this area, it starts to release attractant molecules and forms an attractant concentration profile. The attractant molecules are assumed to be adhesive and not diffusive leading to trails. This concentration gradient pattern makes the follower sensors  be gathered around the target site so that they can perform drug delivery to the target site. The distribution of the leader and follower sensors around the detected abnormality in Fig.\ref{nakano} indicates the correct leading and following process of sensors that lead to the gathering of the sensors around the target site. 
Fig.~\ref{nakano} (top) shows the time evolution of the leader sensors freely distributed in the environment to sense the abnormality. Fig.~\ref{nakano} (middle) demonstrates the concentration profile generated by the activated leaders. Fig.~\ref{nakano} (bottom) depicts the time evolution of the followers towards the activated leaders which lead to a reliable localization.

%------------------------------------------------
\begin{figure*}
	\centering
	\includegraphics[scale=0.37]{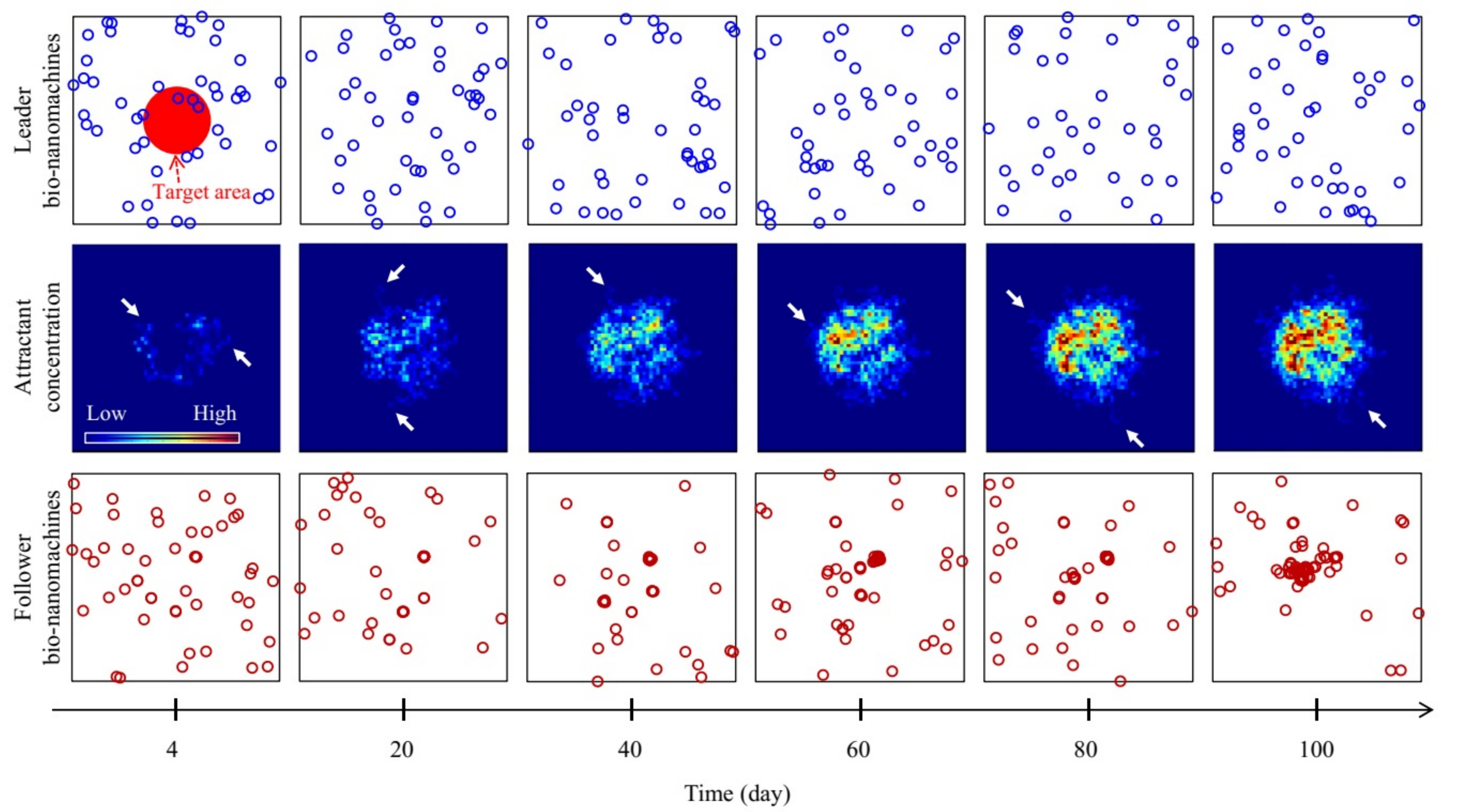}
	\caption{	Distribution of the leader and follower sensors around detected abnormality \cite{nakano2016performance}.  (Top) The leader sensors freely  distributed in the environment, (Middle) the concentration generated due to the activated leader sensors,  (Bottom) the distribution of the followers  attracted towards the abnormality.}
	\label{nakano}
	\vspace{-1em}
\end{figure*}
%------------------------------------------------

In \cite{okaie2018leader}, an MC system is developed which comprises leaders, followers, and amplifying bio-nano-machines, as the sensors. The idea behind the amplifying sensors is to strengthen the attractant field of the leaders in a leader-follower-based framework. The followers migrate towards the abnormality using a discretized space model. In contrast to \cite{nakano2016performance}, which lies on a non-diffusion-based model, the authors in \cite{okaie2018leader} consider a diffusion-based MC with the aid of amplifying sensors. The amplifying sensors follow a simple random-walk in the environment to distribute the attractant molecules switching between reactive and non-reactive phases. The concentration of the attractant molecules released by the leader sensors in the environment is compared to a threshold and if it is higher than a threshold, the amplifying agents react by releasing the same molecules to enhance the gradient field.

In a more recent work \cite{ishiyama2020cooperative}, as a complementary to \cite{okaie2018leader}, the collective migration of the bio-nano-machines (as self-propelled sensors) towards the abnormality in a 2-D unbounded environment is studied. Unlike \cite{okaie2014cooperative, nakano2016performance} where the sensors do not relay the perceived concentration signal, the sensors in \cite{ishiyama2020cooperative} perform depending on the region where they are located.
The sensors outside  the abnormality region use a stop-and-relay strategy, i.e., they relay the perceived signaling molecules and stop for some time to coordinate their motion towards the abnormality. 
Each sensor acts according to its internal state, e.g., the number of activated cell-surface receptors or internal signaling pathways.
The rate at which the sensor releases the molecules in the environment depends on its internal state and how it is located inside or outside of the target region. The release function is designed to support repetitive pulse dispersions if the sensors are located inside the target region. The sensors outside the target region release  molecules  whenever their perceived concentration  increases.
{The locomotion of the relaying self-propelled sensors (as followers) towards the activated sensors (as leaders) is investigated by extensive simulations}  with 99 relay sensors, which shows that only 10 nearest sensors react to the concentration fluctuations due to the leader sensor. The results show that only the 4 nearest sensors can reliably localize and reach the leader. Also, the disruptive effect of the signal relaying by other sensors is confirmed in the simulations which is the main idea behind developing the relay-and-stop strategy to compensate  the unintended motion fluctuations of the sensors.
Finally, the relay-and-stop strategy is examined in a practical setup (including abnormality), and its efficiency to localize the abnormality is approved.

In \cite{vergassola2007infotaxis}, a non-gradient-based method based on 
the Fisher information matrix referred to as infotaxis, is proposed to localize the concentration source. This strategy locally maximizes the expected rate of  information gained to find the best route %\redsout{toward}
to the abnormality source. The idea behind this approach is that entropy decreases faster in the proximity of the source.

From a more practical perspective, \cite{huang2018bioinspired} introduces self-propelled (autonomous) shape-morphing micro-machines consisting of stimuli-responsive hydrogels for drug-delivery applications.
Inspired by the locomotion of the Leukocytes, synthetic thermo-responsive hydrogels (as self-folding nano-sensors) are proposed to localize the abnormality for drug delivery applications.
The abnormality, e.g., tumor site, is primarily localized using conventional medical imaging approaches in an \emph{in-vitro} environment. 
Then, the abnormality is targeted with an external near-infrared light gradient which can be sensed by the injected nano-sensors.
Finally, a rotating magnetic field is applied to steer the micro-machines to explore the monitoring environment to find the abnormality.
The results reveal that the sensors move up along the gradient by increasing velocity and remain at the target site by unfolding their shape to release the drug whenever they arrive. This is a special kind of optotaxis where autonomous localization is conducted using sensory information without external intervention (e.g., imaging feedback). 

For monitoring wide environment  or  slowly-varying diffusion field, cooperative multi-sensor ADL is of high interest. 
In \cite{fabbiano2016distributed}, a stand-alone\footnote{Our aim by using the term \emph{stand-alone} is to highlight that the sensor does not need its own coordinate to localize the source. This is an inherent assumption considered in many bio-inspired localization algorithms, e.g., chemotaxis.} control-driven approach is proposed to steer an equi-spaced circular formation of the sensors towards the abnormality (e.g., a macro-scale heat source located in a room) using distributed gradient-ascent algorithm. The sensors can solely measure the relative angles concerning their neighbors. 
The method of Poisson integral is used to find the concentration gradient which can be approximated by a finite summation of measurements perceived by the sensors. 
Notably, the center of the formed circle by the sensors moves along the estimated gradient field.
The main drawback of the proposed scheme is to use the gradient-ascent algorithm which necessitates convexity assumption over the spatial simulation domain.
The provided steering control algorithms are according to a ring-shaped communication topology where required information can be handed over S2S links using the linear average-consensus algorithm. 
The minimum number of sensors required for the proposed scheme is three without restraining the maximum limit. The {efficiency} of the localization scheme using an illustrative example was approved and theoretically investigated.

Recently, machine learning  approaches have been investigated for OSL problems considering self-propelled sensors in complex environments with time-varying conditions. In \cite{hu2019plume}, due to the spatio-temporal changes of the environment, a continuous state-space and action space model based on a partially observable Markov decision process (POMDP) is proposed.
A deep reinforcement learning method was used to enable sensors to optimize their interactions with their surrounding environment to localize abnormality. 
The POMDP is designed as a long short-term memory (LSTM) structure. Also, a belief-state model is used instead of the conventional state-space model and selects its future action (i.e., path planning) based on its history and current measurement. To maximize the long-term cumulative reward function, an LSTM-based  deterministic policy gradient (DPG) algorithm is proposed.
The training stage of the proposed machine learning approach is accelerated by employing supervised policies (i.e., features inherent in dynamic programming methods).  
A simulation-based abnormality localization setup is considered for a turbulent time-varying environment. To learn a searching strategy, the agent requires to record and use the local concentration measurements, local flow velocity, and the related motion information.
Reynolds-averaged  Navier-Stokes equations are used to make the simulation environment. The results reveals that the LSTM-based DPG algorithm outperforms the deep DPG algorithm  both in efficiency and effectiveness. Also, it is revealed that the LSTM-based DPG supplemented by supervised policy accelerates the localization efficiency.

\subsubsection{Externally-propelled sensors}
Externally-propelled and controllable sensors promise non-invasive health monitoring and interventions \cite{nakano2014externally}.
As the feedback system  in Fig.~\ref{localg2} shows, wireless links (e.g., RF, magnetic, or acoustic) are  {the} only 
mechanisms for external messaging to control the nano-network or {stimulate} the externally-propelled sensors \cite{lemic2021survey, nakano2014externally}. 
Linking the nano-networks (based on MC/THz links) to external communication systems (based on wireless links) is an important ongoing research direction \cite{dressler2015connecting}. The bottleneck is to design appropriate communication links (e.g., FC2GW or S2GW links) to interact with nano-scale externally-propelled sensors. 
Particularly, the hybrid interconnectivity of THz-based nano-networks and MC-based nano-networks was recently reviewed in \cite{yang2020comprehensive}. 
In the following, we review the proof-of-concept models which still demand more theoretical research and investigations.

The authors in \cite{chen2015touch} propose a touch communication (referred to as TouchCom) framework which employs externally-controllable sensors for the transportation of pharmaceutical compounds to the target (e.g., tumor) site inside the body. To this aim, a cross-scale (micro-to-macro) design of communication and control is proposed. The sensors are transient in the sense that they remain in the environment for a medically-useful time duration before any dissolving and resorption by the body and comprise three compartments namely diffusion, branching, and degeneration until they reach the tumor. 
Both silicon-based (transient electronics) and bacteria-based (magnetotactic bacteria) structures are appropriate as sensors for the TouchCom framework. 
The TouchCom can be generalized to other applications including quality control of water and food, environmental pollution, etc. 
The important part of the TouchCom is the external control system (e.g., a GW cooperating with a local cloud computing system) which can partially coordinate the movements of the externally-propelled sensors using a feedback system; see Fig.~\ref{localg2}. The GW intelligently optimizes the localization {strategy} using previous trajectories of the sensors combined with learning methods. The feedback loop can be {implemented} using GW2S links. The GW2S link is a visual communication link that enables the GW to control the localization by touching the tangible maneuvering space using an external field. The propagation delay and attenuation loss minimizations are important measures to evaluate the {efficiency} of the localization system. The experimental setup is based on the blood vessels and capillaries in which the sensors are conveyed and can be seen using an angiogram.
The probability distributions corresponding to the propagation time from the injection site to the tumor, the angle of arrival, and the path loss are investigated. Moreover, the delay spectrum and azimuth spectrum, as crucial parameters for drug-delivery applications are studied.

Recent studies demonstrate the importance of the soft untethered grippers \cite{ongaro2017autonomous} to pick-and-place biological materials in dynamic environments for minimally-invasive surgery and lab-on-a-chip systems.  
Unlike traditional approaches which use imaging as a mechanism to control the in-body systems, ultrasound is more bio-compatible with medical applications. 
Ultrasound provides a high frame rate, which is crucial to realize real-time interventions which can also be the best choice for externally-controllable self-propelled sensors \cite{sanchez2014magnetic}.
An external device, e.g., a GW, can be used to emit wireless signals (e.g., ultrasound signals) to directly accomplish the localization or steering process. 
In \cite{scheggi2017magnetic}, magnetic closed-loop motion control of a soft miniaturized hydrogel-based gripper towards a target site using feedback from ultrasound images is reported. The results demonstrates that the proposed system employing ultrasound images can track the gripper with an error in the order of $0.4~\textrm{mm}$ without payload and $0.36~\textrm{mm}$ conveying a payload. 
The provided system enables the control of miniaturized grippers in the absence of visual feedback cameras.

In \cite{odysseos2021bionanomachine}, a more developed MC-based externally-controllable framework 
based on \cite{nakano2014externally} is proposed
that combines RF communication and MC to realize externally controllable MC systems to apply therapeutic interventions inside the body.  
Furthermore, a proof-of-concept model for targeting cell and exosome-mediated theranostic systems accompanying phenotypic switching and network formation is proposed \cite{odysseos2021bionanomachine}. The authors introduce a novel brain-machine interface to detect, localize, and control molecular activities bidirectionally. The proposed framework consists of an external device (e.g., GW or the local cloud as described in Fig.~\ref{fig_network}) incorporating a processing unit,  storage, and corresponding feedback mechanisms. However, the authors in \cite{odysseos2021bionanomachine} declare that the migration mechanisms of the cells to reach the intended sites need further investigation and research.

In \cite{shi2020nanorobots}, a novel sequential multifocal tumor localization approach using the history of the biological gradient field (BGF) is proposed. The BGF concept is related to the biological characteristics of high-risk tissues in the body. For instance, the capillaries around the healthy tissues exhibit a uniformly-distributed pattern while for the tumor environment, the capillaries are tortuous and highly dense. This leads to varied BGFs which are captured by the GW due to the changed trajectories and magnetic variations of sensors in the multi-focal tumor environment.
The localization is established by externally-propelled sensors, e.g.,  Janus nanoparticles, which are maneuvered by a GW (i.e., external actuating device) using magnetic-based S2GW links. 
The efficiency of the non-intervening delivery methods depends on the body's natural circulatory system leading to  extremely poor delivery performance (about $0.7\%$ of injected particles).  
The sequential localization strategy modifies the BGF and helps the sensors to be divided into multiple clusters to find the tumors one after another.
The sensors are maneuvered by an external actuating device (e.g., a rotating magnetic field generated by a coil) and steered in the fluid using a swarm intelligence algorithm. The typical imaging systems can be deployed to follow the trajectory of the sensors and the BGF is estimated using the sensors' characteristics.
A multi-modal optimization problem to minimize the BGF  is formulated whose solution is the global optimal location of the tumors which is sequentially explored by a share of swarm sensors. 
It is shown that the proposed imaging-and-control-assisted method localizes the tumor sites and outperforms the conventional multi-modal optimization algorithms \cite{shi2020nanorobots}.

In addition to magnetic resonance imaging (MRI) and ultrasound, other technologies may be utilized for the S2GW and FC2GW links involving optical tracking, magnetic particle imaging (MPI), X-ray fluoroscopy, and fluorescence.
In optical tracking, the 3-D localization of the sensor is hard and it is easily disrupted by noise or other environmental optical sources. Currently, this approach is only applicable in transparent micro-environments like vitreous humor (also known as vitreous fluid) in eyes. The MRI is based on the magnetic field, which is bio-compatible non-ionizing radiation causing excellent tissue contrast. This approach is not only employed extensively for imaging but also used to manipulate nano-machines which enables 3-D localization. However, real-time data acquisition is challenging and costly. The MPI has the same advantages as the MRI with high spectral resolution and fast scanning speed. The drawback of the MPI is its limited application to super-paramagnetic nano-particles. The X-ray family has the advantage of deep penetration into the human body. 
As previously described, an interesting approach is ultrasound which is real-time imaging with low cost. Among the technologies mentioned, ultrasound has minimum adverse health effects making it a proper solution for both imaging and nano-sensor control.

\section{Challenges and Directions for Future Works}\label{sec:challenges}

In this section, we discuss the open challenges and directions linked to the ADL problem. We classify them into four main categories: implementation, system design, modeling, and methods in Fig.~\ref{direction}. 
Further, we extensively discuss the mentioned challenges for the internet of things (IoT) systems as an important part of beyond 5G networks, including the internet of nano-things (IoNTs) and internet of bio-nano-things (IoBNTs).

 \subsection{Implementation}\label{practical}
Implementation of MC systems requires multidisciplinary 
collaborative research at the intersection of biology, physics, communications, and computer sciences. Developing multifunctional sensory systems in complex environments is a challenging task that needs more investigation. Here, we provide the challenges and recent trends in 
sensor design, developing interfaces, and creation of testbeds and platforms devoted to the intersection of IoNTs and the future beyond 5G networks.

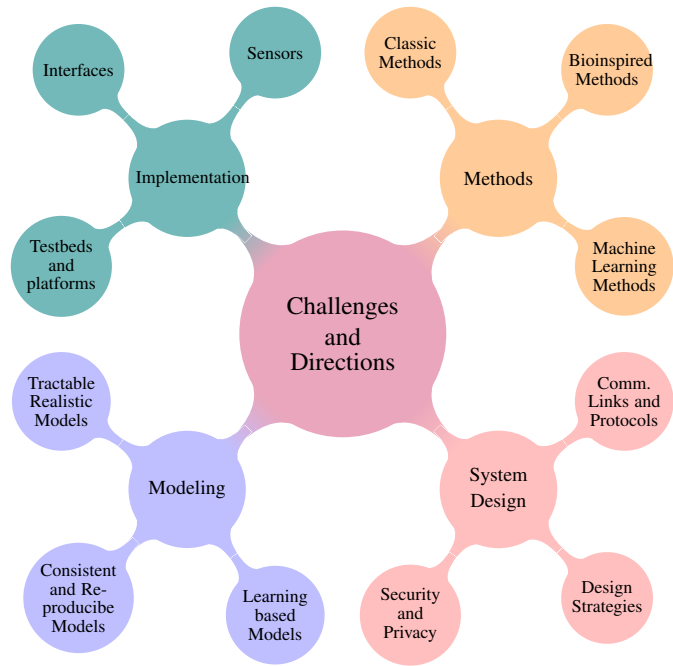
\begin{figure}
	\centering
	\resizebox{\columnwidth}{!}{
	\begin{tikzpicture}	[mindmap, grow cyclic, every node/.style=concept,concept color=purple!35, 
	level 1/.append style={font=\large, level distance=4.3cm,sibling angle=90},
	level 2/.append style={level distance=3cm,sibling angle=45},font=\large]
	\node{\Large{Challenges \\and\\ Directions}}
	child [concept color=blue!25,level 2/.append style={level distance=3cm,sibling angle=80}]{ node {\normalsize{Modeling}}
		child { node {\small{Tractable Realistic Models}}}
		child { node {\small{Consistent and Reproducibe Models}}}
		child { node {\small{Learning based Models}}}
	}
		child [concept color=red!25, level 2/.append style={level distance=3cm,sibling angle=80},]{ node {\normalsize{System Design}}
		child { node {\small{Security and Privacy}}}
		child { node {\small{Design Strategies}}}
		child { node {\small{Comm. Links and Protocols}} }
	}
	child [concept color=orange!40,level 2/.append style={level distance=3cm,sibling angle=80}]{ node {\normalsize{Methods}}
		child { node {\small{Machine Learning Methods}}}
		child { node {\small{Bioinspired Methods}}}
		child { node {\small{Classic Methods}}}
	}
	child [concept color=teal!55,level 2/.append style={level distance=3cm,sibling angle=80}]{ node{\small{Implementation}}
		child { node {\small{Sensors}}}
		child { node {\small{Interfaces}}}
		child { node {\small{Testbeds and platforms}}}
	};
	\end{tikzpicture}
 }
	\caption{Schematic of the research challenges and directions for {MC-based ADL}.}
	\label{direction}
	\vspace{-1em}	
\end{figure}

\textbf{Sensors:}
Developing multi-functional sensors with different physical and technical capabilities such as mobility mechanisms, detection schemes, storage capabilities, release techniques, sampling methods, reception processes, and synchronization are very critical, particularly for nano-scale applications which need great attention from multidisciplinary perspectives. 
While various constructions have been provided for sensors in macro-scale applications, the sensors cannot be necessarily scaled down to nano- and micro-scale dimensions to be operated in a micro-environment. Designing nano-sensors is an ongoing research trend in the physical chemistry field \cite{kuscu2016physical}. 
A promising method to manufacturing nano-sensors is based on using biological elements.
Sensors can be electrical, optical, or mechanical based on the converted signal in the transducer unit. However, electrical transducers gain more attention since they do not need macro-scale elements for detecting optical and mechanical signals. 
To realize electrical transducers, designing miniaturized signal processing units at the nano-scale is urgent  \cite{kuscu2016physical}. The design of biocompatible nano-sensors having potential applications in nano-scale sensory systems and synthetic biology is another challenge that embraces attention from the engineering field.

FET-based bio-sensors \cite{curreli2008real} provide another promising method to manufacturing bio-nano-sensors. The electrical transducers in these sensors consist of a source, a drain, and a semiconductor channel between them covered by multiple bio-receptors for receiving molecules, which is an important element of the transducer. 
Semiconductor fabrication at the nano-scale faces reproducibility problem due to the randomness in the bio-reception process. Also, imperfect fabrications in the system parameters result in extra noise in the analytical model.

\textbf{Interfaces:}
Efficient control of the nano-networks requires more reliable and sensitive implantable interfaces operating in FCs and GWs. Realizing highly-accurate real-time interfaces, e.g., FCs and micro-GWs for nano- and micro-scales is imperative, particularly for healthcare applications. At the nano-scale, fabrication of signal conversion entities, e.g., opto-genomic \cite{jornet2019optogenomic} or electro-chemical \cite{maduraiveeran2018electrochemical} interfaces are of high importance. Typical interfaces need an indirect use of electronically-controlled devices, e.g., the drug injecting systems \cite{chahibi2013molecular} which may trigger the in-body nano-machines. The low probability of triggering a chemically-activated nano-machine from external systems due to the diffusion is the main drawback of such interfaces. A more interesting form of direct external intervention is the use of electroencephalogram (EEG) signals as a carrier of information in brain-machine interfaces. Light-assisted control of the genomic processes, referred to as opto-genetics, has been introduced to control biological processes.
Plasmonic nano-lasers enable the integration of nano-antennas and single-photon detectors functioning as a nano-actuator of light-controlled processes. Energy harvesting and performance of signal conversion are two desired parameters for the adoption of biological micro-electromechanical systems (bio-MEMS).
Despite progress in external control of biological environments, there is still a long way to design efficient bio-MEMS. 

\textbf{Testbeds and Platforms:}
The efforts to emulate biological environments \cite{grebenstein2019molecular, krishnaswamy2013time, de2013communications, abbasi2018controlled}, mostly have focused  to develop microfluidic and macro-scale platforms \cite{farsad2013tabletop, koo2016molecular}. 
Similar to the innovative technologies which need to incentivize the industry for investment, developing application-centric prototypes and platforms for MC systems are also necessary for IoBNT applications \cite{tataria20216g}.

\subsection{System Design}
 \textbf{Communication Links and Protocols:}
The sensors can send the sensing information, i.e., the local decisions, the test statistics, or the unprocessed data centrally to the FC. While the FC is responsible for making global decisions, it may also take the role of central coordinator among  sensors for efficient localization.
 As depicted in Fig.~\ref{fig_network}, the leader-follower framework (which considers  mobile sensors as  self-propelled agents) is  highly reliant on the {efficiency} of MC-based intra-sensor and S2FC communication links. 
Despite extensive research in the area of mobile MC systems, still much effort is needed to realize efficient communication and control among sensors. One challenge is to probabilistically model the mobile MC channels  for {ADL} in complex environments with obstacles, turbulences, or reactive boundaries.

Optimal path planning using resource-constrained leader sensors is of high importance to localize the abnormality.  As previously described in Section~\ref{practical}, the rate-constrained MC channels limit the efficiency of the real-time machine-learning-based detection and localization algorithms for IoNT systems. In addition, mobility of the sensors or  turbulences inherent in the environment may lead to outdated channel state information (CSI) which deteriorates the connectivity for MC systems. THz communication is an alternative for both S2S and S2FC links.
THz communication suffers from shadowing which implicates the proper modulation schemes for reliable data transmission.

If the FC itself is unable to handle the decision-making or coordination processes, it may collaborate with  GW via FC2GW communication link \cite{dressler2015connecting}.
 The GW can be implanted near the intended monitoring environment (referred to as micro-GWs), or external to it (e.g., on-body GWs in healthcare applications). 
The wireless technology (e.g., acoustic, magnetic, and RF) can be employed for both communication and coordination of sensors and FC; see Fig.~\ref{localg2}.
 More accurate statistical channel models are required to implement a reliable communication {link}. In the specific case of RF communications for WBAN, the channel model follows a log-normal distribution \cite{abdulfattah2019performance}. However, developing THz-based FC2GW and GW2S links using nano-scale sensors is more challenging and needs further research. 
This technology uses sensors equipped with THz nano-antennas for IoNT applications where the GW is located outside of the sensors' micro-environment \cite{lee2015highly}.

Bio-compatible FC2GW links are essential for healthcare applications.  
To date, many experimentally-tested options for externally-propelled sensors {have been} proposed including magnetic-assisted mechanisms \cite{kisseleff2016magnetic, ulbrich2016targeted}, floresent-assisted mechanisms \cite{osmekhina2018controlled}, Optofluidics\footnote{Optofluidics is a promising technology to monitor biological \emph{in-vitro} systems which combines optical microscopy with microfluidics \cite{bi2021survey} invasively.} \cite{bi2021survey}, and ultrasound. Besides, imaging methods based on contrast-aided MRI and ultrasonic have been revealed to be an appropriate candidates for \emph{in-vivo} monitoring \cite{maresca2018biomolecular}.
Utilizing more reliable external control of the nano-scale is an ongoing research trend.

There are extensive applications ranging from monitoring to control of inaccessible environments (e.g., in-body networks) which require external interventions. The emergence of 5G networks also provides ultra-reliable low-latency communications to a remote health monitoring center.
When external interventions are needed (e.g., an urgent decision made by a physician), the local decisions may be sent to a remote health center for a precise external intervention. The communication can be {realized} using fixed wireless access  and  fiber-to-the-home  as the last-mile technologies or even mobile connections.
It is envisioned that IoNT applications to be standardized in future 6G networks by 2026  \cite{tataria20216g}. Thereby, more research should be conducted at the intersection of IoNT and 6G networks.

\textbf{Design Strategies:}
The abnormality sensing links and S2FC communication links in Fig.~\ref{localg2} are highly susceptible to the spatio-temporal dynamics of the fluids incurred due to the turbulence.
 The turbulence is a prevalent phenomenon in fluids which makes the environment unpredictable, making the  dispersal map a non-convex function. Designing efficient path planning rules in the presence of turbulence is an  urgent research direction. Prolonging static measurements at specific positions may help to average out this effect  at the cost of reduced efficiency.
Importantly, the problem of avoiding being trapped in local optima when trying to locate the abnormality is a fundamental challenge in source localization. Therefore, developing adaptive optimization algorithms from {the} localization perspective to avoid this problem is of high interest. Moreover, {employing} appropriate strategies to locate multiple abnormality sources, e.g., multiple leakages or sources of diffusion, in real micro-environments needs further research.

Despite many potential  applications, diffusive processes decelerate information dissemination in fluidic environments leading to a low communication rate. 
This limits the efficiency of the more complex localization algorithms (e.g., {machine learning} algorithms) since they may need a huge amount of information transmission to reliably characterize the diffusive channels. To simplify the processing resources of sensors, the computation burden may be shifted outward to FC or GW making them significant for future externally-controllable systems. 
Moreover, cross-layer design and implementing appropriate navigation strategies constrained by limited resources of the sensors is another interesting research direction in this area.

\textbf{Security and Privacy:}
The requirements for security and privacy are application-dependent. Particularly, intrusion into the body through IoNT systems (e.g., MC-based IoBNT systems and WBAN) may lead to bio-cyber attacks \cite{zafar2021systematic, usman2018security, al2020intelligence}. The challenge is to design stringent security-related measures that minimize the probability of intrusion (also as a potential source of abnormality) into the networks. Authors in \cite{zafar2021systematic} also discuss challenges in designing interfaces for secure communications at the intersection of IoNTs and conventional communication systems. As previously mentioned in Subsection~\ref{non}, the attacks on intra-sensor and S2FC links may launch spoofing, sentry, blackhole, and eavesdropping attacks \cite{giaretta2015security}. 
In addition, the sensors and FC should be identifiable as legitimate nodes by the human immune system to prevent deliberate apoptosis by immune cells. 
Importantly, the FC2GW link faces critical challenges substantiated by WBAN systems which include eavesdropping, man-in-the-middle attack, resource depletion, injection attack, device tampering, denial-of-service attack, and malware and firmware attacks \cite{zafar2021systematic}. 
Hackers can launch {several} life-threatening attacks by intruding into external entities, e.g., GW or a local processing unit. To this end, developing stringent secure and privacy-preserving algorithms (i.e., appropriate encryption and authentication methods) constrained on the resource-limited environment of IoNT and IoBNT is crucial which opens new research direction in the intersection of IoNT with commercial beyond 5G networks.
Noteworthily, the IoNT and IoBNT systems are supported by IoT backhauling infrastructure. So, the cross-layer security solutions of IoT systems can be equally employed for IoNT and IoBNT systems.

\subsection{Modeling}
The development of comprehensive models is indispensable for realistic MC systems. Most of the papers in the MC literature solely focus on macroscopic and mesoscopic models which are based on the effective and averaged fluctuations of the particles, respectively. However, molecular-level interactions assuming particular trajectories for the particles are also of high importance for accurate modeling of realistic environments, which need further research from an MC perspective.
In the sequel, we clarify other challenges in the field of modeling MC systems in complex environments.

\textbf{Development of Realistic Models:}
Both natural and synthetic systems are much more complicated than the existing models introduced in the MC literature. More realistic models, but tractable, are required for \emph{in-vivo} micro-environments. 
The complex physiological conditions due to the presence of interfering objects and disruptive flows followed by temperature variations need to be considered in the models.
For most proposed schemes so far, ideal and independent models and functions have
been considered for entities and functions including the molecule generation and reception, channel, and sensors. For instance, the size of the sensors and the molecule generation and reception entities may affect the propagation environment.

A well understanding of the biological systems enables the researchers from communication engineering to robustly model the biological entities and the corresponding environments, to use the biological components as a building block and integrate them into a unified MC network. The challenge is to model biological environments considering signal transduction, signaling pathways,  reciprocal interactions of molecules to each other, etc.

\textbf{Resolving Irreproducibility and Inconsistency Issues:}
Of important challenges inherited from the experimental studies is the lack of models for design procedure, which leads to irreproducibility and inconsistency of experimental results. Particularly, in the area of sensory systems, conducting experimentally verified sensor design can pave the way for realistic nano-scale implementations.

 \textbf{Learning-based Models  for Complex Environments:}
For complex environments, analytical channel modeling may be cumbersome or intractable. Machine learning approaches are widely employed in modeling wireless communication systems and are applicable for MC systems, too \cite{lee2017machine, mohamed2019model}.
Developing learning-based models can obviate the need for analytical methods and realize more practical systems. However, collecting and processing a huge amount of spatio-temporal data which is inevitable for training  the MC systems makes it more challenging. Also, numerical methods for complex environments \cite{zoofaghari2021semi} may need prior information about the environment which cannot be obtained solely by the MC systems and may need external interventions.

\subsection{Methods}
\textbf{Classic Methods:}
Most of the MC-based detection and localization approaches are probabilistic,   
{which} comprise a wide range of theoretical methods including Bayesian approaches, hidden Markov models, information-driven approaches like Fisher information matrix, and filtering approaches. The basic idea behind probabilistic methods is to estimate the posterior distribution of the parameters based on the measurements.

Developing mathematical models for multi-agent distributed and sequential detection and localization schemes in realistic molecular environments needs further research.
Improving probabilistic methods to enhance the efficiency of the detection and localization algorithms is an ongoing research direction \cite{jing2021recent} for both OSL and IoNT systems.   	  

\textbf{Bio-inspired Methods:}
These localization schemes rely on a finite-state model inspired by the locomotion of biological entities to decide on the optimal path towards the abnormality.  
Taxis-based approaches are well-studied approaches in this area which is still of high interest in localization problems, particularly in biology.

Engineered micro-organisms based on efficient bio-inspired locomotion algorithms are imperative for fast and  accurate abnormality localization. Moreover, external interventions augment another degree of freedom  to compensate for intrinsic shortcomings of self-propulsion mechanisms using external actuating systems. 
For externally-propelled sensors, e.g., magnetic sensors \cite{zhang2009characterizing}, the external actuation field can be simply adjusted to attain the desired velocity of the motion. Designing synthetic sensors equipped with more efficient bio-inspired propulsion capabilities is a research trend for IoBNT systems.

\textbf{Machine Learning Methods:}
The machine learning approaches are popular  for parametric and non-parametric estimation in model-free environments.
Non-parametric methods using matrix factorization are a technique for model-free multi-source localization schemes in the literature \cite{chen2019unimodality}.
Recent progress in the area of machine-learning-assisted motion control of micro-robots using self-propelled or externally-propelled micro-agents promises  substantial role in environments with spatio-temporal variations \cite{yu2021recent}. 
Important {machine learning} approaches comprise several schemes from deep reinforcement learning to imitation learning \cite{kober2010imitation}. 
Combining machine learning approaches with probabilistic methods may lead to more efficient localization.

We note that the machine learning approaches in the literature mostly contribute to stationary sensors, while the mobility of sensors can make the problem more challenging due to the outdated CSI which complicates the training mechanism. The development of an appropriate framework for localizing abnormalities based on learning methods is an important and attractive research direction.

Inspired by the collective behavior of natural systems, heuristic methods have been proposed to solve model-free engineering problems. Some of these approaches for macro-scale localization problems include genetic algorithm, particle swarm optimization, and ant colony optimization \cite{mac2016heuristic}. From a macro-scale perspective, 
abnormality localization in dynamic environments with moving obstacles and abnormality sources is still challenging.

The swarm-like behavior of the heuristic approaches makes it more consistent for IoBNT applications. 
In a dynamic environment, adopting hybrid methods, i.e., a combination of classification methods with particle swarm optimization revealed successful outcomes \cite{chen2006smooth}. A combination of heuristic methods with conventional or machine learning approaches customized for nano-scale localization is a future research direction for IoNBTs.

\section{Conclusion}
In this paper, we have presented a comprehensive survey on the ADL schemes using MC systems. 
We presented an end-to-end system model for MC-based ADL which consists of multiple tiers: the abnormality sensing link in the first tier, and the S2FC communication link in the second tier. There may be other tiers in the system for sending the information to an external processing/controlling unit through a GW. We considered different abnormality recognition methods, including molecule release, medium effect, and molecule absorption. Moreover, we described the functional units of the sensors and different sensor features. We also explained possible constructions for interfaces (FC and GW) to convert the internal signals to the external signals including wireless to wireless and molecular to wireless interfaces.
A unified propagation model was also developed to describe the molecular and non-molecular channels in the ADL systems. 
We categorized the abnormality detection schemes based on the sensor mobility, and cooperative detection and sensing/activation, and classified the abnormality localization approaches based on the sensor mobility and propulsion mechanisms. In addition, a general framework for the externally-controllable localization systems was  presented.
Finally, we presented the important challenges of the MC-based ADL systems and introduced some directions for future research.

%%%%%%%%%%%%%%%%%%%%%%%%%%%%
\bibliographystyle{ieeetr}
\bibliography{ref22}

\begin{thebibliography}{100}

\bibitem{abdallah2016fraud}
A.~Abdallah, M.~A. Maarof, and A.~Zainal, ``Fraud detection system: A survey,''
  {\em J. of Netw. and Comput. Appl.}, vol.~68, pp.~90--113, 2016.

\bibitem{khraisat2019survey}
A.~Khraisat, I.~Gondal, P.~Vamplew, and J.~Kamruzzaman, ``Survey of intrusion
  detection systems: techniques, datasets and challenges,'' {\em
  Cybersecurity}, vol.~2, no.~1, 2019.
\newblock {A}rt. no. 20, doi:
  {\href{https://doi.org/10.1186/s42400-019-0038-7}{10.1186/s42400-019-0038-7}}.

\bibitem{miljkovic2011fault}
D.~Miljkovi{\'c}, ``Fault detection methods: A literature survey,'' in {\em
  Proc. of the 34th Int. Convention MIPRO}, pp.~750--755, 2011.

\bibitem{thatte2008detection}
G.~Thatte, U.~Mitra, and J.~Heidemann, ``Detection of low-rate attacks in
  computer networks,'' in {\em IEEE INFOCOM Workshops}, pp.~1--6, 2008.
\newblock doi:
  {\href{https://doi.org/10.1109/INFOCOM.2008.4544638}{10.1109/INFOCOM.2008.4544638}}.

\bibitem{thatte2010parametric}
G.~Thatte, U.~Mitra, and J.~Heidemann, ``Parametric methods for anomaly
  detection in aggregate traffic,'' {\em IEEE/ACM Trans. On Networking},
  vol.~19, no.~2, pp.~512--525, 2010.

\bibitem{chattopadhyay2018attack}
A.~Chattopadhyay and U.~Mitra, ``Attack detection and secure estimation under
  false data injection attack in cyber-physical systems,'' in {\em 52nd Annu.
  Conf. on Inf. Sci. and Sys. (CISS)}, pp.~1--6, 2018.
\newblock doi:
  {\href{https://doi.org/10.1109/CISS.2018.8362307}{10.1109/CISS.2018.8362307}}.

\bibitem{levorato2012fast}
M.~Levorato and U.~Mitra, ``Fast anomaly detection in smartgrids via sparse
  approximation theory,'' in {\em IEEE 7th Sensor Array and Multichannel Sig.
  Process. Workshop (SAM)}, pp.~5--8, 2012.
\newblock doi:
  {\href{https://doi.org/10.1109/SAM.2012.6250561}{10.1109/SAM.2012.6250561}}.

\bibitem{hu2020decentralized}
L.~Hu, X.~Wang, and S.~Wang, ``Decentralized underwater target detection and
  localization,'' {\em IEEE Sensors J.}, vol.~21, no.~2, pp.~2385--2399, 2020.

\bibitem{chen2017underwater}
J.~Chen and U.~Mitra, ``Underwater acoustic source localization using
  unimodal-constrained matrix factorization,'' in {\em 51st Asilomar Conf. on
  Signals, Sys., and Comput.}, pp.~2002--2006, 2017.
\newblock doi:
  {\href{https://doi.org/10.1109/ACSSC.2017.8335718}{10.1109/ACSSC.2017.8335718}}.

\bibitem{sanderson1975early}
D.~R. Sanderson and R.~S. Fontana, ``Early lung cancer detection and
  localization,'' {\em Annals of Otology, Rhinology \& Laryngology}, vol.~84,
  no.~5, pp.~583--588, 1975.

\bibitem{islam2013survey}
M.~S. Islam, N.~Kaabouch, and W.~C. Hu, ``A survey of medical imaging
  techniques used for breast cancer detection,'' in {\em IEEE Int. Conf. on
  Electro-Inf. Tech. (EIT)}, pp.~1--5, 2013.
\newblock doi:
  {\href{https://doi.org/10.1109/EIT.2013.6632694}{10.1109/EIT.2013.6632694}}.

\bibitem{matthes2005source}
J.~Matthes, L.~Groll, and H.~B. Keller, ``Source localization by spatially
  distributed electronic noses for advection and diffusion,'' {\em IEEE Trans.
  on Sig. Process.}, vol.~53, no.~5, pp.~1711--1719, 2005.

\bibitem{kuzu2008survey}
A.~Kuzu, S.~Bogosyan, and M.~Gokasan, ``Survey: detection, recognition and
  source localization of odor,'' {\em WSEAS Trans. on Sys.}, vol.~7, no.~6,
  pp.~611--621, 2008.

\bibitem{chandola2009anomaly}
V.~Chandola, A.~Banerjee, and V.~Kumar, ``Anomaly detection: A survey,'' {\em
  ACM Computing Surveys (CSUR)}, vol.~41, no.~3, 2009.
\newblock {A}rt. no. 15, doi:
  {\href{https://doi.org/10.1145/1541880.1541882}{10.1145/1541880.1541882}}.

\bibitem{zhu2021anomaly}
F.~Zhu, R.~Shankaran, B.~Zhao, Y.~Zhao, Y.~Gang, and X.~Chen, ``Anomaly
  detection-based intelligent computing in internet of things and network
  applications,'' {\em Internet Tech. Letters}, vol.~4, no.~3, 2021.
\newblock {A}rt. no. e293, doi:
  {\href{https://doi.org/10.1002/itl2.293}{10.1002/itl2.293}}.

\bibitem{xie2011anomaly}
M.~Xie, S.~Han, B.~Tian, and S.~Parvin, ``Anomaly detection in wireless sensor
  networks: A survey,'' {\em J. of netw. and comput. Appl.}, vol.~34, no.~4,
  pp.~1302--1325, 2011.

\bibitem{karapistoli2013anomaly}
E.~Karapistoli and A.~A. Economides, ``Anomaly detection and localization in
  uwb wireless sensor networks,'' in {\em IEEE 24th Annu. Int. Symp. on
  Personal, Indoor, and Mobile Radio Comm. (PIMRC)}, pp.~2326--2330, 2013.

\bibitem{fu2011wireless}
X.~Fu, W.~Chen, S.~Ye, Y.~Tu, Y.~Tang, D.~Li, H.~Chen, and K.~Jiang, ``A
  wireless implantable sensor network system for in vivo monitoring of
  physiological signals,'' {\em IEEE Trans. on Inf. Tech. in Biomedicine},
  vol.~15, no.~4, pp.~577--584, 2011.

\bibitem{meesookho2007energy}
C.~Meesookho, U.~Mitra, and S.~Narayanan, ``On energy-based acoustic source
  localization for sensor networks,'' {\em IEEE Trans. on Sig. Process.},
  vol.~56, no.~1, pp.~365--377, 2007.

\bibitem{akyildiz2010wireless}
I.~F. Akyildiz and M.~C. Vuran, {\em Wireless sensor networks}.
\newblock John Wiley \& Sons, 2010.

\bibitem{yuce2012introduction}
M.~Yuce and J.~Khan, ``Introduction to wireless body area network,'' in {\em
  Wireless Body Area Netw.-Tech., Implementation and Appl.}, pp.~1--17, Pan
  Stanford Publishing, 2012.

\bibitem{movassaghi2014wireless}
S.~Movassaghi, M.~Abolhasan, J.~Lipman, D.~Smith, and A.~Jamalipour, ``Wireless
  body area networks: A survey,'' {\em IEEE Comm. Surveys \& Tutorials},
  vol.~16, no.~3, pp.~1658--1686, 2014.

\bibitem{hasan2019comprehensive}
K.~Hasan, K.~Biswas, K.~Ahmed, N.~S. Nafi, and M.~S. Islam, ``A comprehensive
  review of wireless body area network,'' {\em J. of Netw. and Comput. Appl.},
  vol.~143, pp.~178--198, 2019.

\bibitem{negra2016wireless}
R.~Negra, I.~Jemili, and A.~Belghith, ``Wireless body area networks:
  Applications and technologies,'' {\em Procedia Comput. Sci.}, vol.~83,
  pp.~1274--1281, 2016.

\bibitem{anwar2017wireless}
M.~Anwar, A.~H. Abdullah, K.~N. Qureshi, and A.~H. Majid, ``Wireless body area
  networks for healthcare applications: An overview,'' {\em Telkomnika},
  vol.~15, no.~3, pp.~1088--1095, 2017.

\bibitem{thatte2009energy}
G.~Thatte, M.~Li, A.~Emken, U.~Mitra, S.~Narayanan, M.~Annavaram, and
  D.~Spruijt-Metz, ``Energy-efficient multihypothesis activity-detection for
  health-monitoring applications,'' in {\em Annu. Int. Conf. of the IEEE Eng.
  in Medicine and Biology Society}, vol.~1, pp.~4678--4681, 2009.

\bibitem{thatte2009optimal}
G.~Thatte, V.~Rozgic, M.~Li, S.~Ghosh, U.~Mitra, S.~Narayanan, M.~Annavaram,
  and D.~Spruijt-Metz, ``Optimal allocation of time-resources for
  multihypothesis activity-level detection,'' in {\em Int. Conf. on Distributed
  Computing in Sensor Sys.}, pp.~273--286, 2009.

\bibitem{thatte2011optimal}
G.~Thatte, M.~Li, S.~Lee, B.~A. Emken, M.~Annavaram, S.~Narayanan,
  D.~Spruijt-Metz, and U.~Mitra, ``Optimal time-resource allocation for
  energy-efficient physical activity detection,'' {\em IEEE Trans. on Sig.
  Process.}, vol.~59, no.~4, pp.~1843--1857, 2011.

\bibitem{zois2013energy}
D.-S. Zois, M.~Levorato, and U.~Mitra, ``Energy-efficient, heterogeneous sensor
  selection for physical activity detection in wireless body area networks,''
  {\em IEEE Trans. on Sig. Process.}, vol.~61, no.~7, pp.~1581--1594, 2013.

\bibitem{farsad2016comprehensive}
N.~Farsad, H.~B. Yilmaz, A.~Eckford, C.-B. Chae, and W.~Guo, ``A comprehensive
  survey of recent advancements in molecular communication,'' {\em IEEE Comm.
  Surveys \& Tutorials}, vol.~18, no.~3, pp.~1887--1919, 2016.

\bibitem{mosayebi2018early}
R.~Mosayebi, A.~Ahmadzadeh, W.~Wicke, V.~Jamali, R.~Schober, and
  M.~Nasiri-Kenari, ``Early cancer detection in blood vessels using mobile
  nanosensors,'' {\em IEEE Trans. on NanoBiosci.}, vol.~18, no.~2,
  pp.~103--116, 2018.

\bibitem{tarro2005early}
G.~Tarro, A.~Perna, and C.~Esposito, ``Early diagnosis of lung cancer by
  detection of tumor liberated protein,'' {\em J. of Cellular Physiology},
  vol.~203, no.~1, pp.~1--5, 2005.
\newblock doi: {\href{https://doi.org/10.1002/jcp.20195}{10.1002/jcp.20195}}.

\bibitem{qiu2014molecular}
S.~Qiu, W.~Guo, S.~Wang, N.~Farsad, and A.~Eckford, ``A molecular communication
  link for monitoring in confined environments,'' in {\em IEEE Int. Conf. on
  Comm. (ICC)}, pp.~718--723, 2014.
\newblock doi:
  {\href{https://doi.org/10.1109/ICCW.2014.6881284}{10.1109/ICCW.2014.6881284}}.

\bibitem{nakano2013molecular}
T.~Nakano, A.~W. Eckford, and T.~Haraguchi, {\em Molecular communication}.
\newblock Cambridge University Press, 2013.

\bibitem{AtakanBook}
B.~Atakan, {\em Molecular communication Among Nanomachines}.
\newblock Springer, 2014.

\bibitem{felicetti2016applications}
L.~Felicetti, M.~Femminella, G.~Reali, and P.~Li{\`o}, ``Applications of
  molecular communications to medicine: A survey,'' {\em Nano Comm. Netw.},
  vol.~7, pp.~27--45, 2016.

\bibitem{Akyildiz2008}
I.~F. Akyildiz, F.~Brunetti, and C.~Blázquez, ``Nanonetworks: A new
  communication paradigm,'' {\em Comput. Netw.: The Int. J. of Comput. and
  TeleComm. Networking}, vol.~52, no.~12, pp.~2260--2279, 2008.

\bibitem{felicetti2014molecular}
L.~Felicetti, M.~Femminella, G.~Reali, and P.~Li{\`o}, ``A molecular
  communication system in blood vessels for tumor detection,'' in {\em Proc. of
  ACM The First Annu. Int. Conf. on Nanoscale Computing and Comm.}, pp.~1--9,
  2014.
\newblock {A}rt. no. 21, doi:
  {\href{https://doi.org/10.1145/2619955.2619978}{10.1145/2619955.2619978}}.

\bibitem{amin2021viral}
O.~Amin, H.~Dahrouj, N.~Almayouf, T.~Y. Al-Naffouri, B.~Shihada, and M.-S.
  Alouini, ``Viral aerosol concentration characterization and detection in
  bounded environments,'' {\em IEEE Trans. on Mol., Bio. and Multi-Scale
  Comm.}, vol.~7, no.~3, pp.~185--199, 2021.

\bibitem{khaloopour2021theoretical}
L.~Khaloopour, M.~Mirmohseni, and M.~Nasiri-Kenari, ``Theoretical concept study
  of cooperative abnormality detection and localization in fluidic-medium
  molecular communication,'' {\em IEEE Sensors J.}, vol.~21, no.~15,
  pp.~17118--17130, 2021.

\bibitem{khalid2020modeling}
M.~Khalid, O.~Amin, S.~Ahmed, B.~Shihada, and M.-S. Alouini, ``Modeling of
  viral aerosol transmission and detection,'' {\em IEEE Trans. on Comm.},
  vol.~68, no.~8, pp.~4859--4873, 2020.

\bibitem{ghavami2017abnormality}
S.~Ghavami and F.~Lahouti, ``Abnormality detection in correlated gaussian
  molecular nano-networks: Design and analysis,'' {\em IEEE Trans. on
  NanoBiosci.}, vol.~16, no.~3, pp.~189--202, 2017.

\bibitem{varshney2018abnormality}
N.~Varshney, A.~Patel, Y.~Deng, W.~Haselmayr, P.~K. Varshney, and
  A.~Nallanathan, ``Abnormality detection inside blood vessels with mobile
  nanomachines,'' {\em IEEE Trans. on Mol., Bio. and Multi-Scale Comm.},
  vol.~4, no.~3, pp.~189--194, 2018.

\bibitem{gomez2021machine}
J.~T. G{\'o}mez, A.~Kuestner, K.~Pitke, J.~Simonjan, B.~D. Unluturk, and
  F.~Dressler, ``A machine learning approach for abnormality detection in blood
  vessels via mobile nanosensors,'' in {\em Proc. of the 19th ACM Conf. on
  Embedded Networked Sensor Sys.}, pp.~596--602, 2021.
\newblock doi:
  {\href{https://doi.org/10.1145/3485730.3494037}{10.1145/3485730.3494037}}.

\bibitem{stelzner2016precise}
M.~Stelzner, F.-L. Lau, K.~Freundt, F.~B{\"u}ther, M.~L. Nguyen, C.~Stamme, and
  S.~Ebers, ``Precise detection and treatment of human diseases based on nano
  networking,'' in {\em Proc. of the 11th EAI Int. Conf. on Body Area Netw.},
  pp.~58--64, 2016.

\bibitem{nakano2014externally}
T.~Nakano, S.~Kobayashi, T.~Suda, Y.~Okaie, Y.~Hiraoka, and T.~Haraguchi,
  ``Externally controllable molecular communication,'' {\em IEEE J. on Selected
  Areas in Comm.}, vol.~32, no.~12, pp.~2417--2431, 2014.

\bibitem{raz2015bioinspired}
N.~R. Raz, M.-R. Akbarzadeh-T, and M.~Tafaghodi, ``Bioinspired nanonetworks for
  targeted cancer drug delivery,'' {\em IEEE Trans. on NanoBiosci.}, vol.~14,
  no.~8, pp.~894--906, 2015.

\bibitem{nakano2016performance}
T.~Nakano, Y.~Okaie, S.~Kobayashi, T.~Koujin, C.-H. Chan, Y.-H. Hsu, T.~Obuchi,
  T.~Hara, Y.~Hiraoka, and T.~Haraguchi, ``Performance evaluation of
  leader--follower-based mobile molecular communication networks for target
  detection applications,'' {\em IEEE Trans. on Comm.}, vol.~65, no.~2,
  pp.~663--676, 2016.

\bibitem{okaie2014cooperative}
Y.~Okaie, T.~Nakano, T.~Hara, T.~Obuchi, K.~Hosoda, Y.~Hiraoka, and S.~Nishio,
  ``Cooperative target tracking by a mobile bionanosensor network,'' {\em IEEE
  Trans. on NanoBiosci.}, vol.~13, no.~3, pp.~267--277, 2014.

\bibitem{jamali2019channel}
V.~Jamali, A.~Ahmadzadeh, W.~Wicke, A.~Noel, and R.~Schober, ``Channel modeling
  for diffusive molecular communication-a tutorial review,'' {\em Proc. of the
  IEEE}, vol.~107, no.~7, pp.~1256--1301, 2019.

\bibitem{hoelen1999new}
C.~Hoelen and F.~De~Mul, ``A new theoretical approach to photoacoustic signal
  generation,'' {\em The J. of the Acoustical Society of America}, vol.~106,
  no.~2, pp.~695--706, 1999.

\bibitem{santagati2013opto}
H.~Kim, M.~Park, C.~W. Kim, and D.~Shin, ``Opto-ultrasonic communications in
  wireless body area nanonetworks,'' in {\em Asilomar Conf. on Sig., Sys. and
  Comput.}, pp.~1066--1070, 2013.
\newblock doi:
  {\href{https://doi.org/10.1109/ACSSC.2013.6810455}{10.1109/ACSSC.2013.6810455}}.

\bibitem{khalid2018system}
M.~Khalid, O.~Amin, S.~Ahmed, and M.-S. Alouini, ``System modeling of virus
  transmission and detection in molecular communication channels,'' in {\em
  IEEE Int. Conf. on Comm. (ICC)}, pp.~1--6, 2018.
\newblock doi:
  {\href{https://doi.org/10.1109/ICC.2018.8422665}{10.1109/ICC.2018.8422665}}.

\bibitem{mosayebi2018advanced}
R.~Mosayebi, W.~Wicke, V.~Jamali, A.~Ahmadzadeh, R.~Schober, and
  M.~Nasiri-Kenari, ``Advanced target detection via molecular communication,''
  in {\em IEEE Glob. Comm. Conf. (GLOBECOM)}, pp.~1--7, 2018.
\newblock doi:
  {\href{https://doi.org/10.1109/GLOCOM.2018.8647734}{10.1109/GLOCOM.2018.8647734}}.

\bibitem{kumar2020nanomachine}
S.~Kumar, ``Nanomachine localization in a diffusive molecular communication
  system,'' {\em IEEE Sys. J.}, vol.~14, no.~2, pp.~3011--3014, 2020.

\bibitem{baidoo2020channel}
H.~E. Baidoo-Williams, M.~M.~U. Rahman, and Q.~H. Abbasi, ``Channel impulse
  response-based source localization in a diffusion-based molecular
  communication system,'' {\em Internet Tech. Letters}, vol.~3, no.~2, 2020.
\newblock {A}rt. no. e143, doi:
  {\href{https://doi.org/10.1002/itl2.143}{10.1002/itl2.143}}.

\bibitem{zhu2021target}
D.~Zhu, X.~Bao, and W.~Zhang, ``Target recognition in turbulent diffusion
  channels,'' {\em IEEE Comm. Letters}, vol.~25, no.~11, pp.~3694--3698, 2021.

\bibitem{miao2019cooperative}
Y.~Miao, W.~Zhang, and X.~Bao, ``Cooperative source positioning for simo
  molecular communication via diffusion,'' in {\em IEEE 19th Int. Conf. on
  Comm. Tech. (ICCT)}, pp.~495--499, 2019.
\newblock doi:
  {\href{https://doi.org/10.1109/ICCT46805.2019.8947256}{10.1109/ICCT46805.2019.8947256}}.

\bibitem{yetimoglu2021multiple}
O.~Yetimoglu, M.~K. Avci, B.~C. Akdeniz, H.~B. Yilmaz, A.~E. Pusane, and
  T.~Tugcu, ``Multiple transmitter localization via single receiver in 3-d
  molecular communication via diffusion,'' {\em Digital Sig. Process.},
  vol.~124, 2021.
\newblock {A}rt. no. 103185, doi:
  {\href{https://doi.org/10.1016/j.dsp.2021.103185}{10.1016/j.dsp.2021.103185}}.

\bibitem{chahibi2013molecular}
Y.~Chahibi, M.~Pierobon, S.~O. Song, and I.~F. Akyildiz, ``A molecular
  communication system model for particulate drug delivery systems,'' {\em IEEE
  Trans. on Biomedical Eng.}, vol.~60, no.~12, pp.~3468--3483, 2013.

\bibitem{dunbabin2012robots}
M.~Dunbabin and L.~Marques, ``Robots for environmental monitoring: Significant
  advancements and applications,'' {\em IEEE Robotics \& Automation Mag.},
  vol.~19, no.~1, pp.~24--39, 2012.

\bibitem{tsang2020roads}
A.~C. Tsang, E.~Demir, Y.~Ding, and O.~S. Pak, ``Roads to smart artificial
  microswimmers,'' {\em Advanced Intelligent Sys.}, vol.~2, no.~8, 2020.
\newblock {A}rt. no. 1900137, doi:
  {\href{https://doi.org/10.1002/aisy.201900137}{10.1002/aisy.201900137}}.

\bibitem{roussos2011chemotaxis}
E.~T. Roussos, J.~S. Condeelis, and A.~Patsialou, ``Chemotaxis in cancer,''
  {\em Nature Reviews Cancer}, vol.~11, no.~8, pp.~573--587, 2011.

\bibitem{de2018optogenetic}
S.~de~Beco, K.~Vaid{\v{z}}iulyt{\.e}, J.~Manzi, F.~Dalier, F.~Di~Federico,
  G.~Cornilleau, M.~Dahan, and M.~Coppey, ``Optogenetic dissection of rac1 and
  cdc42 gradient shaping,'' {\em Nature Comm.}, vol.~9, no.~1, 2018.
\newblock {A}rt. no. 4816, doi:
  {\href{https://doi.org/10.1038/s41467-018-07286-8}{10.1038/s41467-018-07286-8}}.

\bibitem{jatmiko2007pso}
W.~Jatmiko, K.~Sekiyama, and T.~Fukuda, ``A pso-based mobile robot for odor
  source localization in dynamic advection-diffusion with obstacles
  environment: theory, simulation and measurement,'' {\em IEEE Computational
  Intelligence Mag.}, vol.~2, no.~2, pp.~37--51, 2007.

\bibitem{vergassola2007infotaxis}
M.~Vergassola, E.~Villermaux, and B.~I. Shraiman, ```{I}nfotaxis' as a strategy
  for searching without gradients,'' {\em Nature}, vol.~445, no.~7126,
  pp.~406--409, 2007.

\bibitem{sinha2018consensus}
A.~Sinha, R.~Kumar, R.~Kaur, and A.~P. Bhondekar, ``Consensus-based odor source
  localization by multiagent systems,'' {\em IEEE Trans. on Cybernetics},
  vol.~49, no.~12, pp.~4450--4459, 2018.

\bibitem{chen2015touch}
Y.~Chen, P.~Kosmas, P.~S. Anwar, and L.~Huang, ``A touch-communication
  framework for drug delivery based on a transient microbot system,'' {\em IEEE
  Trans. on NanoBiosci.}, vol.~14, no.~4, pp.~397--408, 2015.

\bibitem{shi2020nanorobots}
S.~Shi, Y.~Yan, J.~Xiong, U.~K. Cheang, X.~Yao, and Y.~Chen,
  ``Nanorobots-assisted natural computation for multifocal tumor sensitization
  and targeting,'' {\em IEEE Trans. on NanoBiosci.}, vol.~20, no.~2,
  pp.~154--165, 2020.

\bibitem{solak2020rnn}
S.~N. Solak and M.~Oner, ``Rnn based abnormality detection with nanoscale
  sensor networks using molecular communications,'' in {\em Proc. of the 7th
  ACM Int. Conf. on Nanoscale Computing and Comm.}, pp.~1--6, 2020.
\newblock {A}rt. no. 17, doi:
  {\href{https://doi.org/10.1145/3411295.3411313}{10.1145/3411295.3411313}}.

\bibitem{kerry2021comprehensive}
R.~G. Kerry, K.~E. Ukhurebor, S.~Kumari, G.~K. Maurya, S.~Patra, B.~Panigrahi,
  S.~Majhi, J.~R. Rout, M.~del Pilar Rodr{\'\i}guez-Torres, G.~Das, {\em
  et~al.}, ``A comprehensive review on the applications of nano-biosensor based
  approaches for non-communicable and communicable disease detection,'' {\em
  Biomaterials Sci.}, vol.~9, no.~10, pp.~3576--3602, 2021.

\bibitem{ghavami2020anomaly}
S.~Ghavami, ``Anomaly detection in molecular communications with applications
  to health monitoring networks,'' {\em IEEE Trans. on Mol., Bio. and
  Multi-Scale Comm.}, vol.~6, no.~1, pp.~50--59, 2020.

\bibitem{hori2011mathematical}
S.~S. Hori and S.~S. Gambhir, ``Mathematical model identifies blood
  biomarker--based early cancer detection strategies and limitations,'' {\em
  Sci. Translational Medicine}, vol.~3, no.~109, pp.~109--116, 2011.

\bibitem{norton1976predicting}
L.~Norton, R.~Simon, H.~D. Brereton, and A.~E. Bogden, ``Predicting the course
  of gompertzian growth,'' {\em Nature}, vol.~264, no.~5586, pp.~542--545,
  1976.

\bibitem{fouron2003changes}
J.-C. Fouron, F.~Absi, A.~Skoll, F.~Proulx, and J.~Gosselin, ``Changes in flow
  velocity patterns of the superior and inferior venae cavae during placental
  circulatory insufficiency,'' {\em Ultrasound in Obstetrics and Gynecology:
  The Official J. of the Int. Society of Ultrasound in Obstetrics and
  Gynecology}, vol.~21, no.~1, pp.~53--56, 2003.

\bibitem{zhu2017gas}
J.~Zhu, L.~Ren, S.-C. Ho, Z.~Jia, and G.~Song, ``Gas pipeline leakage detection
  based on pzt sensors,'' {\em Smart Materials and Structures}, vol.~26, no.~2,
  2017.
\newblock {A}rt. no. 025022, doi:
  {\href{https://doi.org/10.1088/1361-665X/26/2/025022}{10.1088/1361-665X/26/2/025022}}.

\bibitem{jafari2020cell}
M.~Jafari and M.~Hasanzadeh, ``Cell-specific frequency as a new hallmark to
  early detection of cancer and efficient therapy: Recording of cancer voice as
  a new horizon,'' {\em Biomedicine \& Pharmacotherapy}, vol.~122, 2020.
\newblock {A}rt. no. 109770, doi:
  {\href{https://doi.org/10.1016/j.biopha.2019.109770}{10.1016/j.biopha.2019.109770}}.

\bibitem{liu2020detection}
M.~Liu, Y.~Sun, and Y.~Chen, ``Detection of atherosclerotic lesions based on
  molecular communication,'' in {\em Int. Conf. on Bio-inspired Inf. and Comm.
  Tech.}, pp.~221--225, 2020.
\newblock doi:
  {\href{https://doi.org/10.1007/978-3-030-57115-3_19}{10.1007/978-3-030-57115-3\_19}}
  %organization={Springer}.

\bibitem{guo2016eavesdropper}
W.~Guo, Y.~Deng, B.~Li, C.~Zhao, and A.~Nallanathan, ``Eavesdropper
  localization in random walk channels,'' {\em IEEE Comm. Letters}, vol.~20,
  no.~9, pp.~1776--1779, 2016.

\bibitem{agrawal2011designing}
D.~P. Agrawal, ``Designing wireless sensor networks: From theory to
  applications,'' {\em Central European J. of Comput. Sci.}, vol.~1, no.~1,
  pp.~2--18, 2011.

\bibitem{chaudhary2020twin}
V.~S. Chaudhary, D.~Kumar, R.~Mishra, and S.~Sharma, ``Twin core photonic
  crystal fiber for temperature sensing,'' {\em Materials Today: Proc.},
  vol.~33, no.~5, pp.~2289--2292, 2020.

\bibitem{xu2018recent}
F.~Xu, X.~Li, Y.~Shi, L.~Li, W.~Wang, L.~He, and R.~Liu, ``Recent developments
  for flexible pressure sensors: A review,'' {\em Micromachines}, vol.~9,
  no.~11, p.~580, 2018.

\bibitem{mishra2021design}
G.~P. Mishra, D.~Kumar, V.~S. Chaudhary, and S.~Kumar, ``Design and sensitivity
  improvement of microstructured-core photonic crystal fiber based sensor for
  methane and hydrogen fluoride detection,'' {\em IEEE Sensors J.}, vol.~22,
  no.~2, pp.~1265--1272, 2021.

\bibitem{chaudhary2020topas}
V.~S. Chaudhary and D.~Kumar, ``Topas based porous core photonic crystal fiber
  for terahertz chemical sensor,'' {\em Optik}, vol.~223, 2020.
\newblock {A}rt. no. 165562, doi:
  {\href{https://doi.org/10.1016/j.ijleo.2020.165562}{10.1016/j.ijleo.2020.165562}}.

\bibitem{kuscu2016physical}
M.~Kuscu and O.~B. Akan, ``On the physical design of molecular communication
  receiver based on nanoscale biosensors,'' {\em IEEE Sensors J.}, vol.~16,
  no.~8, pp.~2228--2243, 2016.

\bibitem{grieshaber2008electrochemical}
D.~Grieshaber, R.~MacKenzie, J.~V{\"o}r{\"o}s, and E.~Reimhult,
  ``Electrochemical biosensors-sensor principles and architectures,'' {\em
  Sensors}, vol.~8, no.~3, pp.~1400--1458, 2008.

\bibitem{tamayo2013biosensors}
J.~Tamayo, P.~M. Kosaka, J.~J. Ruz, {\'A}.~San~Paulo, and M.~Calleja,
  ``Biosensors based on nanomechanical systems,'' {\em Chem. Society Reviews},
  vol.~42, no.~3, pp.~1287--1311, 2013.

\bibitem{chaudhary2021gold}
V.~S. Chaudhary, D.~Kumar, and S.~Kumar, ``Gold-immobilized photonic crystal
  fiber-based spr biosensor for detection of malaria disease in human body,''
  {\em IEEE Sensors J.}, vol.~21, no.~16, pp.~17800--17807, 2021.

\bibitem{mishra2021terahertz}
G.~P. Mishra, D.~Kumar, V.~S. Chaudhary, and S.~Sharma, ``Terahertz refractive
  index sensor with high sensitivity based on two-core photonic crystal
  fiber,'' {\em Microwave and Optical Tech. Letters}, vol.~63, no.~1,
  pp.~24--31, 2021.

\bibitem{solak2020neural}
S.~N. Solak and M.~Oner, ``Neural network based decision fusion for abnormality
  detection via molecular communications,'' in {\em IEEE Workshop on Sig.
  Process. Sys. (SiPS)}, pp.~1--5, 2020.
\newblock doi:
  {\href{https://doi.org/10.1109/SiPS50750.2020.9195212}{10.1109/SiPS50750.2020.9195212}}.

\bibitem{ganesh2021rf}
P.~Ganesh and H.~Venkataraman, ``Rf-based wireless communication for shallow
  water networks: Survey and analysis,'' {\em Wireless Personal Comm.},
  vol.~120, no.~4, pp.~3415--3441, 2021.

\bibitem{lemic2021survey}
F.~Lemic, S.~Abadal, W.~Tavernier, P.~Stroobant, D.~Colle, E.~Alarc{\'o}n,
  J.~Marquez-Barja, and J.~Famaey, ``Survey on terahertz nanocommunication and
  networking: A top-down perspective,'' {\em IEEE J. on Selected Areas in
  Comm.}, vol.~39, no.~6, pp.~1506--1543, 2021.

\bibitem{guo2015intra}
H.~Guo, P.~Johari, J.~M. Jornet, and Z.~Sun, ``Intra-body optical channel
  modeling for in vivo wireless nanosensor networks,'' {\em IEEE Trans. on
  NanoBiosci.}, vol.~15, no.~1, pp.~41--52, 2015.

\bibitem{santagati2014medium}
G.~E. Santagati, T.~Melodia, L.~Galluccio, and S.~Palazzo, ``Medium access
  control and rate adaptation for ultrasonic intrabody sensor networks,'' {\em
  IEEE/ACM Trans. on Networking}, vol.~23, no.~4, pp.~1121--1134, 2014.

\bibitem{dressler2015connecting}
F.~Dressler and S.~Fischer, ``Connecting in-body nano communication with body
  area networks: Challenges and opportunities of the internet of nano things,''
  {\em Nano Comm. Netw.}, vol.~6, no.~2, pp.~29--38, 2015.

\bibitem{arjmandi2016ionchannel}
H.~Arjmandi, A.~Ahmadzadeh, R.~Schober, and M.~N. Kenari, ``Ion channel based
  bio-synthetic modulator for diffusive molecular communication,'' {\em IEEE
  Trans. on NanoBiosci.}, vol.~15, no.~5, pp.~418--432, 2016.

\bibitem{arjmandi2016ionpump}
H.~Arjmandi, V.~Jamali, A.~Ahmadzadeh, A.~Burkovski, R.~Schober, and M.~N.
  Kenari, ``Ion pump based bio-synthetic modulator model for diffusive
  molecular communications,'' in {\em IEEE 17th Int. Workshop on Sig. Process.
  Advances in Wireless Comm. (SPAWC)}, pp.~1--6, 2016.
\newblock doi:
  {\href{https://doi.org/10.1109/SPAWC.2016.7536836}{10.1109/SPAWC.2016.7536836}}.

\bibitem{rodrigues2020skin}
D.~Rodrigues, A.~I. Barbosa, R.~Rebelo, I.~K. Kwon, R.~L. Reis, and V.~M.
  Correlo, ``Skin-integrated wearable systems and implantable biosensors: A
  comprehensive review,'' {\em Biosensors}, vol.~10, no.~7, 2020.
\newblock {A}rt. no. 79, doi:
  {\href{https://doi.org/10.3390/bios10070079}{10.3390/bios10070079}}.

\bibitem{koo2020deep}
B.-H. Koo, H.~J. Kim, J.-Y. Kwon, and C.-B. Chae, ``Deep learning-based human
  implantable nano molecular communications,'' in {\em IEEE Int. Conf. on Comm.
  (ICC)}, pp.~1--7, 2020.
\newblock doi:
  {\href{https://doi.org/10.1109/ICC40277.2020.9148818}{10.1109/ICC40277.2020.9148818}}.

\bibitem{wisniewski2017tissue}
N.~A. Wisniewski, S.~P. Nichols, S.~J. Gamsey, S.~Pullins, K.~Y. Au-Yeung,
  B.~Klitzman, and K.~L. Helton, ``Tissue-integrating oxygen sensors:
  continuous tracking of tissue hypoxia,'' in {\em Oxygen transport to tissue
  XXXIX}, pp.~377--383, Springer, 2017.

\bibitem{soto2020medical}
F.~Soto, J.~Wang, R.~Ahmed, and U.~Demirci, ``Medical micro/nanorobots in
  precision medicine,'' {\em Advanced Sci.}, vol.~7, no.~21, p.~2002203, 2020.

\bibitem{sun2020cooperative}
Y.~Sun, Y.~Hsiang, Y.~Chen, and Y.~Zhou, ``A cooperative molecular
  communication for targeted drug delivery,'' in {\em Int. Conf. on
  Bio-inspired Inf. and Comm. Tech.}, pp.~16--26, 2020.
\newblock doi:
  {\href{https://doi.org/10.1007/978-3-030-57115-3_2}{10.1007/978-3-030-57115-3\_2}}.

\bibitem{katic2015efficient}
J.~Katic, {\em Efficient energy harvesting interface for implantable
  biosensors}.
\newblock PhD thesis, KTH, 2015.

\bibitem{lu2010thermal}
X.~Lu and S.-H. Yang, ``Thermal energy harvesting for wsns,'' in {\em IEEE Int.
  Conf. on Sys., Man and Cybernetics}, pp.~3045--3052, 2010.

\bibitem{wei2017comprehensive}
C.~Wei and X.~Jing, ``A comprehensive review on vibration energy harvesting:
  Modelling and realization,'' {\em Renewable and Sustainable Energy Reviews},
  vol.~74, pp.~1--18, 2017.

\bibitem{park2020nanofluidic}
C.~H. Park, H.~Bae, C.-s. Kim, D.-H. Peck, and J.~Lee, ``Nanofluidic energy
  harvesting through a biological 1d protein-embedded nanofilm membrane by
  interfacial polymerization,'' {\em Nano Energy}, vol.~74, 2020.
\newblock {A}rt. no. 104906, doi:
  {\href{https://doi.org/10.1016/j.nanoen.2020.104906}{10.1016/j.nanoen.2020.104906}}.

\bibitem{phillips2021energy}
J.~D. Phillips, ``Energy harvesting in nanosystems: Powering the next
  generation of the internet of things,'' {\em Frontiers in Nanotech.}, vol.~3,
  2021.
\newblock {A}rt. no. 633931, doi:
  {\href{https://doi.org/10.3389/fnano.2021.633931}{10.3389/fnano.2021.633931}}.

\bibitem{yang2020comprehensive}
K.~Yang, D.~Bi, Y.~Deng, R.~Zhang, M.~M.~U. Rahman, N.~A. Ali, M.~A. Imran,
  J.~M. Jornet, Q.~H. Abbasi, and A.~Alomainy, ``A comprehensive survey on
  hybrid communication in context of molecular communication and terahertz
  communication for body-centric nanonetworks,'' {\em IEEE Trans. on Mol., Bio.
  and Multi-Scale Comm.}, vol.~6, no.~2, pp.~107--133, 2020.

\bibitem{islam2016catch}
N.~Islam and S.~Misra, ````{C}atch the pendulum": The problem of asymmetric
  data delivery in electromagnetic nanonetworks,'' {\em IEEE Trans. on
  NanoBiosci.}, vol.~15, no.~6, pp.~576--584, 2016.

\bibitem{canovas2018nanoscale}
S.~Canovas-Carrasco, A.-J. Garcia-Sanchez, and J.~Garcia-Haro, ``A nanoscale
  communication network scheme and energy model for a human hand scenario,''
  {\em Nano Comm. Netw.}, vol.~15, pp.~17--27, 2018.

\bibitem{wu2009periodic}
Y.~Wu, A.~D. Kaiser, Y.~Jiang, and M.~S. Alber, ``Periodic reversal of
  direction allows myxobacteria to swarm,'' {\em Proc. of the National Academy
  of Sci.}, vol.~106, no.~4, pp.~1222--1227, 2009.

\bibitem{ozawa2013advances}
T.~Ozawa, H.~Yoshimura, and S.~B. Kim, ``Advances in fluorescence and
  bioluminescence imaging,'' {\em Analytical Chemistry}, vol.~85, no.~2,
  pp.~590--609, 2013.

\bibitem{roman2004single}
C.~Roman, F.~Ciontu, and B.~Courtois, ``Single molecule detection and
  macromolecular weighting using an all-carbon-nanotube nanoelectromechanical
  sensor,'' in {\em 4th IEEE Conf. on Nanotech.}, pp.~263--266, 2004.
\newblock doi:
  {\href{https://doi.org/10.1109/NANO.2004.1392318}{10.1109/NANO.2004.1392318}}.

\bibitem{lazar2013adsorption}
P.~Lazar, F.~Karlicky, P.~Jurecka, M.~Kocman, E.~Otyepkov{\'a},
  K.~\v{S}af{\'a}\v{r}ov{\'a}, and M.~Otyepka, ``Adsorption of small organic
  molecules on graphene,'' {\em J. of the American Chem. Society}, vol.~135,
  no.~16, pp.~6372--6377, 2013.

\bibitem{luchko2013fractional}
Y.~Luchko, ``Fractional wave equation and damped waves,'' {\em J. of
  Mathematical Physics}, vol.~54, no.~3, 2013.
\newblock {A}rt. no. 031505, doi:
  {\href{https://doi.org/10.1063/1.4794076}{10.1063/1.4794076}}.

\bibitem{mainardi2010fractional}
F.~Mainardi, {\em Fractional calculus and waves in linear viscoelasticity: an
  introduction to mathematical models}.
\newblock World Scientific, 2010.

\bibitem{arjmandi2020mathematical}
H.~Arjmandi, M.~Zoofaghari, S.~V. Rouzegar, M.~Veleti{\'c}, and I.~Balasingham,
  ``On mathematical analysis of active drug transport coupled with flow-induced
  diffusion in blood vessels,'' {\em IEEE Trans. on NanoBiosci}, vol.~20,
  no.~1, pp.~105--115, 2020.

\bibitem{mosayebi2017cooperative}
R.~Mosayebi, V.~Jamali, N.~Ghoroghchian, R.~Schober, M.~Nasiri-Kenari, and
  M.~Mehrabi, ``Cooperative abnormality detection via diffusive molecular
  communications,'' {\em IEEE Trans. on NanoBiosci.}, vol.~16, no.~8,
  pp.~828--842, 2017.

\bibitem{ghoroghchian2019abnormality}
N.~Ghoroghchian, M.~Mirmohseni, and M.~Nasiri-Kenari, ``Abnormality detection
  and monitoring in multi-sensor molecular communication,'' {\em IEEE Trans. on
  Mol., Bio. and Multi-Scale Comm.}, vol.~5, no.~2, pp.~68--83, 2019.

\bibitem{solak2020sequential}
S.~Solak and M.~{\"O}ner, ``Sequential decision fusion for abnormality
  detection via diffusive molecular communications,'' {\em IEEE Comm. Letters},
  vol.~25, no.~3, pp.~825--829, 2020.

\bibitem{mai2017event}
T.~C. Mai, M.~Egan, T.~Q. Duong, and M.~Di~Renzo, ``Event detection in
  molecular communication networks with anomalous diffusion,'' {\em IEEE Comm.
  Letters}, vol.~21, no.~6, pp.~1249--1252, 2017.

\bibitem{rogers2016parallel}
U.~Rogers and M.-S. Koh, ``Parallel molecular distributed detection with
  brownian motion,'' {\em IEEE Trans. on NanoBiosci.}, vol.~15, no.~8,
  pp.~871--880, 2016.

\bibitem{gulec2020localization}
F.~Gulec and B.~Atakan, ``Localization of a passive molecular transmitter with
  a sensor network,'' in {\em Int. Conf. on Bio-inspired Inf. and Comm. Tech.},
  pp.~317--335, 2020.
\newblock doi:
  {\href{https://doi.org/10.1007/978-3-030-57115-3_28}{10.1007/978-3-030-57115-3\_28}}.

\bibitem{qiu2015long}
S.~Qiu, W.~Guo, B.~Li, Y.~Wu, X.~Chu, S.~Wang, and Y.~Y. Dong, ``Long range and
  long duration underwater localization using molecular messaging,'' {\em IEEE
  Trans. on Mol., Bio. and Multi-Scale Comm.}, vol.~1, no.~4, pp.~363--370,
  2015.

\bibitem{kim2019source}
H.~Kim, M.~Park, C.~W. Kim, and D.~Shin, ``Source localization for hazardous
  material release in an outdoor chemical plant via a combination of lstm-rnn
  and cfd simulation,'' {\em Comput. \& Chem. Eng.}, vol.~125, pp.~476--489,
  2019.

\bibitem{vijayakumaran2007maximum}
S.~Vijayakumaran, Y.~Levinbook, and T.~F. Wong, ``Maximum likelihood
  localization of a diffusive point source using binary observations,'' {\em
  IEEE Trans. on Sig. Process.}, vol.~55, no.~2, pp.~665--676, 2007.

\bibitem{murray2015estimating}
J.~Murray-Bruce and P.~L. Dragotti, ``Estimating localized sources of diffusion
  fields using spatiotemporal sensor measurements,'' {\em IEEE Trans. on Sig.
  Process.}, vol.~63, no.~12, pp.~3018--3031, 2015.

\bibitem{li2020inference}
J.~Li, W.~Zhang, X.~Bao, M.~Abbaszadeh, and W.~Guo, ``Inference in turbulent
  molecular information channels using support vector machine,'' {\em IEEE
  Trans. on Mol., Bio. and Multi-Scale Comm.}, vol.~6, no.~1, pp.~25--35, 2020.

\bibitem{yang2020recent}
Q.~Yang, L.~Xu, W.~Zhong, Q.~Yan, Y.~Gao, W.~Hong, Y.~She, and G.~Yang,
  ``Recent advances in motion control of micro/nanomotors,'' {\em Advanced
  Intelligent Sys.}, vol.~2, no.~8, 2020.
\newblock {A}rt. no. 2000049, doi:
  {\href{https://doi.org/10.1002/aisy.202000049}{10.1002/aisy.202000049}}.

\bibitem{fabbiano2016distributed}
R.~Fabbiano, F.~Garin, and C.~Canudas-de Wit, ``Distributed source seeking
  without global position information,'' {\em IEEE Trans. on Cont. of Netw.
  Sys.}, vol.~5, no.~1, pp.~228--238, 2016.

\bibitem{giaretta2015security}
A.~Giaretta, S.~Balasubramaniam, and M.~Conti, ``Security vulnerabilities and
  countermeasures for target localization in bio-nanothings communication
  networks,'' {\em IEEE Trans. on Inf. Forensics and Security}, vol.~11, no.~4,
  pp.~665--676, 2015.

\bibitem{tran2014localization}
H.~Tran-Dang, N.~Krommenacker, and P.~Charpentier, ``Localization algorithms
  based on hop counting for wireless nano-sensor networks,'' in {\em Int. Conf.
  on Indoor Positioning and Indoor Navigation (IPIN)}, pp.~300--306, 2014.
\newblock doi:
  {\href{https://doi.org/10.1109/IPIN.2014.7275496}{10.1109/IPIN.2014.7275496}}.

\bibitem{zhou2017pulse}
L.~Zhou, G.~Han, and L.~Liu, ``Pulse-based distance accumulation localization
  algorithm for wireless nanosensor networks,'' {\em IEEE Access}, vol.~5,
  pp.~14380--14390, 2017.

\bibitem{el2018high}
M.~El-Absi, A.~A. Abbas, A.~Abuelhaija, F.~Zheng, K.~Solbach, and T.~Kaiser,
  ``High-accuracy trans.localization based on chipless rfid systems at thz
  band,'' {\em IEEE Access}, vol.~6, pp.~54355--54368, 2018.

\bibitem{lemic2021toward}
F.~Lemic, S.~Abadal, E.~Alarc{\'o}n, and J.~Famaey, ``Toward location-aware
  in-body terahertz nanonetworks with energy harvesting,'' {\em
  arXiv:2101.01952}, 2021.

\bibitem{rady2020biosensors}
N.~Rady~Raz, M.~R. Akbarzadeh~Totonchi, {\em et~al.}, ``Biosensors localization
  for swarm medical delivery,'' in {\em 4th Conf. on Swarm Intelligence and
  Evolutionary Computation}, pp.~010--014, 2020.
\newblock doi:
  {\href{https://doi.org/10.1109/CSIEC49655.2020.9237305}{10.1109/CSIEC49655.2020.9237305}}.

\bibitem{odysseos2021bionanomachine}
A.~D. Odysseos and T.~Nakano, ``Bionanomachine diagnostics and nanonetwork
  therapeutic in brain malignancies with bionanodevice interfaces,'' {\em IEEE
  Trans. on Mol., Bio. and Multi-Scale Comm.}, vol.~8, no.~1, pp.~28--35, 2022.

\bibitem{ishiyama2020cooperative}
S.~Ishiyama, T.~Nakano, Y.~Okaie, T.~Hara, and K.~Harumoto, ``Cooperative
  signaling and directed migration of bio-nanomachines in mobile molecular
  communication,'' in {\em Proc. of the 7th ACM Int. Conf. on Nanoscale
  Computing and Comm.}, pp.~1--7, 2020.
\newblock {A}rt. no. 3, doi:
  {\href{https://doi.org/10.1145/3411295.3411299}{10.1145/3411295.3411299}}.

\bibitem{okaie2018leader}
Y.~Okaie, S.~Ishiyama, and T.~Hara, ``Leader-follower-amplifier based mobile
  molecular communication systems for cooperative drug delivery,'' in {\em IEEE
  Glob. Comm. Conf. (GLOBECOM)}, pp.~206--212, 2018.
\newblock doi:
  {\href{https://doi.org/10.1109/GLOCOM.2018.8647185}{10.1109/GLOCOM.2018.8647185}}.

\bibitem{poor1994elements}
H.~V. Poor, ``Elements of hypothesis testing,'' in {\em An introduction to
  signal detection and estimation}, pp.~5--44, Springer, 1994.
\newblock doi:
  {\href{https://doi.org/10.1007/978-1-4757-2341-0_2}{10.1007/978-1-4757-2341-0\_2}}.

\bibitem{wald2004sequential}
A.~Wald, {\em Sequential analysis}.
\newblock Courier Corporation, 2004.

\bibitem{young1998sequential}
L.~J. Young and J.~H. Young, ``Sequential hypothesis testing,'' in {\em
  Statistical Ecology}, pp.~153--190, Springer, 1998.

\bibitem{naghshvar2013active}
M.~Naghshvar and T.~Javidi, ``Active sequential hypothesis testing,'' {\em The
  Annals of Statistics}, vol.~41, no.~6, pp.~2703--2738, 2013.

\bibitem{kartik2019active}
D.~Kartik, A.~Nayyar, and U.~Mitra, ``Active hypothesis testing: beyond
  chernoff-stein,'' in {\em IEEE Int. Symp. on Inf. Theory (ISIT)},
  pp.~897--901, 2019.
\newblock doi:
  {\href{https://doi.org/10.1109/ISIT.2019.8849436}{10.1109/ISIT.2019.8849436}}.

\bibitem{chernoff1959sequential}
H.~Chernoff, ``Sequential design of experiments,'' {\em The Annals of
  Mathematical Statistics}, vol.~30, no.~3, pp.~755--770, 1959.

\bibitem{nitinawarat2012controlled}
S.~Nitinawarat, G.~K. Atia, and V.~V. Veeravalli, ``Controlled sensing for
  hypothesis testing,'' in {\em IEEE Int. Conf. on Acoustics, Speech and Sig.
  Process. (ICASSP)}, pp.~5277--5280, 2012.
\newblock doi:
  {\href{https://doi.org/10.1109/ICASSP.2012.6289111}{10.1109/ICASSP.2012.6289111}}.

\bibitem{kartik2020testing}
D.~Kartik, A.~Nayyar, and U.~Mitra, ``Testing for anomalies: Active strategies
  and non-asymptotic analysis,'' in {\em IEEE Int. Symp. on Inf. Theory
  (ISIT)}, pp.~1277--1282, 2020.
\newblock doi:
  {\href{https://doi.org/10.1109/ISIT44484.2020.9174399}{10.1109/ISIT44484.2020.9174399}}.

\bibitem{cohen2015active}
K.~Cohen and Q.~Zhao, ``Active hypothesis testing for anomaly detection,'' {\em
  IEEE Trans. on Inf. Theory}, vol.~61, no.~3, pp.~1432--1450, 2015.

\bibitem{davies1987hypothesis}
R.~B. Davies, ``Hypothesis testing when a nuisance parameter is present only
  under the alternative,'' {\em Biometrika}, vol.~74, no.~1, pp.~33--43, 1987.

\bibitem{geyer2018bloodvoyagers}
R.~Geyer, M.~Stelzner, F.~B{\"u}ther, and S.~Ebers, ``Bloodvoyagers: Simulation
  of the work environment of medical nanobots,'' in {\em Proc. of the 5th ACM
  Int. Conf. on Nanoscale Computing and Comm.}, pp.~1--6, 2018.
\newblock {A}rt. no. 5, doi:
  {\href{https://doi.org/10.1145/3233188.3233196}{10.1145/3233188.3233196}}.

\bibitem{ishiyama2018epidemic}
S.~Ishiyama, T.~Nakano, Y.~Okaie, and T.~Hara, ``Epidemic information
  dissemination in mobile molecular communication systems,'' in {\em IEEE Glob.
  Comm. Conf. (GLOBECOM)}, pp.~1--7, 2018.
\newblock doi:
  {\href{https://doi.org/10.1109/GLOCOM.2018.8648028}{10.1109/GLOCOM.2018.8648028}}.

\bibitem{mesquita2012jump}
A.~R. Mesquita and J.~P. Hespanha, ``Jump control of probability densities with
  applications to autonomous vehicle motion,'' {\em IEEE Trans. on Automatic
  Cont.}, vol.~57, no.~10, pp.~2588--2598, 2012.

\bibitem{huang2020channel}
X.~Huang, Y.~Fang, A.~Noel, and N.~Yang, ``Channel characterization for 1-d
  molecular communication with two absorbing receivers,'' {\em IEEE Comm.
  Letters}, vol.~24, no.~6, pp.~1150--1154, 2020.

\bibitem{bao2021relative}
X.~Bao, Q.~Shen, Y.~Zhu, and W.~Zhang, ``Relative localization for silent
  absorbing target in diffusive molecular communication system,'' {\em IEEE
  Internet of Things J.}, vol.~9, no.~7, pp.~5009--5018, 2022.

\bibitem{qiu2015under}
S.~Qiu, N.~Farsad, Y.~Dong, A.~Eckford, and W.~Guo, ``Under-water molecular
  signalling: A hidden transmitter and absent receivers problem,'' in {\em IEEE
  Int. Conf. on Comm. (ICC)}, pp.~1085--1090, 2015.
\newblock doi:
  {\href{https://doi.org/10.1109/ICC.2015.7248467}{10.1109/ICC.2015.7248467}}.

\bibitem{farsad2017novel}
N.~Farsad, D.~Pan, and A.~Goldsmith, ``A novel experimental platform for
  in-vessel multi-chemical molecular communications,'' in {\em IEEE Glob. Comm.
  Conf. (GLOBECOM)}, pp.~1--6, 2017.
\newblock doi:
  {\href{https://doi.org/10.1109/GLOCOM.2017.8255058}{10.1109/GLOCOM.2017.8255058}}.

\bibitem{regonesi2020relative}
E.~Regonesi, M.~Rapisarda, M.~Magarini, and M.~Ferrari, ``Relative angle
  estimation of an interferer in a diffusion-based molecular communication
  system,'' in {\em Proc. of the 7th ACM Int. Conf. on Nanoscale Computing and
  Comm.}, pp.~1--7, 2020.
\newblock {A}rt. no. 14, doi:
  {\href{https://doi.org/10.1145/3411295.3411310}{10.1145/3411295.3411310}}.

\bibitem{alpay2000model}
M.~E. Alpay and M.~H. Shor, ``Model-based solution techniques for the source
  localization problem,'' {\em IEEE Trans. on Cont. Sys. Tech.}, vol.~8, no.~6,
  pp.~895--904, 2000.

\bibitem{zoofaghari2021semi}
M.~Zoofaghari, H.~Arjmandi, A.~Etemadi, and I.~Balasingham, ``A semi-analytical
  method for channel modeling in diffusion-based molecular communication
  networks,'' {\em IEEE Trans. on Comm.}, vol.~69, no.~6, pp.~3957--3970, 2021.

\bibitem{jornet2012joint}
J.~M. Jornet and I.~F. Akyildiz, ``Joint energy harvesting and communication
  analysis for perpetual wireless nanosensor networks in the terahertz band,''
  {\em IEEE Trans. on Nanotech.}, vol.~11, no.~3, pp.~570--580, 2012.

\bibitem{berg2000motile}
H.~Berg, ``Motile behavior of bacteria,'' {\em Physics today}, vol.~53, no.~1,
  pp.~24--29, 2000.

\bibitem{bazylinski2004magnetosome}
D.~A. Bazylinski and R.~B. Frankel, ``Magnetosome formation in prokaryotes,''
  {\em Nature Reviews Microbiology}, vol.~2, no.~3, pp.~217--230, 2004.

\bibitem{dai2016programmable}
B.~Dai, J.~Wang, Z.~Xiong, X.~Zhan, W.~Dai, C.-C. Li, S.-P. Feng, and J.~Tang,
  ``Programmable artificial phototactic microswimmer,'' {\em Nature Nanotech.},
  vol.~11, no.~12, pp.~1087--1092, 2016.

\bibitem{zhang2016human}
Z.~Zhang, J.~Liu, J.~Meriano, C.~Ru, S.~Xie, J.~Luo, and Y.~Sun, ``Human sperm
  rheotaxis: a passive physical process,'' {\em Scientific Reports}, vol.~6,
  no.~1, 2016.
\newblock {A}rt. no. 23553, doi:
  {\href{https://doi.org/10.1038/srep23553}{10.1038/srep23553}}.

\bibitem{liebchen2018viscotaxis}
B.~Liebchen, P.~Monderkamp, B.~Ten~Hagen, and H.~L{\"o}wen, ``Viscotaxis:
  Microswimmer navigation in viscosity gradients,'' {\em Physical Review
  Letters}, vol.~120, no.~20, 2018.
\newblock {A}rt. no. 208002, doi:
  {\href{https://doi.org/10.1103/PhysRevLett.120.208002}{10.1103/PhysRevLett.120.208002}}.

\bibitem{pane2019imaging}
S.~Pan{\'e}, J.~Puigmart{\'\i}-Luis, C.~Bergeles, X.-Z. Chen, E.~Pellicer,
  J.~Sort, V.~Po{\v{c}}epcov{\'a}, A.~Ferreira, and B.~J. Nelson, ``Imaging
  technologies for biomedical micro-and nanoswimmers,'' {\em Advanced Materials
  Tech.}, vol.~4, no.~4, 2019.
\newblock {A}rt. no. 1800575, doi:
  {\href{https://doi.org/10.1002/admt.201800575}{10.1002/admt.201800575}}.

\bibitem{francis2015self}
W.~Francis, C.~Fay, L.~Florea, and D.~Diamond, ``Self-propelled chemotactic
  ionic liquid droplets,'' {\em Chem. Comm.}, vol.~51, no.~12, pp.~2342--2344,
  2015.

\bibitem{kim2011extracellular}
S.-H. Kim, J.~Turnbull, and S.~Guimond, ``Extracellular matrix and cell
  signalling: the dynamic cooperation of integrin, proteoglycan and growth
  factor receptor.,'' {\em The J. of Endocrinology}, vol.~209, no.~2,
  pp.~139--151, 2011.

\bibitem{huang2018bioinspired}
H.-W. Huang, S.~Lyttle, and B.~J. Nelson, ``Bioinspired navigation in shape
  morphing micromachines for autonomous targeted drug delivery,'' in {\em IEEE
  Int. Conf. on Soft Robotics (RoboSoft)}, pp.~13--18, 2018.
\newblock doi:
  {\href{https://doi.org/10.1109/ROBOSOFT.2018.8404890}{10.1109/ROBOSOFT.2018.8404890}}.

\bibitem{hu2019plume}
H.~Hu, S.~Song, and C.~P. Chen, ``Plume tracing via model-free reinforcement
  learning method,'' {\em IEEE Trans. on Neural Netw. and Learning Sys.},
  vol.~30, no.~8, pp.~2515--2527, 2019.

\bibitem{ongaro2017autonomous}
F.~Ongaro, S.~Scheggi, C.~Yoon, F.~Van~den Brink, S.~H. Oh, D.~H. Gracias, and
  S.~Misra, ``Autonomous planning and control of soft untethered grippers in
  unstructured environments,'' {\em J. of micro-bio robotics}, vol.~12, no.~1,
  pp.~45--52, 2017.

\bibitem{sanchez2014magnetic}
A.~S{\'a}nchez, V.~Magdanz, O.~G. Schmidt, and S.~Misra, ``Magnetic control of
  self-propelled microjets under ultrasound image guidance,'' in {\em 5th IEEE
  RAS/EMBS Int. Conf. on Biomedical Robotics and Biomechatronics},
  pp.~169--174, 2014.
\newblock doi:
  {\href{https://doi.org/10.1109/BIOROB.2014.6913771}{10.1109/BIOROB.2014.6913771}}.

\bibitem{scheggi2017magnetic}
S.~Scheggi, K.~K.~T. Chandrasekar, C.~Yoon, B.~Sawaryn, G.~van~de Steeg, D.~H.
  Gracias, and S.~Misra, ``Magnetic motion control and planning of untethered
  soft grippers using ultrasound image feedback,'' in {\em IEEE Int. Conf. on
  Robotics and Automation (ICRA)}, pp.~6156--6161, 2017.
\newblock doi:
  {\href{https://doi.org/10.1109/ICRA.2017.7989730}{10.1109/ICRA.2017.7989730}}.

\bibitem{curreli2008real}
M.~Curreli, R.~Zhang, F.~N. Ishikawa, H.-K. Chang, R.~J. Cote, C.~Zhou, and
  M.~E. Thompson, ``Real-time, label-free detection of biological entities
  using nanowire-based fets,'' {\em IEEE Trans. on Nanotech.}, vol.~7, no.~6,
  pp.~651--667, 2008.

\bibitem{jornet2019optogenomic}
J.~M. Jornet, Y.~Bae, C.~R. Handelmann, B.~Decker, A.~Balcerak, A.~Sangwan,
  P.~Miao, A.~Desai, L.~Feng, E.~K. Stachowiak, {\em et~al.}, ``Optogenomic
  interfaces: Bridging biological networks with the electronic digital world,''
  {\em Proc. of the IEEE}, vol.~107, no.~7, pp.~1387--1401, 2019.

\bibitem{maduraiveeran2018electrochemical}
G.~Maduraiveeran, M.~Sasidharan, and V.~Ganesan, ``Electrochemical sensor and
  biosensor platforms based on advanced nanomaterials for biological and
  biomedical applications,'' {\em Biosensors and Bioelectronics}, vol.~103,
  pp.~113--129, 2018.

\bibitem{grebenstein2019molecular}
L.~Grebenstein, J.~Kirchner, W.~Wicke, A.~Ahmadzadeh, V.~Jamali, G.~Fischer,
  R.~Weigel, A.~Burkovski, and R.~Schober, ``A molecular communication testbed
  based on proton pumping bacteria: Methods and data,'' {\em IEEE Trans. on
  Mol., Bio. and Multi-Scale Comm.}, vol.~5, no.~1, pp.~56--62, 2019.

\bibitem{krishnaswamy2013time}
B.~Krishnaswamy, C.~M. Austin, J.~P. Bardill, D.~Russakow, G.~L. Holst, B.~K.
  Hammer, C.~R. Forest, and R.~Sivakumar, ``Time-elapse communication:
  Bacterial communication on a microfluidic chip,'' {\em IEEE Trans. on Comm.},
  vol.~61, no.~12, pp.~5139--5151, 2013.

\bibitem{de2013communications}
E.~De~Leo, L.~Donvito, L.~Galluccio, A.~Lombardo, G.~Morabito, and L.~M.
  Zanoli, ``Communications and switching in microfluidic systems: Pure
  hydrodynamic control for networking labs-on-a-chip,'' {\em IEEE Trans. on
  Comm.}, vol.~61, no.~11, pp.~4663--4677, 2013.

\bibitem{abbasi2018controlled}
N.~A. Abbasi, D.~Lafci, and O.~B. Akan, ``Controlled information transfer
  through an in vivo nervous system,'' {\em Scientific reports}, vol.~8, no.~1,
  2018.
\newblock {A}rt. no. 2298, doi:
  {\href{https://doi.org/10.1038/s41598-018-20725-2}{10.1038/s41598-018-20725-2}}.

\bibitem{farsad2013tabletop}
N.~Farsad, W.~Guo, and A.~W. Eckford, ``Tabletop molecular communication: Text
  messages through chemical signals,'' {\em PloS One}, vol.~8, no.~12, 2013.
\newblock {A}rt. no. e82935, doi:
  {\href{https://doi.org/10.1371/journal.pone.0082935}{10.1371/journal.pone.0082935}}.

\bibitem{koo2016molecular}
B.-H. Koo, C.~Lee, H.~B. Yilmaz, N.~Farsad, A.~Eckford, and C.-B. Chae,
  ``Molecular mimo: From theory to prototype,'' {\em IEEE J. on Selected Areas
  in Comm.}, vol.~34, no.~3, pp.~600--614, 2016.

\bibitem{tataria20216g}
H.~Tataria, M.~Shafi, A.~F. Molisch, M.~Dohler, H.~Sj{\"o}land, and
  F.~Tufvesson, ``6g wireless systems: Vision, requirements, challenges,
  insights, and opportunities,'' {\em Proc. of the IEEE}, vol.~109, no.~7,
  pp.~1166--1199, 2021.

\bibitem{abdulfattah2019performance}
A.~N. Abdulfattah, C.~C. Tsimenidis, B.~Z. Al-Jewad, and A.~Yakovlev,
  ``Performance analysis of mics-based rf wireless power transfer system for
  implantable medical devices,'' {\em IEEE Access}, vol.~7, pp.~11775--11784,
  2019.

\bibitem{lee2015highly}
D.-K. Lee, J.-H. Kang, J.-S. Lee, H.-S. Kim, C.~Kim, J.~H. Kim, T.~Lee, J.-H.
  Son, Q.-H. Park, and M.~Seo, ``Highly sensitive and selective sugar detection
  by terahertz nano-antennas,'' {\em Scientific Reports}, vol.~5, no.~1, 2015.
\newblock {A}rt. no. 15459, doi:
  {\href{https://doi.org/10.1038/srep15459}{10.1038/srep15459}}.

\bibitem{kisseleff2016magnetic}
S.~Kisseleff, R.~Schober, and W.~H. Gerstacker, ``Magnetic nanoparticle based
  interface for molecular communication systems,'' {\em IEEE Comm. Letters},
  vol.~21, no.~2, pp.~258--261, 2016.

\bibitem{ulbrich2016targeted}
K.~Ulbrich, K.~Hola, V.~Subr, A.~Bakandritsos, J.~Tucek, and R.~Zboril,
  ``Targeted drug delivery with polymers and magnetic nanoparticles: covalent
  and noncovalent approaches, release control, and clinical studies,'' {\em
  Chem. Reviews}, vol.~116, no.~9, pp.~5338--5431, 2016.

\bibitem{osmekhina2018controlled}
E.~Osmekhina, C.~Jonkergouw, G.~Schmidt, F.~Jahangiri, V.~Jokinen,
  S.~Franssila, and M.~B. Linder, ``Controlled communication between physically
  separated bacterial populations in a microfluidic device,'' {\em Comm.
  Biology}, vol.~1, no.~1, 2018.
\newblock {A}rt. no. 97, doi:
  {\href{https://doi.org/10.1038/s42003-018-0102-y}{10.1038/s42003-018-0102-y}}.

\bibitem{bi2021survey}
D.~Bi, A.~Almpanis, A.~Noel, Y.~Deng, and R.~Schober, ``A survey of molecular
  communication in cell biology: Establishing a new hierarchy for
  interdisciplinary applications,'' {\em IEEE Comm. Surveys \& Tutorials},
  vol.~23, no.~3, pp.~1494--1545, 2021.

\bibitem{maresca2018biomolecular}
D.~Maresca, A.~Lakshmanan, M.~Abedi, A.~Bar-Zion, A.~Farhadi, G.~J. Lu, J.~O.
  Szablowski, D.~Wu, S.~Yoo, and M.~G. Shapiro, ``Biomolecular ultrasound and
  sonogenetics,'' {\em Annu. Review of Chem. and Biomolecular Eng.}, vol.~9,
  pp.~229--252, 2018.

\bibitem{zafar2021systematic}
S.~Zafar, M.~Nazir, T.~Bakhshi, H.~A. Khattak, S.~Khan, M.~Bilal, K.-K.~R.
  Choo, K.-S. Kwak, and A.~Sabah, ``A systematic review of bio-cyber interface
  technologies and security issues for internet of bio-nano things,'' {\em IEEE
  Access}, vol.~9, pp.~93529--93566, 2021.

\bibitem{usman2018security}
M.~Usman, M.~R. Asghar, I.~S. Ansari, and M.~Qaraqe, ``Security in wireless
  body area networks: From in-body to off-body communications,'' {\em IEEE
  Access}, vol.~6, pp.~58064--58074, 2018.

\bibitem{al2020intelligence}
F.~Al-Turjman, ``Intelligence and security in big 5g-oriented iont: An
  overview,'' {\em Future Generation Comput. Sys.}, vol.~102, pp.~357--368,
  2020.

\bibitem{lee2017machine}
C.~Lee, H.~B. Yilmaz, C.-B. Chae, N.~Farsad, and A.~Goldsmith, ``Machine
  learning based channel modeling for molecular mimo communications,'' in {\em
  IEEE 18th Int. Workshop on Sig. Process. Advances in Wireless Comm. (SPAWC)},
  pp.~1--5, 2017.
\newblock doi:
  {\href{https://doi.org/10.1109/SPAWC.2017.8227765}{10.1109/SPAWC.2017.8227765}}.

\bibitem{mohamed2019model}
S.~Mohamed, J.~Dong, A.~Junejo, {\em et~al.}, ``Model-based: End-to-end
  molecular communication system through deep reinforcement learning auto
  encoder,'' {\em IEEE Access}, vol.~7, pp.~70279--70286, 2019.

\bibitem{jing2021recent}
T.~Jing, Q.-H. Meng, and H.~Ishida, ``Recent progress and trend of robot odor
  source localization,'' {\em IEEJ Trans. on Electrical and Electronic Eng.},
  vol.~16, no.~7, pp.~938--953, 2021.

\bibitem{zhang2009characterizing}
L.~Zhang, J.~J. Abbott, L.~Dong, K.~E. Peyer, B.~E. Kratochvil, H.~Zhang,
  C.~Bergeles, and B.~J. Nelson, ``Characterizing the swimming properties of
  artificial bacterial flagella,'' {\em Nano Letters}, vol.~9, no.~10,
  pp.~3663--3667, 2009.

\bibitem{chen2019unimodality}
J.~Chen and U.~Mitra, ``Unimodality-constrained matrix factorization for
  non-parametric source localization,'' {\em IEEE Trans. on Sig. Process.},
  vol.~67, no.~9, pp.~2371--2386, 2019.

\bibitem{yu2021recent}
S.~Yu, Y.~Cai, Z.~Wu, and Q.~He, ``Recent progress on motion control of
  swimming micro/nanorobots,'' {\em View}, vol.~2, no.~5, 2021.
\newblock {A}rt. no. 20200113, doi:
  {\href{https://doi.org/10.1002/VIW.20200113}{10.1002/VIW.20200113}}.

\bibitem{kober2010imitation}
J.~Kober and J.~Peters, ``Imitation and reinforcement learning,'' {\em IEEE
  Robotics \& Automation Mag.}, vol.~17, no.~2, pp.~55--62, 2010.

\bibitem{mac2016heuristic}
T.~T. Mac, C.~Copot, D.~T. Tran, and R.~De~Keyser, ``Heuristic approaches in
  robot path planning: A survey,'' {\em Robotics and Autonomous Sys.}, vol.~86,
  pp.~13--28, 2016.

\bibitem{chen2006smooth}
X.~Chen and Y.~Li, ``Smooth formation navigation of multiple mobile robots for
  avoiding moving obstacles,'' {\em Int. J. of Cont., Automation, and Sys.},
  vol.~4, no.~4, pp.~466--479, 2006.

\end{thebibliography}
	
	\cleardoublepage
\end{document}